\documentclass[fleqn,usenatbib]{mnras}

\usepackage[T1]{fontenc}

\DeclareRobustCommand{\VAN}[3]{#2}
\let\VANthebibliography\thebibliography
\def\thebibliography{\DeclareRobustCommand{\VAN}[3]{##3}\VANthebibliography}

\usepackage{graphicx}	
\usepackage{amsmath}	
\usepackage{amssymb}	
\usepackage{subcaption}
\usepackage{array} 
\usepackage{newtxtext,newtxmath}

\newcommand{\mjup}{\,$M_{\rm J}$}
\newcommand{\rjup}{\,$R_{\rm J}$}
\newcommand{\msun}{\,$M_{\odot}$}	
\newcommand{\rsun}{\,$R_{\odot}$}
\newcommand{\tess}{{\it TESS}}
\newcommand{\gaia}{{\it Gaia}}
\newcommand{\ngts}{{NGTS}}
\newcommand{\saao}{{SAAO}}
\newcommand{\speculooss}{{SPECULOOS-S}}
\newcommand{\trappists}{{TRAPPIST-S}}
\newcommand{\harps}{{HARPS}}

\newcommand{\PAR}{Paranal Observatory}
\newcommand{\teff}{{T$_{\rm eff}$}}
\newcommand{\logg}{{$\log$ g}}
\newcommand{\feh}{[Fe/H]}
\newcommand{\system}{{\rm NGTS-28AB}}
\newcommand{\systemt}{{\rm NGTS-28Ab}}
\newcommand{\systemA}{{\rm NGTS-28A}}
\newcommand{\systemB}{{\rm NGTS-28B}}

\title[\systemt]{NGTS-28Ab: A short period transiting brown dwarf}
\author[Henderson et. al.]{Beth A. Henderson$^{1}$ \thanks{E-mail: bah26@leicester.ac.uk}, 
Sarah L. Casewell$^{1}$, 
Michael R. Goad$^{1}$,
Jack S. Acton$^{1}$,
Maximilian N. G\"unther$^{2}$,
\newauthor Louise D. Nielsen$^{3}$
Matthew R. Burleigh$^{1}$,
Claudia Belardi$^{1}$,
Rosanna H. Tilbrook$^{1}$,
Oliver Turner$^{4}$,
\newauthor Steve B. Howell$^{5}$,
Catherine A. Clark$^{6,7}$,
Colin Littlefield$^{5,8}$,
Khalid Barkaoui$^{9,10,11}$
Douglas R. Alves$^{12,13}$,
\newauthor David R. Anderson$^{14,15}$, 
Daniel Bayliss$^{14,15}$,
Francois Bouchy$^{4}$,
Edward~M.~Bryant$^{16}$,
George Dransfield$^{17}$,
\newauthor Elsa Ducrot$^{18,19}$,
Philipp Eigm\"uller$^{20}$,
Samuel Gill$^{14,15}$, 
Edward Gillen$^{21,22}$,
Michaël Gillon$^{9}$,
Faith~Hawthorn$^{14,15}$,
\newauthor Matthew J. Hooton$^{32}$,
James A. G. Jackman$^{23,14,15}$,
Emmanuel Jehin$^{24}$,
James S. Jenkins$^{25,13}$,
Alicia Kendall$^{1}$,
\newauthor Monika Lendl$^{4}$,
James McCormac$^{14,15}$,
Maximiliano Moyano$^{26}$,
Peter Pihlmann Pedersen$^{27,32}$,
\newauthor Francisco J. Pozuelos$^{28}$,
Gavin Ramsay$^{29}$,
Ramotholo~R.~Sefako$^{30}$,
Mathilde Timmermans$^{9}$,
\newauthor Amaury H. M. J. Triaud$^{17}$,
Stephane Udry$^{4}$,
Jose I. Vines$^{26}$,
Christopher A. Watson$^{31}$,
Richard~G.~West$^{14,15}$, 
\newauthor Peter J. Wheatley$^{14,15}$,
Sebastián Zúñiga-Fernández$^{9}$\\
$^{1}$ School of Physics and Astronomy, University of Leicester, University Road, Leicester LE1 7RH, UK\\
$^{2}$European Space Agency (ESA), European Space Research and Technology Centre (ESTEC), Keplerlaan 1, 2201 AZ Noordwijk, The Netherlands\\
$^{3}$ European Southern Observatory, Karl-Schwarzschildstr. 2, D-85748 Garching bei  M{\"u}nchen, Germany\\
$^{4}$ D\'epartement d'astronomie, Universit\'e de Genève, 51 chemin Pegasi, 1290 Sauverny, Switzerland\\
$^{5}$ NASA Ames Research Center, Moffett Field, CA 94035, USA\\
$^{6}$ Jet Propulsion Laboratory, California Institute of Technology, Pasadena, CA 91109 USA\\
$^{7}$ NASA Exoplanet Science Institute, IPAC, California Institute of Technology, Pasadena, CA 91125 USA\\
$^{8}$ Bay Area Environmental Research Institute, Moffett Field, CA 94035, USA\\
$^{9}$Astrobiology Research Unit, Universit\'e de Li\`ege, All\'ee du 6 Ao\^ut 19C, B-4000 Li\`ege, Belgium\\
$^{10}$Department of Earth, Atmospheric and Planetary Science, Massachusetts Institute of Technology, 77 Massachusetts Avenue, Cambridge, MA 02139, USA\\
$^{11}$Instituto de Astrof\'isica de Canarias (IAC), Calle V\'ia L\'actea s/n, 38200, La Laguna, Tenerife, Spain\\
$^{12}$Departamento de Astronom\'ia, Universidad de Chile, Casilla 36-D, Santiago, Chile\\
$^{13}$Centro de Astrof\'isica y Tecnolog\'ias Afines (CATA), Casilla 36-D, Santiago, Chile\\
$^{14}$Centre for Exoplanets and Habitability, University of Warwick, Gibbet Hill Road, Coventry, CV4 7AL, UK\\
$^{15}$Dept. of Physics, University of Warwick, Gibbet Hill Road, Coventry, CV4 7AL, UK\\
$^{16}$Mullard Space Science Laboratory, University College London, Holmbury St Mary, Dorking, Surrey, RH5 6NT, UK \\
$^{17}$School of Physics \& Astronomy, University of Birmingham, Edgbaston, Birmingham B15 2TT, UK\\
$^{18}$Paris Region Fellow, Marie Sklodowska-Curie Action\\
$^{19}$AIM, CEA, CNRS, Universit\'e Paris-Saclay, Universit\'e de Paris, F-91191 Gif-sur-Yvette, France\\
$^{20}$Institute of Planetary Research, German Aerospace Center, Rutherfordstrasse 2, 12489 Berlin, Germany\\
$^{21}$Astronomy Unit, Queen Mary University of London, Mile End Road, London E1 4NS, UK\\
$^{22}$Astrophysics Group, Cavendish Laboratory, J.J. Thomson Avenue, Cambridge, CB3 0HE, UK\\
$^{23}$School of Earth and Space Exploration, Arizona State University, Tempe, AZ, 85287, USA\\
$^{24}$STAR Institute, University of Liège, Allée du 6 Août 19c, 4000 Liège, Belgium\\
$^{25}$N\'ucleo de Astronom\'ia, Facultad de Ingenier\'ia y Ciencias, Universidad Diego Portales, Av. Ej\'ercito 441, Santiago, Chile\\
$^{26}$Instituto de Astronom\'ia, Universidad Cat\'olica del Norte,
Angamos 0610, 1270709, Antofagasta, Chile\\
$^{27}$Department of Physics, ETH Zurich, Wolfgang-Pauli-Strasse 2, CH-8093 Zurich, Switzerland\\
$^{28}$Instituto de Astrof\'isica de Andaluc\'ia (IAA-CSIC), Glorieta de la Astronom\'ia s/n, 18008 Granada, Spain\\
$^{29}$Armagh Observatory \& Planetarium, College Hill, Armagh, BT61 9DG, UK\\
$^{30}$South African Astronomical Observatory, P.O Box 9, Observatory 7935, Cape Town, South Africa\\
$^{31}$Astrophysics Research Centre, School of Mathematics and Physics, Queen’s University Belfast, BT7 1NN Belfast, UK \\
$^{32}$ Cavendish Laboratory, JJ Thomson Avenue, Cambridge CB3 0HE, UK\\
}

\date{Accepted XXX. Received YYY; in original form ZZZ}

\pubyear{2022}

\begin{document}
\label{firstpage}
\pagerange{\pageref{firstpage}--\pageref{lastpage}}
\maketitle

\begin{abstract}

We report the discovery of a brown dwarf orbiting a M1 host star. We first identified the brown dwarf within the Next Generation Transit Survey data, with supporting observations found in \tess\ sectors 11 and 38. We confirmed the discovery with follow-up photometry from the South African Astronomical Observatory, \speculooss, and \trappists, and radial velocity measurements from \harps, which allowed us to characterise the system. We find an orbital period of $\sim{1.25}$~d, a mass of $69.0^{+5.3}_{-4.8}$ \mjup, close to the Hydrogen burning limit, and a radius of $0.95\pm0.05$ \rjup. We determine the age to be  $>$0.5 Gyr, using model isochrones, which is found to be in agreement with SED fitting within errors. \systemt\ is one of the shortest period systems found within the brown dwarf desert, as well as one of the highest mass brown dwarfs that transits an M dwarf. This makes \systemt\ another important discovery within this scarcely populated region. 

\end{abstract}




\section{Introduction}

Brown dwarfs (BDs) are intriguing objects, located between high mass exoplanets and low mass main sequence stars. Their masses are canonically defined to lie in the 13 \mjup\ to 80 \mjup\ range (\citealt{burgasser, burrows93, baraffe02, triaud17}). BDs never become massive enough to burn hydrogen in their core, supporting themselves through electron degeneracy pressure \citep{whitworth18} although some BDs are massive enough to burn deuterium (\citealt{whitworth18,bate02}). The first BDs were not discovered until 1995 (Gliese 229B: \citealp{gliese}; Teide 1: \citealp{teide}) but candidates were found with radial velocity methods previous to this (HD114762: \citealp{rvbd}). However, thousands have since been observed with the majority being isolated BDs. 

Due to their degenerate nature, BDs exhibit an age-radius-mass-temperature degeneracy: their radii contract over their lifetimes as they cool \citep{burrows93}. If the radius can be determined, how the BD evolves over their lifetimes can be predicted and therefore can be used as an age estimate (e.g. \citealt{marley21, baraffe03, baraffe15}). However, most BDs do not have accurately measured radii due to being isolated objects. BDs found in eclipsing binary systems therefore offer an opportunity to directly measure their radii, breaking this degeneracy.

Only two BD-BD eclipsing binary systems are known to date (\citealt{stassun06, triaud20}), and searching for BDs around main sequence stars is an obvious place to look.  Fortunately, BD radii are similar to that of Jupiter \citep{burrows93} making them as detectable as Jupiter-sized exoplanets in exoplanet surveys (e.g. \citealt{19b,  grieves21, carmichael21}).

However, the radii of BDs can also be affected by exterior factors to the BD, such as interactions with their host star through irradiation \citep{casewell20}. Irradiation will affect the BD atmosphere by heating it, and is also thought to cause a BD's radius to be inflated (e.g. Kelt-1b: \citet{kelt1b}, WD1032+11: \citet{wd1032}) in the same way that is seen for hot Jupiters (e.g. \citealt{thorngren, tilbrook,alves22}).

To date, 34 transiting BDs have been discovered orbiting within 3~AU of a main sequence star (e.g. \citealt{carmichael22} and references therein) which is small compared to the thousands of isolated BDs. This number is also much lower than would be expected when compared to the population of low mass eclipsing binary systems or the hot Jupiter exoplanet population. This local minima between these populations is called the `brown dwarf desert' \citep{grether06}. The desert is thought to be caused by the differences in formation mechanisms between exoplanets and stars (\citealt{grieves21,mage14}) but is still not well understood.  

Since it was first identified, the BD desert is still scarcely populated, even including recent works using large photometric surveys (see Figure 7 of \citealp{triaud17}). Multiple studies have looked into characteristics of objects within the BD desert. \citet{mage14} discussed whether there is a relationship between mass of the BD and eccentricity. They considered all known exoplanets (at the time), along with all BD candidates around FGK stars (spanning 0-100 \mjup) and determined that they could separate BDs at a mass of 42.5 \mjup\ into two populations. Those which are more massive ($> 42.5$ \mjup) appeared to display a period-eccentricity distribution with a circularisation limit of $\sim$12 d, similar to stellar binaries. This is due to objects with shorter periods (up to 12 d) having circular orbits, and those with longer orbits having a flat but wide distribution of eccentricities (as seen in \citealt{mage14,Raghavan10}). The lower mass population displayed a period-eccentricity trend similar to exoplanets, where higher mass objects (up to the 42.5 \mjup\ split) have lower eccentricities. \citet{mage14} determined that these trends point to objects with masses above 42.5 \mjup\ forming via stellar formation mechanisms, instead of via protoplanetary discs \citep{mage14}. 

Other studies have since sought to find this split population in eccentricity distributions within their observed sample of BDs. \citet{grieves21} state that they do not see this trend or evidence of \citet{mage14}'s split population at 42.5 \mjup. Although they note their small sample size as they only include transiting BDs and low mass stars (12.9 \mjup\ to 149.8 \mjup, \citealt{grieves21}) and not high mass exoplanets. \citet{carmichael19} also does not find a split at 42.5 \mjup. However, they conclude it is due to there not being a split in the population at all, as opposed to a lack of objects. This illustrates a clear need for more objects to populate this region, to aid in the understanding of whether there is a mass split at 42.5 \mjup\ due to formation mechanisms.

Metallicity (\feh) has also been looked at within the brown dwarf sample. \citet{mage14} did not find any evidence of a split population at 42.5 \mjup\ when comparing \feh, and found a mean \feh\ of $-0.05$. \citet{jenkins15} compared the mean \feh\ values for BDs, stellar binaries and exoplanets. They found that BDs and stellar binaries have comparable sub-solar mean \feh\ values, where as exoplanets have an above solar mean \feh. The brown dwarfs within the \citet{jenkins15} sample were mainly found with radial velocity methods, rather than the transiting sample as is done in this work.

In this paper we present the discovery of \systemt, a transiting BD lying within the `BD desert'. In section \ref{sec:obs} we detail the various photometric and spectral observations obtained. In Section \ref{sec:analysis}, the methods used to analyse both the host star and BD are described, along with the probable common proper motion companion, \systemB. Section \ref{sec:discussion} discusses the importance of this discovery in terms of its position within the BD desert and analyses the eccentricity and age of \systemt\ further.

\section{Observations}\label{sec:obs}

\begin{figure}
    \centering
    \includegraphics[width=\linewidth, trim =2.5cm 8cm 3cm 8cm, clip]{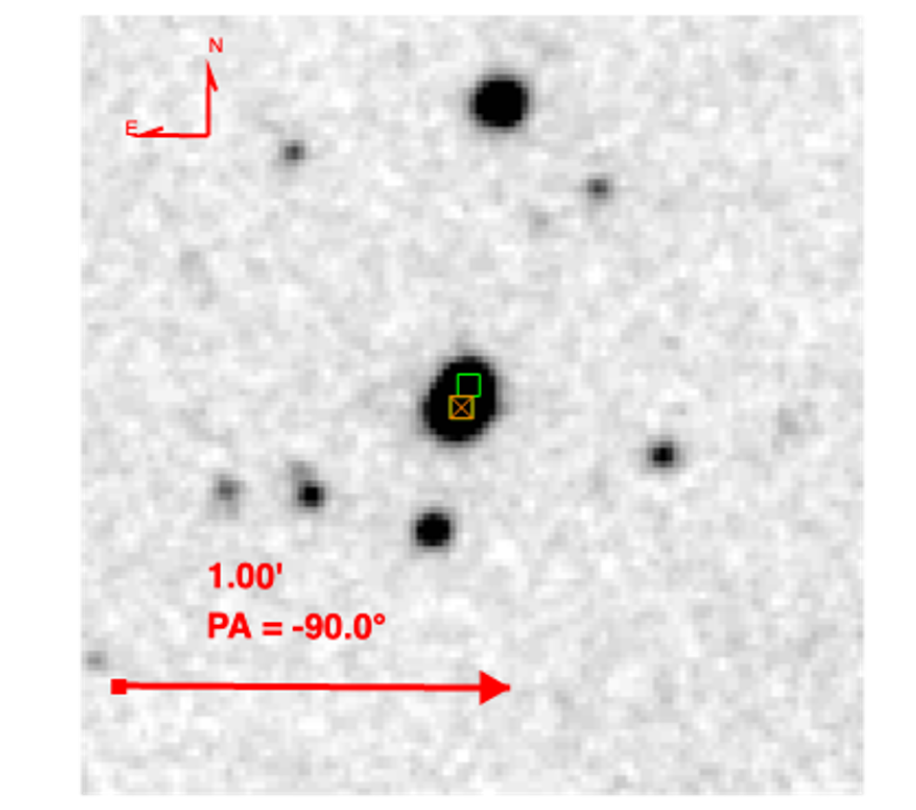}
    \caption{DSS image showing \systemA\ (orange box with cross) and \systemB\ (green box) blended on sky, along with other nearby objects. North and East are shown on the figure in the top left corner as well as a 1~arcmin position angle (PA) arrow for scale.}
    \label{fig:dss}
\end{figure}

There is a star located $\sim4$ arcseconds from \systemA, meaning the photometry of \systemA\ was blended with that nearby object, hereafter referred to as \systemB. \systemB\ is of similar magnitude and is likely a common proper motion companion to \systemA\ (see Section \ref{sec:cpmc}). \system\ refers to the combination of both \systemA\ and \systemB\, where both stars are present in a photometric aperture. \system\ was first observed with the Next Generation Transit Survey, hereafter \ngts\ \citep{NGTS} and later the \textit{Transiting Exoplanet Survey Satellite}, \tess\ \citep{tess}. Due to \ngts\ and \tess\ having pixel sizes of 5~arcsec \citep{NGTS} and 21~arcsec \citep{tess}, respectively, the photometry from both these facilities of \systemA\ was blended with that of \systemB\ (Figure \ref{fig:dss}). We obtained follow-up observations in various wavelengths (Table \ref{tab:obs_p}) using SHOC on the 1 m Lesedi telescope at the South African Astronomical Observatory \citep[\saao,][]{saao} to obtain observations that were not blended. We obtained follow up photometry with \speculooss\ and \trappists, most of which were uncontaminated by \systemB. We obtained six spectra in total with \harps\ \citep{harps}, four of the host star \systemA, and two of \systemB. A summary of these observations can be found in Table \ref{tab:obs_p} and \ref{tab:radial_velocities}. 

\begin{table*}
\caption{Photometric observations of \systemt.}              
\label{tab:obs_p}      
\centering

\begin{tabular}{l c c c c }          
\hline\hline                        
Instrument & Bandpass & Cadence (s) & Number  & Observation \\
&&&of transits&date\\
\hline 
\ngts\ & 520-890 nm & 13 & 26 & 2017 Jan 2 to 2017 Aug 21\\
\tess\ Sector 38& 600-1000 nm & 120 & 20 & 2021 Apr 29 to 2021 May 25\\
\tess\ Sector 11& 600-1000 nm & 1800 & 16 & 2019 Apr 22 to 2019 May 21\\
\saao\ & $I$& 10 s& 1 & 2018 July 4 \\
\speculooss\ & $g'$,$r'$,$i'$& 147, 52, 26 & 3 & 2023 Apr 18\\
\speculooss\ & $z'$, $I+z'$ & 19, 12 & 2 & 2023 June 15 \\
\trappists\ & $z'$ & 60 & 1 & 2023 Mar 29 \\
\trappists\ & $I+z'$ & 45 & 1 & 2023 Apr 18 \\
\hline
\end{tabular}
\end{table*}

\subsection{Photometry}

\begin{figure*}
    \centering
    \includegraphics[width=\linewidth, trim =2.5cm 6cm 2.5cm 7cm, clip]{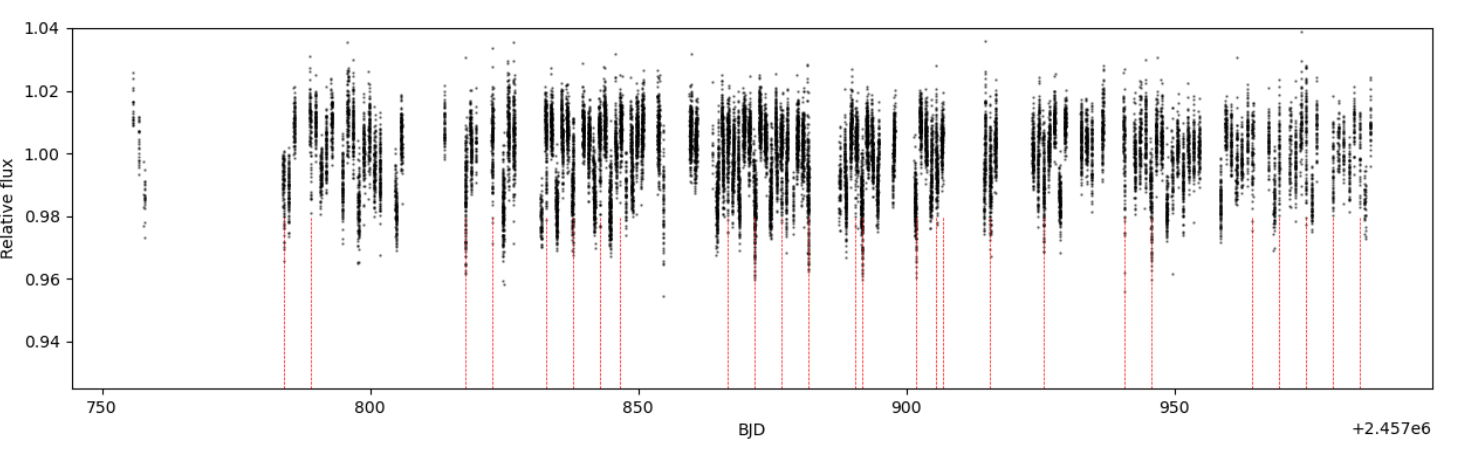}
    \caption{The full \ngts\ lightcurve for \system\ with the transit positions marked by the red dashed lines.}
    \label{fig:ngts_full}
\end{figure*}

\begin{figure}
    \centering
    \includegraphics[width=\linewidth, trim =2.5cm 9cm 3cm 10.1cm, clip]{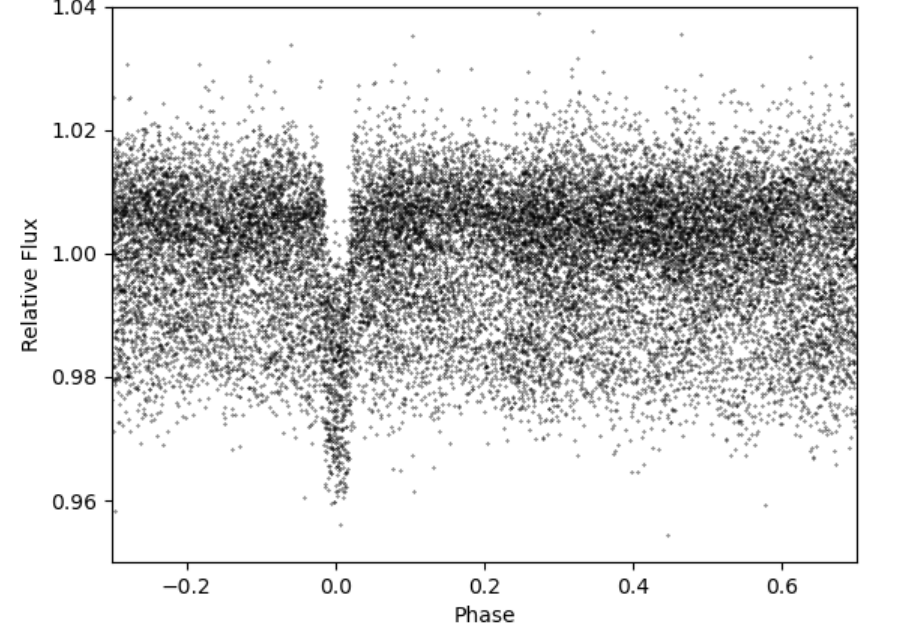}
    \caption{The \ngts\ lighcurve, phase-folded on the 1.254145~d period from \ngts.}
    \label{fig:ngts_phase}
\end{figure}

\begin{figure}
    \centering
    \begin{subfigure}[l]{1\textwidth}
        \includegraphics[width=0.5\linewidth, trim =2.5cm 9cm 3cm 10.1cm, clip]{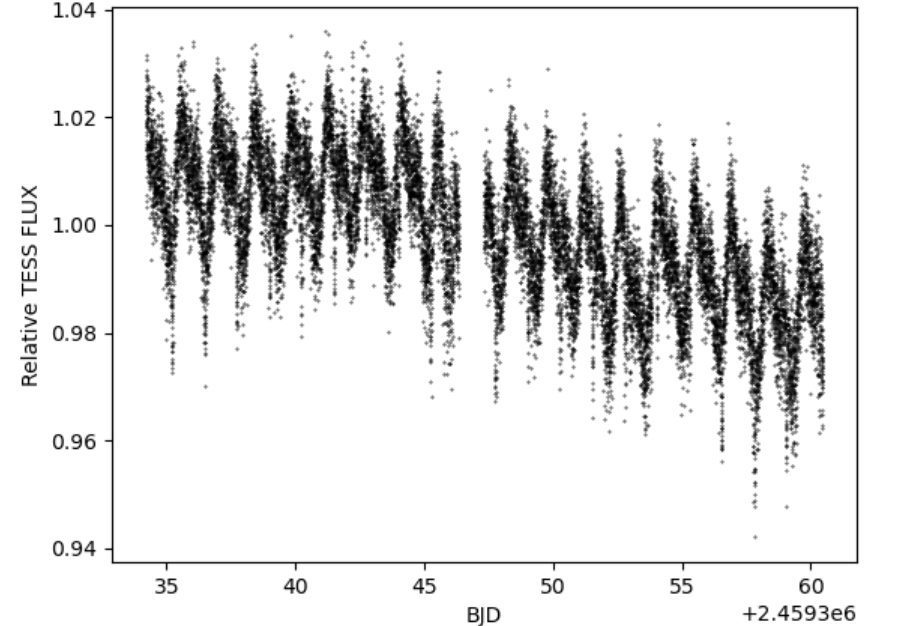}
    \end{subfigure}
    \centering
    \begin{subfigure}[l]{1\textwidth}
        \includegraphics[width=0.5\linewidth, trim =2.5cm 9cm 3cm 10.1cm, clip]{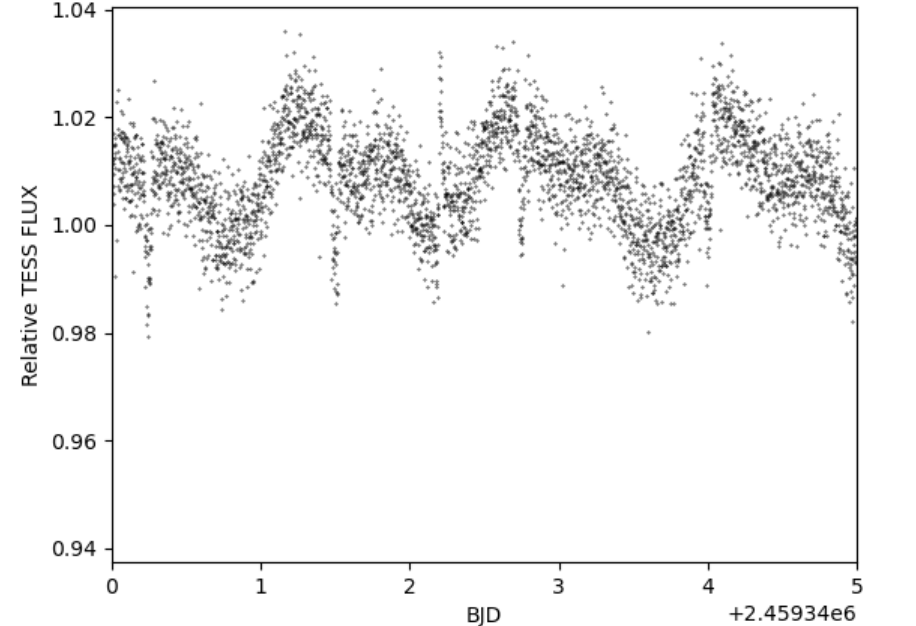}
    \end{subfigure}
    \caption{Top: The full \tess\ lightcurve for \system\ two-minute cadence data in sector 38, showing the transits. An apparent rotation of the object can also be seen. Bottom: A snippet of the top image, showing the apparent rotation of the object and a flare.}
    \label{fig:tess_full}
\end{figure}

\subsubsection{\ngts}\label{sec:ngts}

\systemA\ was initially observed with \ngts\ a twelve-telescope array situated at ESO's \PAR, Chile \citep{NGTS}. The observations taken between the dates 2017 January 2 to 2017 August 21. The twelve, 20~cm  telescopes cover almost 100~deg$^2$ of sky and operate in a custom bandpass of 520 nm to 890 nm \citep{NGTS}. We observed for a total of 135 nights, during which we obtained 199,319 science images with a 13~s cadence. 

The raw images were processed following the method discussed in \citet{NGTS} and the lightcurves were detrended using \textsc{SysRem} \citep{sysrem}. After detrending, transits were detected using a \textsc{Boxed Least Squares} algorithm.  A total of 26 transits were detected, which can be seen in Figure \ref{fig:ngts_full} and a phase-fold of the data on the predicted period can be seen in Figure \ref{fig:ngts_phase}. The star is clearly variable, but we determined an orbital period of 1.25~d and a transit depth of around 1.36 per cent.

\subsubsection{\tess}\label{sec:tess}

\tess\ obtained 30 minute cadence photometry of \system, observing between 2019 April 22 to 2019 May 21 for sector 11 and 2021 April 29 to 2021 May 25 for sector 38. Sector 38 also had two-minute cadence data. The 30 minute full frame image has a precision around 5 mmag \citep{tess}. Figure \ref{fig:tess_full} shows the full SPOC lightcurve, along with the a cropped version showing a flare and the individual transits.

The transits within both \tess\ sectors were independantly identified by the \tess\ team and \systemB\ was initially named TOI (\tess\ Object of Interest) 4339 on 2022 September 09. \systemA\ has since been found to be the true host star. \systemA\ was named TOI 6110 on 2023 May 27. The two-minute cadence data was processed using the SPOC pipeline \citep{spoc}. As with the \ngts\ data, the \tess\ photometry is blended. \tess\ provides two flux values: SAP flux and PDCSAP flux. The PDCSAP flux removes systematics in the lightcurve, as well as correcting for dilution. Within the PDCSAP flux, we noticed the transit depths were much larger than expected. PDCSAP flux lightcurve has been corrected for dilution, assuming the transit is on the companion star, \systemB\ and not \systemA. We therefore chose to use SAP flux  instead and modelled the data using a Gaussian process (GP) fit during the global modelling process to remove systematics in the data. We also performed our own dilution calculation to use within the global modelling. The \tess\ data contains a total of 20 transits in the two-minute cadence data. Unfortunately, the sector 11 30-minute cadence data includes a large (~5 d) data gap in the centre of the observation run, despite this it was used to provide confirmation of the ephemeris.

\subsubsection{\saao}\label{sec:saao}

We obtained $V$ and $I$ band transit light curves with the \saao\ 1.0~m telescope (Table \ref{tab:obs_p}) and one of the frame-transfer SHOC (Sutherland High Speed Photometer, \citep{saao}) CCD cameras, specifically "SHOC'n'awe", between 2018 and 2020. On the 1.0~m, the SHOC cameras have a pixel scale of $0.167$~arcsecs/pixel. Combined with $4 \times 4$ binning, this gave a pixel resolution of $0.67$~arcsecs. Both datasets were de-biassed and flat-fielded using \saao's local Python-based SHOC reduction pipeline, which utilises \textsc{IRAF} \citep{tody86,tody93} and \textsc{PyRAF} \citep{pyraf} tasks. 

We used the Starlink package \textsc{autophotom} \citep{Currie2014} to perform aperture photometry on both our target and several suitable comparison stars in the 2.85\arcmin $\times$2.85\arcmin field of view. Since the two stars \systemA\ and \systemB\ lie $\approx4$~arcsecs apart, a $2$~pixel radius aperture was chosen, together with a background annulus with radius between $12 \text{ and } 14$~pixels, to try to avoid contaminating light. Even so, while the two stars are indeed fully resolved in the \saao\ $V$~band images, they are increasingly blended in the longer wavelength $I$ filter image. 

The measured fluxes of the comparison stars were used to perform differential photometry on each of \systemA\ and \systemB, which confirms that the transit is on \systemA. While the transit was detected within the \saao\ $V$ data, we did not use this in our global modelling. This is due to concerns over the noise levels within the data, with a large amount of scatter seen in the baseline.

\subsubsection{SPECULOOS-South} 
\label{sec:spec_s}
 
We obtained simultaneous, multi-wavelength observations of \systemA\ with the three of the four, 1.0-m `Search for Planets EClipsing Ultra-cOOl Stars' (\speculooss) telescopes at Cerro, Paranal \citep{jehin18,delrez2018,sebastian21}. Each \speculooss\ telescope is equipped with a 2K$\times$2K detector with a pixel scale of 0.35~arcsec and an on-sky field of view of 12$\times$12~arcminutes. On 2023 April 18, we simultaneously observed a transit on \systemA\ in the Sloan-$g'$, -$r'$ and -$i'$ filters on the Io, Europa and Callisto telescopes, respectively. We also obtained transits in $z'$ and $I+z'$ filters on 2023 June 15 using the Io and Europa telescopes. Details on the cadence for each observation can be found in Table~\ref{tab:obs_p}.
 
We processed all of the data following the \textsc{prose} pipeline detailed in \citet{garcia22}. Briefly, \textsc{prose} is a modular and adaptable pipeline, which performs image reduction/calibration, then differential aperture photometry on \systemA. The differential aperture photometry creates a weighted, artificial comparison star from all available objects within the image as described in \citet{broeg05}. The lightcurves are then detrended for airmass, full-width half-maximum and sky background, with \textsc{prose}, producing the normalised, resultant curves seen in Figure \ref{fig:28_new_lc}. This procedure has been successfully used for previous \speculooss\ observations (e.g. \citealt{barkaoui23,dransfield24}).
 
After performing aperture photometry, we confirmed all the \speculooss\ transits, except for the $i'$ filter observation, were uncontaminated. Due to high background in the $g'$ and $r'$ observations, the last few points of each were discarded from global modelling.
 
\subsubsection{TRAPPIST} \label{sec:trap_s}
 
\begin{figure}
    \centering
    \subfloat[]{\includegraphics[scale=0.49]{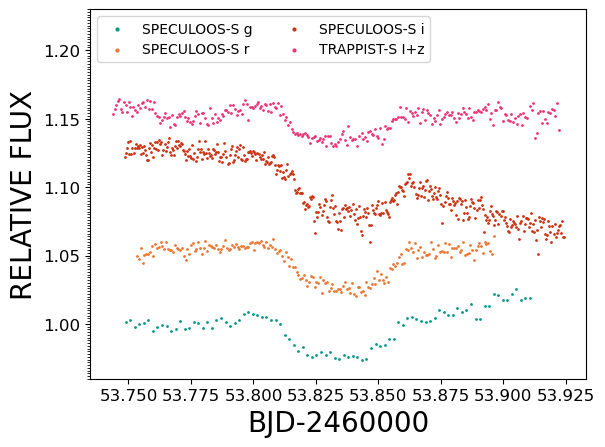}}
    \qquad
    \subfloat[]{\includegraphics[scale=0.49]{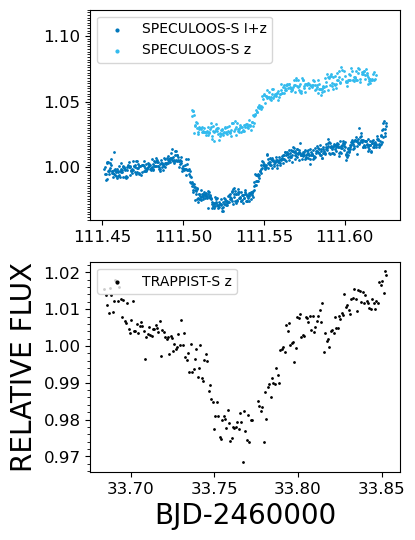}}
    \caption{Normalised lightcurves of the new data from \speculooss\ and \trappists. Offsets have been applied to some of the fluxes for clarity.}
    \label{fig:28_new_lc}
\end{figure}

The TRAnsiting Planets and PlanetIsmals Small Telescope (\trappists, \citealt{gillon11,jehin11}) is a 0.6~m robotic telescope based at ESO's La Silla Observatory. \trappists\ is equipeped with a 2K$\times$2K detector with a 0.6~arcsec pixel scale, and a field of view of 22 $\times$ 22~arcminutes. Two single-transit observations were taken at \trappists. The $z'$ filter observation was made on 2023 March 29, and the $I+z'$ filter observation was made on 2023 April 18. Cadence for each observation can be seen in Table \ref{tab:obs_p}. The \textsc{prose} pipeline \citep{garcia22} was also used for the \trappists\ data reduction, as in Section~\ref{sec:spec_s}. The resultant, relative flux lightcurves can be seen in Figure \ref{fig:28_new_lc}.

\subsection{Radial Velocities}\label{sec:harps}

\subsubsection{\harps} 

\begin{figure}
    \centering
    \includegraphics[width=\linewidth, trim =4cm 11cm 4cm 11cm, clip]{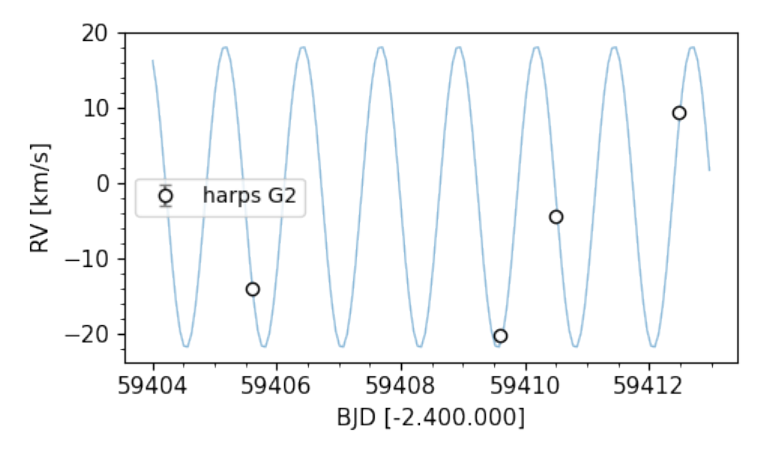}
    \caption{\harps\ time series plot of the radial velocity points from the spectra, with a G2 mask used.}
    \label{fig:harps_timeseries}
\end{figure}

We obtained high resolution spectra of both \systemA\ and \systemB\ using \harps, an echelle spectrograph on ESO's 3.6~m telescope at La Silla Observatory \citep{harps}. \harps\ is fibre-fed with an on-sky diameter of 1 or 1.4~arcsec \citep{harps}. There is also the option of a second fibre simultaneously monitoring the sky or wavelength solution of the instrument. It covers a spectral range of 380 to 690~nm, has a resolution of R = 115000 \citep{harps}. 

For our observations we used the default high accuracy mode with a 1~arcsec science fibre and fibre B on sky (for more information see: \citealt{harps}). We used exposures of 2400 s for each spectrum, except for the final exposure for \systemA\ which has an exposure time of 1800 s. The spectra had signal-to-noise ratios that ranged between 3.4 and 4.7, and was extracted per pixel in order 50 (corresponding to $\sim{550}$~nm). We obtained four spectra of \systemA\ between 2021 July 10 - 2021 July 16 (Table \ref{tab:radial_velocities}). On 2021 June 11 - 2021 June 12, we also obtained two spectra of \systemB. The mean RV precision reached with the four \systemA\ observations was 0.0515~$\rm km\, \rm s^{-1}$.

The data were reduced using the \harps\ pipeline and radial velocities (RVs) measured using a G2V mask and the cross-correlation technique (\citealt{Pepe2002, Baranne1996}). A G2V mask was used as it has been highly optimised for RV measurements but also because the spectra have low signal-to-noise (meaning only the strongest lines are detected). In addition, \systemA\ is rotating fast, meaning many lines are blended. Figure \ref{fig:harps_timeseries} shows the time series for the \systemA\ radial velocity measurements.

The pipeline also produces values for the bisector inverse slope (BIS, \citealt{Dall06,queloz01}) which can be an indicator for variability or a blended eclipsing binary \citep{ngts3Ab,santos02}. The automatic pipeline calculation of the BIS picked up a local minima in the cross-correlation function (CCF) and computed BIS values from that. To mitigate this, we binned the CCF to 5~kms$^{-1}$ in velocity space and re-calculated the BIS. The radial velocity values for \systemB\ and BIS values for both objects are also in Table \ref{tab:radial_velocities}.

\begin{table}
\caption{Radial velocity measurements of \systemA\ and \systemB\ and the associated uncertainties and BIS values from \harps.}              
\label{tab:radial_velocities}      
\centering   
\begin{tabular}{l c c}          
\hline\hline                        
BJD & Radial velocity [$\rm km\, \rm s^{-1}$] & BIS [$\rm km\, \rm s^{-1}$]\\
\hline 
\multicolumn{3}{c}{\textit{\systemA}} \\ 
\hline 
2459405.599495 & $-13.99\pm0.049$ & $-0.34 \pm 0.10$\\
2459409.590870 & $-20.22\pm0.055$ & $1.08\pm 0.11$\\
2459410.492873 & $-4.38\pm0.053$ & $2.87 \pm 0.11$\\
2459412.475762 & $9.46\pm0.049$ & $0.37 \pm 0.10$\\
\hline
\multicolumn{3}{c}{\textit{\systemB}} \\ 
\hline 
2459376.625674 & $-2.01\pm0.048$ & $0.06 \pm 0.10$\\
2459377.543570 & $-1.75\pm0.049$ & $0.76 \pm 0.10$\\
\hline
\end{tabular}
\end{table}

\subsection{High-resolution imaging with ZORRO}\label{sec:zorro}

Nearby companion stars can appear to mimic transit signatures in lightcurves, as well as affecting the detected depth of any real transiting event meaning the derived radii of any secondary object will be incorrect (\citealp{howell11,furlan17}). High resolution imaging of the host star is one way to determine if the host star is actually an unresolved binary.

\begin{figure}
    \centering
    \includegraphics[width=\linewidth, trim =0cm 6.5cm 0cm 7cm, clip]{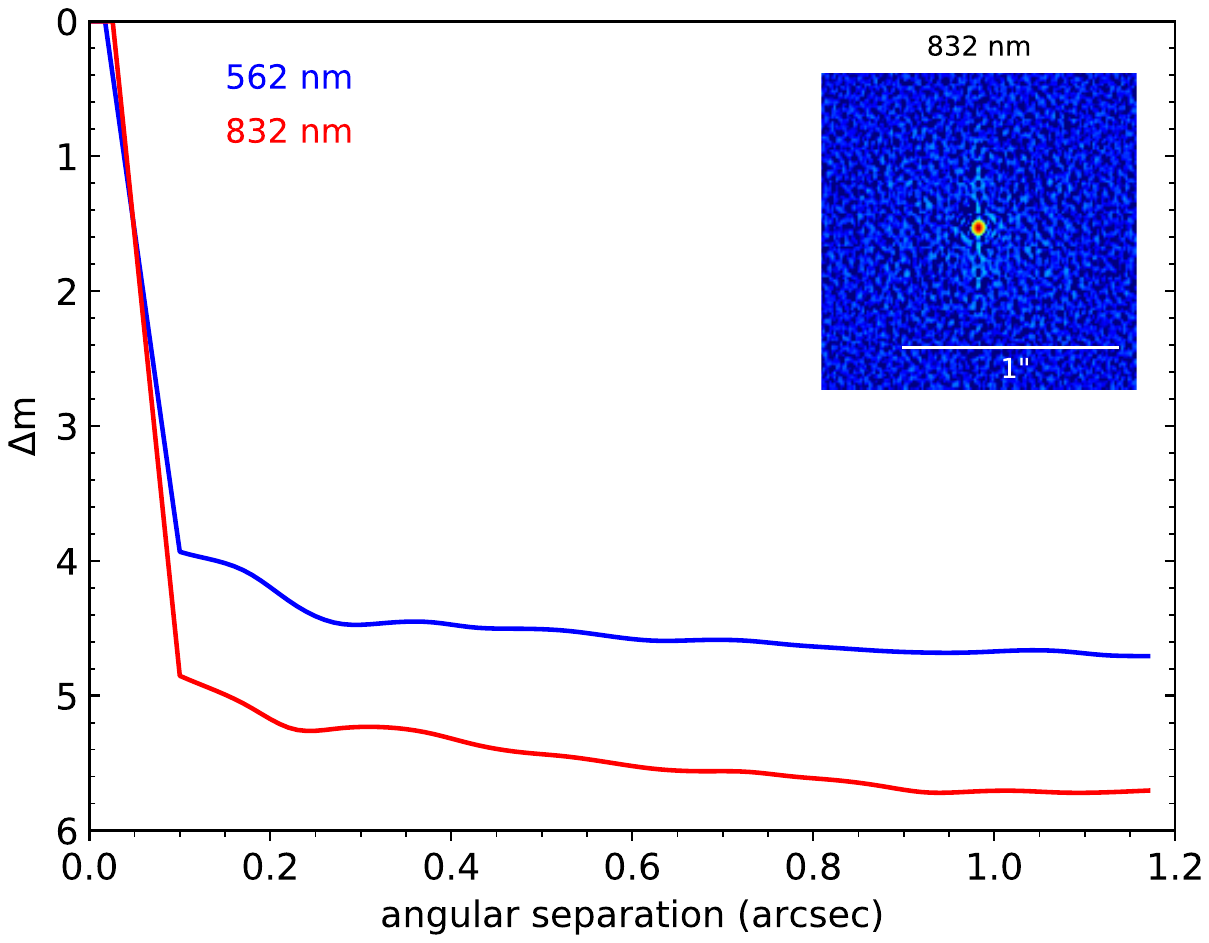}
    \caption{5-$\sigma$ contrast curves for speckle imaging in both 562 nm and 832 nm filters. This is plotted as a function of the angular separation from 0.02 to 1.2 arcsec. The reconstructed image for 832 nm can be seen in the top right of the figure. No close companions have been found within these limits.}
\label{fig:zorro_image}
\end{figure}

We performed speckle imaging with ZORRO on 2023 May 08 (\citealp{howell22, scott21}). The resultant 5-$\sigma$ contrast curve can be seen in Figure \ref{fig:zorro_image}, where the two filters used (562~nm and 832~nm, blue and red respectively) are plotted. We obtained 12 sets of 1000 images with an exposure time of 0.06 seconds, which was processed using the standard pipeline, detailed in \citet{howell11}.

Figure \ref{fig:zorro_image} shows no nearby stellar companions detected near \systemA\ to within the contrast (4 magnitudes at 562~nm and 5.8 magnitudes at 832~nm) and the angular separation (0.02-1.2~arcsec) limits of the observation. Meaning we detected no stellar companion between around 2.5 and 147.9 AU around \systemA. A reconstructed 832~nm image can be seen in the inset of Figure \ref{fig:zorro_image}.

\section{Analysis}\label{sec:analysis}

\subsection{NGTS-28B}\label{sec:cpmc}

Figure \ref{fig:dss} shows the DSS image of \systemt\ and shows \systemA\ is blended with another star (\systemB), with an on sky separation of $\sim{3.6}$~arcsec. We compared the \gaia\ DR3  kinematics (\citealt{gaia,gaiadr3}) for \systemB\ and \systemA\ (Table \ref{tab:systemAparameters}). \systemB\ has a proper motion and parallax in \gaia\ DR3 that is similar to that of \systemA\ although the proper motion is not consistent within the errors. However, \citet{multiplicity} have found \systemA\ and \systemB\ to be co-moving within their multiplicity study. Combining the distance to \systemA\ and angular separation (3.6~arcsec) of \system, the separation is $\sim{443}$ AU. This wide separation also means \systemB\ will not have any substantial gravitational effect on \systemA, and so will not affect orbital radial velocity measurements for \systemt\ over the time span of the observations taken. 

Within 15~arcsec, there is a third object which has a \gaia\ $G$ magnitude of 20.3 \citep{gaiadr3}, but it does not have a parallax or proper motion which is consistent with being a common proper motion companion. This object is too faint to have a significant affect on our photometry. We also searched the wider field within 5' for proper motion companions, but none were found. Within 60 arcseconds, there are 18 objects, but none of which are brighter than the host star. It is unlikely any of these objects have a significant contribution to the photometry of the system due to their separation, especially compared to \systemB. \systemB\ has been taken into consideration within our analysis, as blending within the aperture will affect the depth of the measured transit \citep{dilution}. This is discussed further in Section \ref{sec:dilution}.

\subsection{NGTS-28A: Host star parameters}\label{sec:host}

We analysed \systemA\ using spectral fitting and spectral energy distribution (SED) fitting. Table \ref{tab:systemAparameters} gives the proper motion, parallax and magnitudes from  \gaia\ DR3 (\citealt{gaia, gaiadr3}), \tess, 2MASS \citep{2mass}, and PanSTARRS (\citealt{panstarrs1,panstarrs2}). 

\begin{table}
\caption{Magnitudes, parameters and kinematics of \systemA\ and \systemB. The kinematics of the two objects are from \gaia\ DR3 and the photometry is from \gaia\ DR3 \citep{gaiadr3}, 2MASS \citep{2mass}, PanSTARRS (\citealt{panstarrs1,panstarrs2}). The \tess\ magnitude is from ExoFOP \citep{tess}. We list parameters of \systemA\ and \systemB\ found using \textsc{specmatch-emp} as priors for \textsc{ariadne}. Due to the low SNR of the spectra and the blending in photometry available, the errors in the parameters are likely underestimated and represent numerical errors.}             
\label{tab:systemAparameters}      
\centering   
%

\begin{tabular}{l l l}          
\hline\hline                        
Parameter & \systemA\ & \systemB\ \\
\hline 
\gaia\ Source ID & 6172969891098581120 & 6172969891098581248 \\
RA  & 14:11:43.0 & 14:11:42.9 \\
Dec  & -29:58:28.39 & -29:58:26.32 \\

\\
pmRA [$\rm mas\, \rm yr^{-1}$] & -78.61 $\pm0.02$ & -79.06 $\pm0.04$ \\
pmDec [$\rm mas\, \rm yr^{-1}$] & 24.74 $\pm0.02$ &23.77 $\pm0.03$ \\
Parallax [$\rm mas$] & 8.113 $\pm0.020$ & 8.085 $\pm0.034$ \\
Distance [pc] & 123.25 $\pm0.31$ & 123.86 $\pm0.53$\\ \\

Magnitudes\\
\gaia\ $G$ & 14.137 $\pm0.001$ & 15.249 $\pm0.001$ \\
\gaia\ $BP$ & 15.250 $\pm0.006$ & 16.550 $\pm0.005$ \\
\gaia\ $RP$ & 13.088$\pm0.003$ & 14.097 $\pm0.002$ \\
\tess\ [$T$]  & 13.074 $\pm0.009$ & 14.064 $\pm0.008$\\
2MASS$_{J}$ & 11.664 $\pm0.039$& 12.552 $\pm0.025$\\
2MASS$_{H}$ & 11.057 $\pm0.047$& 11.958 $\pm0.038$\\
2MASS$_{Ks}$ & 10.832 $\pm0.037$& 11.757 $\pm0.027$\\ 
PanSTARRS$_{g}$ & 15.623 $\pm0.004$& 16.923 $\pm0.007$\\ 
PanSTARRS$_{r}$ & 14.373 $\pm0.003$& 15.716 $\pm0.010$\\ 
PanSTARRS$_{i}$ & n/a & 14.568 $\pm0.006$\\ \\

Fitted Parameters\\
$\rm T_{\rm eff}$ $\rm(K)$ &3626$_{-44}^{+47}$ & 3441$_{-52}^{+70}$\\
$\log g$ (dex) & 4.74$_{-0.09}^{+0.10}$ & 4.85$_{-0.14}^{+0.13}$\\
$\rm [Fe/H]$ (dex) &-0.14$_{-0.17}^{+0.16}$& -0.02$_{-0.23}^{+0.26}$\\
$M_{\rm A}$ [$M_{\odot}$] & 0.56$_{-0.02}^{+0.02}$ & 0.43$_{-0.02}^{+0.01}$\\
$R_{\rm A}$ [$R_{\odot}$] & 0.59$_{-0.03}^{+0.03}$ & 0.42$_{-0.03}^{+0.03}$\\
Age [Gyr] & 6.99$_{-6.49}^{+5.09}$&2.06$_{-1.49}^{+10.08}$\\
\hline
\multicolumn{3}{l}{}
\end{tabular}
\end{table}

\subsubsection{Spectral analysis}

We analysed the \harps\ spectra with \textsc{specmatch-emp} \citep{specmatch} for both \systemA\ and \systemB. First, we shifted the spectra into the rest frame to remove the radial velocity variations, then co-added the spectra to obtain a single spectrum with a higher signal-to-noise ratio than the individual spectra, although the total signal-to-noise ratio (SNR) of the co-added spectrum of \systemA\ was still low per pixel (5.5). We then used \textsc{specmatch-emp} to compare our spectra to the KECK/HIRES libraries of template spectra in order to select a few `best fitting' objects to the target spectra. Linearly combined spectra of the comparison objects were then created and compared to the target spectra, to get a best fit. The output parameters are calculated using the weighted averages of the linearly combined spectra. \textsc{specmatch-emp} then uses $\chi^{2}$ statistics to test how good of a fit the estimated parameters are to the data.

We determined that \systemA\ has a mass of 0.56$\pm 0.08$ \msun, radius of 0.53$\pm 0.1$ \rsun\ and a \teff\ of 3630$\pm 70$ K. We also found it has a \logg\ of 4.74$\pm 0.12$ and a \feh\ of 0.36$\pm 0.09$. For \systemB\ we determined a mass of 0.39$\pm 0.08$ \msun\, radius of 0.39$\pm 0.1$ \rsun\ and a \teff\ of 3500$\pm 70$ K. We also found a \logg\ of 4.87$\pm 0.12$ and a \feh\ of 0.00$\pm 0.09$. The errors estimated by \textsc{specmatch-emp} are likely underestimates due to the low signal-to-noise ratio of the spectra, which is not considered within \textsc{specmatch-emp}. The \textsc{specmatch-emp} code uses fixed errors for all results, which depend only on the property's value. For example, the \teff\ will have an adopted error of 70~K if it is less than 4500~K and an adopted error of 110~K if \teff\ is greater than or equal to 5500~K. For more information we refer the reader to the software paper, \citet{specmatch}.

The two objects are of a similar \teff\ with \systemB\ being slightly cooler than \systemA. We also find that \systemB\ is less massive with a  smaller radius than \systemA. These values are  consistent with \systemB\ being fainter than \systemA. The age of both objects are also found to be in agreement within errors, supporting that they are likely common proper motion companions. We used the \logg, radius and \teff\ of \systemA\ and \systemB\ as priors for \textsc{ariadne}.

\subsubsection{SED fitting}

We fit the broadband photometry in Table \ref{tab:systemAparameters} for \systemA\ and \systemB\ using \textsc{ariadne}, a SED fitting tool  \citep{ariadne} which uses six atmospheric models (\textsc{phoenix}: \citet{phoenix}, \textsc{BT-Settl}: \citealt{btsettl}, \textsc{BT-NextGen}: \citealt{btsettl, btnextgen}, \textsc{BT-Cond}: \citealt{btsettl}, \textsc{Castelli \& Kurucz}: \citealt{ck04}, \textsc{Kurucz}: \citealt{kurucz})  to determine various stellar parameters. 

For \systemA\ and \systemB\ we used \teff, radius and \logg\ from \textsc{specmatch-emp} as priors for the \textsc{ariadne} fit. This produced the values in Table \ref{tab:systemAparameters}. It is likely that these values have slightly underestimated errors due to how blended \systemA\ and \systemB\ are in the optical photometry.

\begin{figure}
    \centering
    \includegraphics[width=\linewidth, trim =0cm 1cm 0cm 1cm, clip]{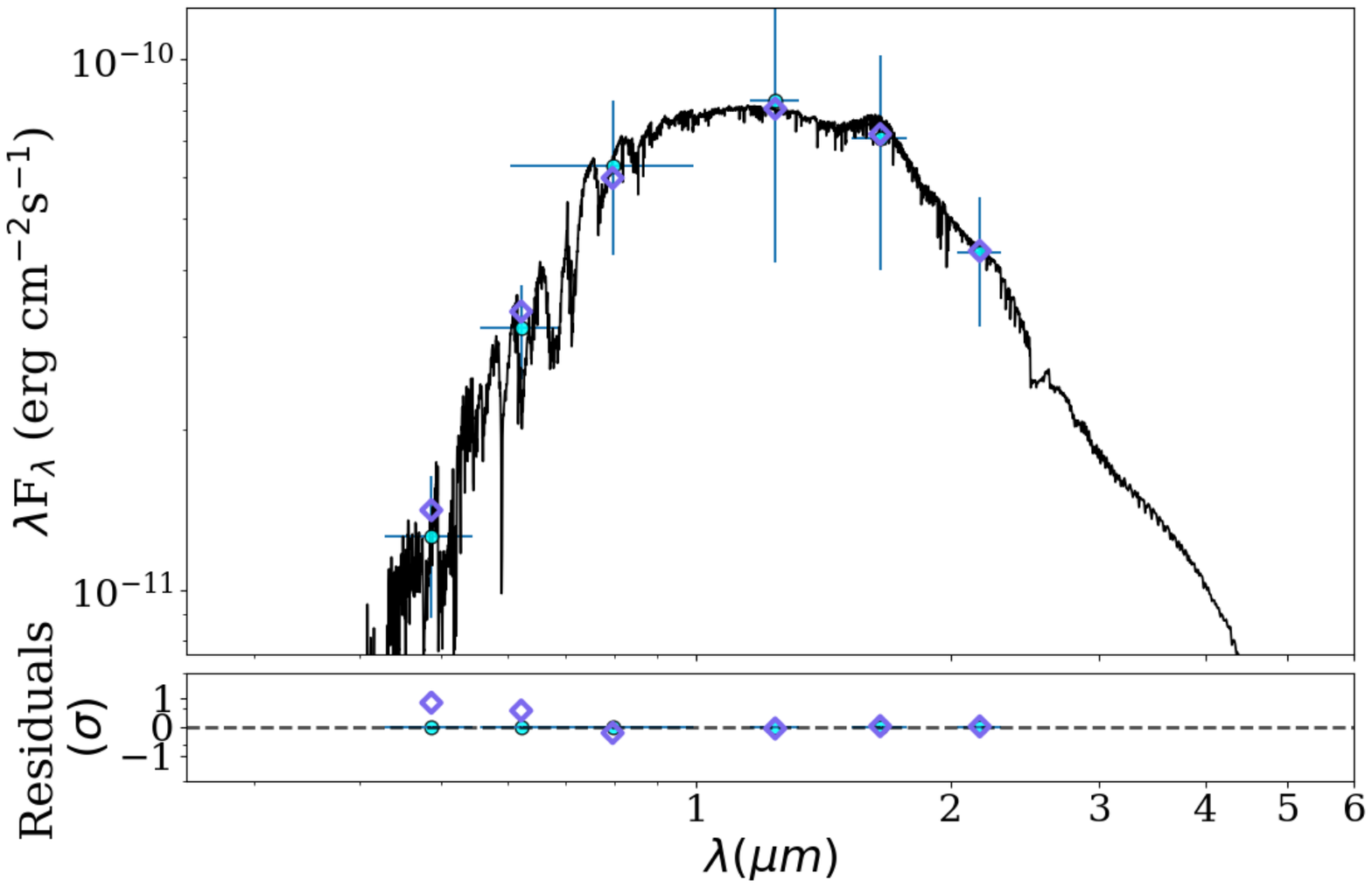}
    \caption{SED fit produced by \textsc{ariadne}. Black line shows the PHOENIX V2 model, light blue points show the various catalogue photometry points for \systemA\ and the purple diamonds show the synthetic photometry fit. Plotted underneath are the residuals for the data and synthetic photometry.}
\label{fig:sed}
\end{figure}

The SED for \systemA\ is shown in Figure \ref{fig:sed} with the \textsc{phoenix v2} model spectra. Synthetic photometry points, and the residuals between the synthetic and real photometry points are also seen in Figure \ref{fig:sed}. The data in Table \ref{tab:systemAparameters} were taken from the Bayesian Model Average output from \textsc{ariadne}.

The ages from \textsc{ariadne} are in agreement with each other but the errors very large. The mass and radius of \systemB\ are smaller than \systemA, as expected when compared to \textsc{specmatch-emp}.

To obtain rough estimates for the spectral types of both \systemA\ and \systemB\, the \teff\ values were compared to Figure 5 of \citet{spec}. In \citet{spec}'s figure, the \teff\ values are plotted with spectral type to show a trend for how \teff\ values change as you progress to later M dwarf types, along with multiple atmospheric models for comparison. Using the \teff\ values for \systemA\ and \systemB, we compared to this figure \citep{spec} to estimate a spectral type for both objects. We estimate a spectral type of $\sim$M1 for \systemA\ and  $\sim$M2 for \systemB. We compared this to the Mamajek spetral types \citep{mamajek} which were estimated with \textsc{ariadne}. \textsc{ariadne} estimates a M1.5 and M3 spectral classification for \systemA\ and \systemB, respectively. Comparing the results from both these methods provides a good estimate for the spectral type of both stars. Although more detailed methods can be used, they are not necessary for understanding \systemt, and are limited by the low signal-to-noise spectra that are available.

\subsection{\systemt\ Parameters}\label{sec:fitted}

Due to the complexity of the system, \systemt's parameters involved a multi-step process to get the most accurate results possible. \systemt\ was analysed using \textsc{Allesfitter} (\citealt{allesfitter-code,allesfitter-paper}). \textsc{Allesfitter} is a statistical tool used to fit lightcurve and radial velocity data to produce estimations for parameters of the system (\citealt{allesfitter-code, allesfitter-paper}). It utilises \textsc{mcmc} and dynamic nested sampling (\textsc{dns}) methods to estimate the best fitting parameters to the data, which can simultaneously be provided from multiple instruments and bandpasses, making use of various software including models such as \textsc{ellc} \citep{ellc} and samplers such as \textsc{Dynesty} \citep{dynesty} and \textsc{emcee} \citep{emcee}. 

\subsubsection{Estimating Dilution Values}\label{sec:dilution}

The first step involved calculating estimates for the dilution values for each instrument. We do this in a similar manner to \citet{ngts3Ab}. To do this, we used \textsc{pysynphot} \citep{pysyn}. We first obtained the bandpass response function for \ngts\ and \tess\ (\ngts: \citealt{ngts3Ab,NGTS}, \tess: \citet{tess}). We then obtained model spectra for both \systemA\ and \systemB\ using the \textsc{phoenix} models \citep{phoenix} within \textsc{pysynphot}. We randomly sampled 10000 \teff\ values between the errors of each object's \teff\ which created 10000 possible spectra for each object.

\begin{figure}
    \centering
    \includegraphics[width=\linewidth, trim =4cm 4cm 4cm 4.5cm, clip]{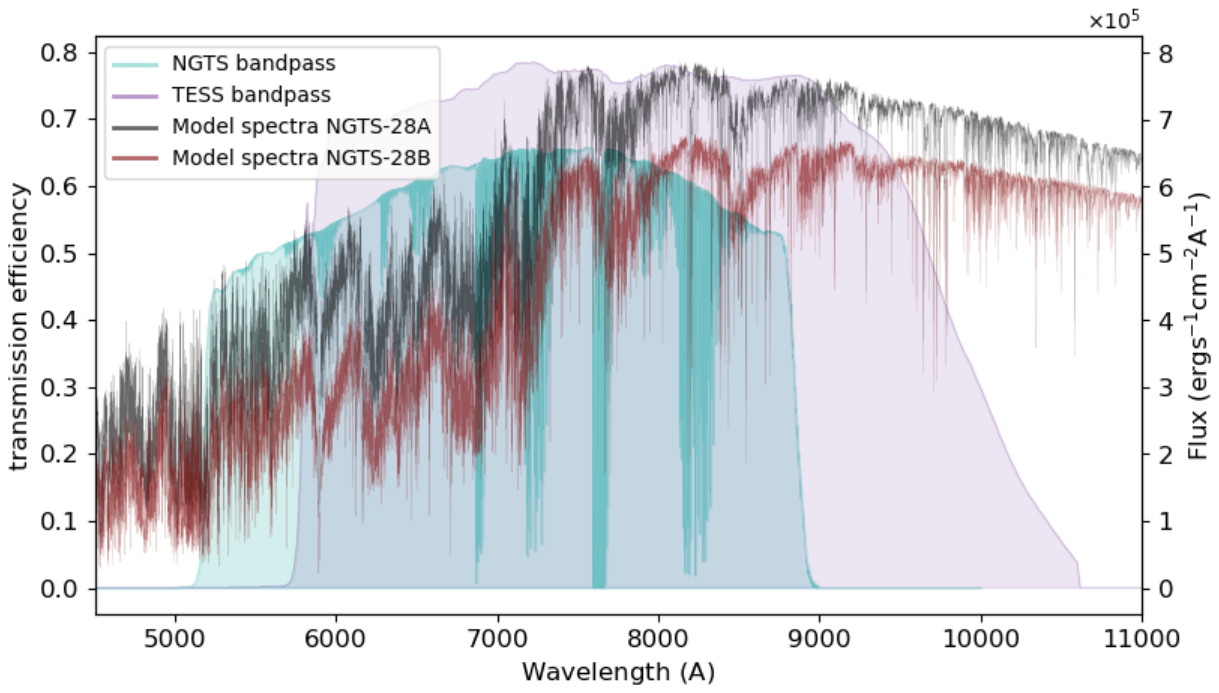}
    \caption{The \ngts\ bandpass (blue) and \tess\ bandpass (indigo) with \systemA\ (black) and \systemB\ (red) model spectra plotted. The \ngts\ bandpass curve is the product of the quantum efficiency curve of the CCD, the filter transmission curve and an atmospheric absorption curve (\citealp{NGTS,ngts3Ab}). The \tess\ bandpass curve is the product of the filter transmission curve and detector quantum efficiency curve \citep{tess}.}
    \label{fig:bandpass}
\end{figure}

The bandpass functions for each instrument and an example model spectra for each object produced by the \textsc{phoenix} models are plotted in Figure \ref{fig:bandpass}. Once we had obtained these, we interpolated the bandpass and spectra, multiplied the stellar spectra by the transmission curve and then integrated over the entire region. This provided flux values for \systemA\ and \systemB\ which could be used in equations \eqref{eq:db} and \eqref{eq:da} from \citet{ngts3Ab} to calculate the dilution for \systemA. $D_{B}$ and $D_{A}$ are the dilution values for \systemB\ and \systemA, respectively. $F_{B}$ and $F_{A}$ are the flux values for \systemB\ and \systemA, respectively.

\begin{equation}\label{eq:db}
    D_{B}= 1 - \frac{F_{B}}{F_{A}+F_{B}}
\end{equation}
\begin{equation}\label{eq:da}
    D_{A}= 1 - D_{B}
\end{equation}

The dilutions were calculated for each of the 10000 spectra, and a mean was calculated. To estimate the error on this value we also calculated a standard deviation for the sample. From this we determined that \systemA\ has a dilution value of $0.4364\pm0.0186$ and $0.4466\pm0.0156$ in \ngts\ and \tess, respectively. \tess\ PDCSAP estimated a dilution of 0.73 which is larger than the modelled dilution for \systemA\ or \systemB. This confirmed that the PDCSAP flux has likely been over corrected for this object.

\subsubsection{Global Modelling}

For the global modelling of the system we modelled the \tess\ two-minute cadence data, the \ngts\ data which we binned to two-minutes, the \saao, \speculooss\ and \trappists\ datasets and the \harps\ radial velocity data simultaneously using \textsc{Allesfitter}. We removed some data in the \saao\ $I$ band data due to variable seeing but kept in as much as possible. We also remove some data from the \speculooss\ $g'$ and $r'$ due to high background noise.

We estimated the baseline parameters $ln_\sigma$ and $ln_\rho$ for \ngts\ and \tess, which describe a Gaussian process (GP) baseline with a Matern-3/2 kernel, as seen in \citet{allesfitter-paper}, using the `out of transit' data in the lightcurve. These were used as priors for the final fit on the in-transit data, varying normally with their errors. \harps\ was fit with a hybrid offset and the rest of the photometric datasets were fit with hybrid splines.

\begin{table}
\caption{Estimate Matern-3/2 baseline parameters from \textsc{Allesfitter}. These were used as priors within the final fit.}              
\label{tab:baseline}      
\centering   
\begin{tabular}{l c c} 
\hline\hline  
Parameter & \ngts & \tess\\
\hline
$ln_\sigma$ & -4.6161$_{-0.0587}^{+0.0587}$&  -4.4589$_{-0.1135}^{+0.1135}$\\
$ln_\rho$ & -1.4840$_{-0.0764}^{+0.0764}$&-0.9483$_{-0.1049}^{+0.1049}$\\
\hline
\end{tabular}
\end{table}

For the dynamic nested sampling (\textsc{dns}) fitting, we fixed the dilution for \speculooss\ $I+z'$, $g'$, $r'$ and $z'$ to 0 due to there not being any contamination within the aperture. We used the dilution values for \ngts\ and \tess\ for \systemA\ from Section \ref{sec:dilution} as priors for the fit, and allowed them to vary normally within their errors. For all other datasets we used an initial prior of 0 and allowed it to vary uniformly between 0 and 1. We did not calculate dilution values due to the method of photometry described in Sections \ref{sec:saao}, \ref{sec:spec_s} and \ref{sec:trap_s}.

The limb darkening values for \tess, \ngts\ and all \speculooss\ and \trappists\ data were calculated using \textsc{ldtk} (\citealt{ldtk,phoenix}) using the \citet{kipping13} limb darkening relations. These were fitted with a normal distribution around the priors within the errors produced by \textsc{ldtk}. For \saao\ $I$ data, the limb darkening constants had initial priors of 0.5, varying uniformly between 0 and 1.

All priors, their parameter spaces and resultant values can be seen in Table \ref{tab:parameters}. The epoch is given a prior value and parameter space but is allowed to shift in order to prevent biasing towards one dataset which would cause correlations between epoch and period. We used 500 live points and a tolerance of 0.001. We fit the data with a 0.4 phase width around the transit.

\begin{figure*}
    \centering
    \subfloat[]{\includegraphics[scale=0.4]{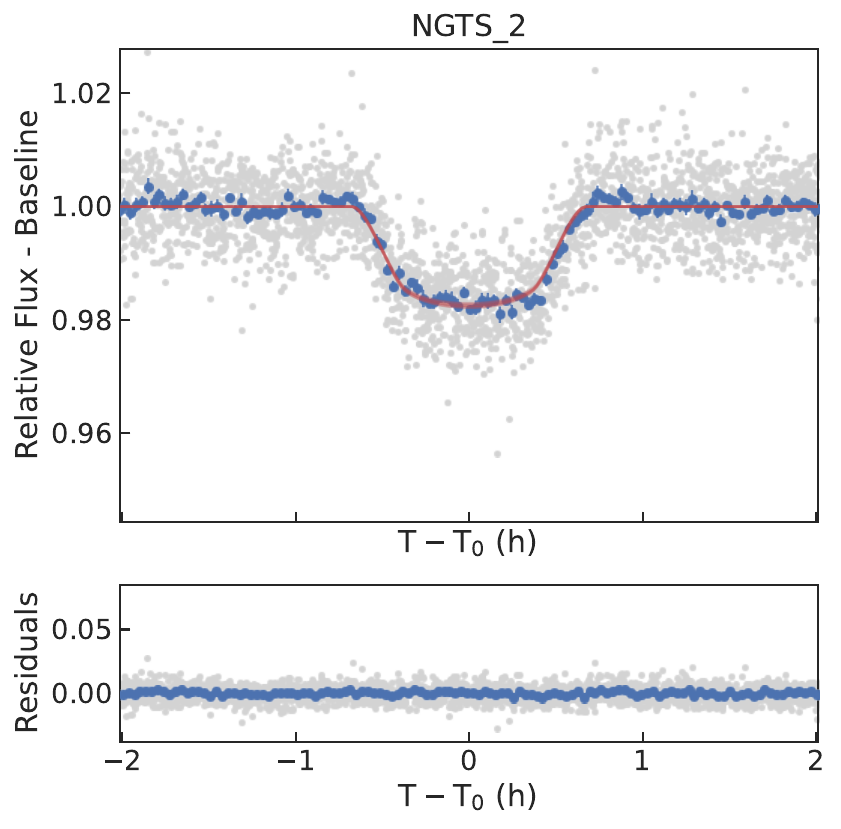}}
    \qquad
    \subfloat[]{\includegraphics[scale=0.4]{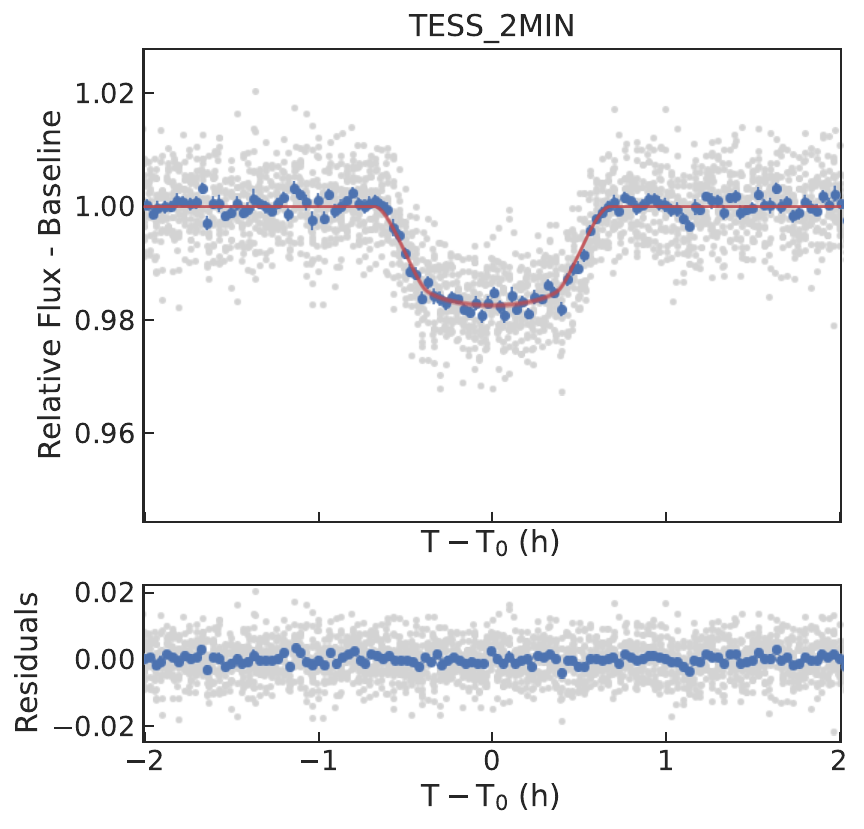}}
    \qquad
    \subfloat[]{\includegraphics[scale=0.4]{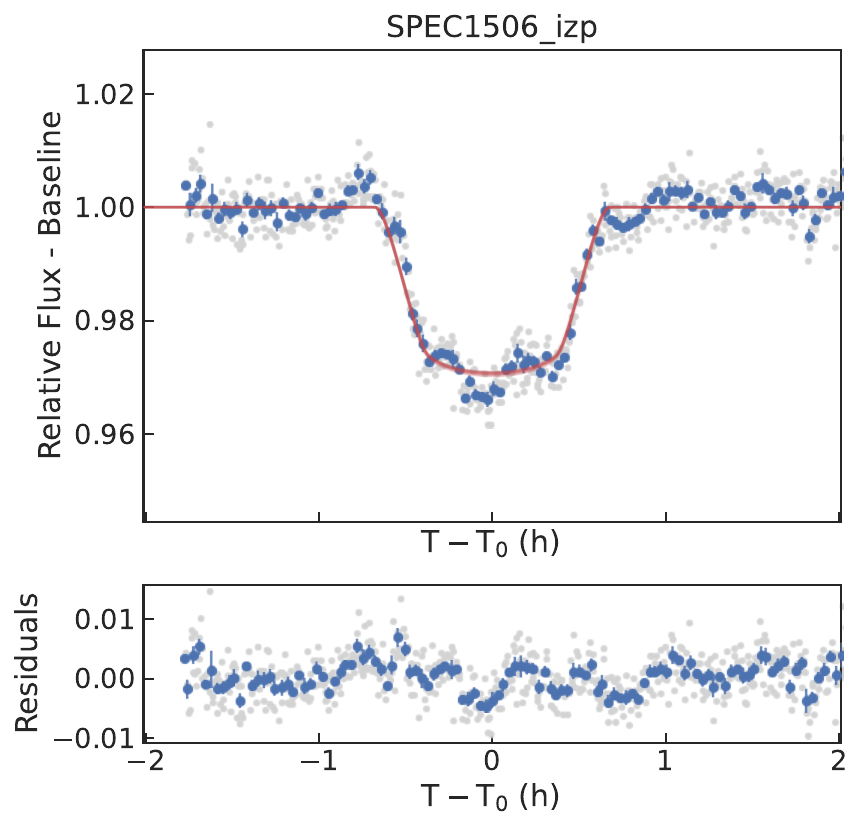}}
    \qquad
    \subfloat[]{\includegraphics[scale=0.4]{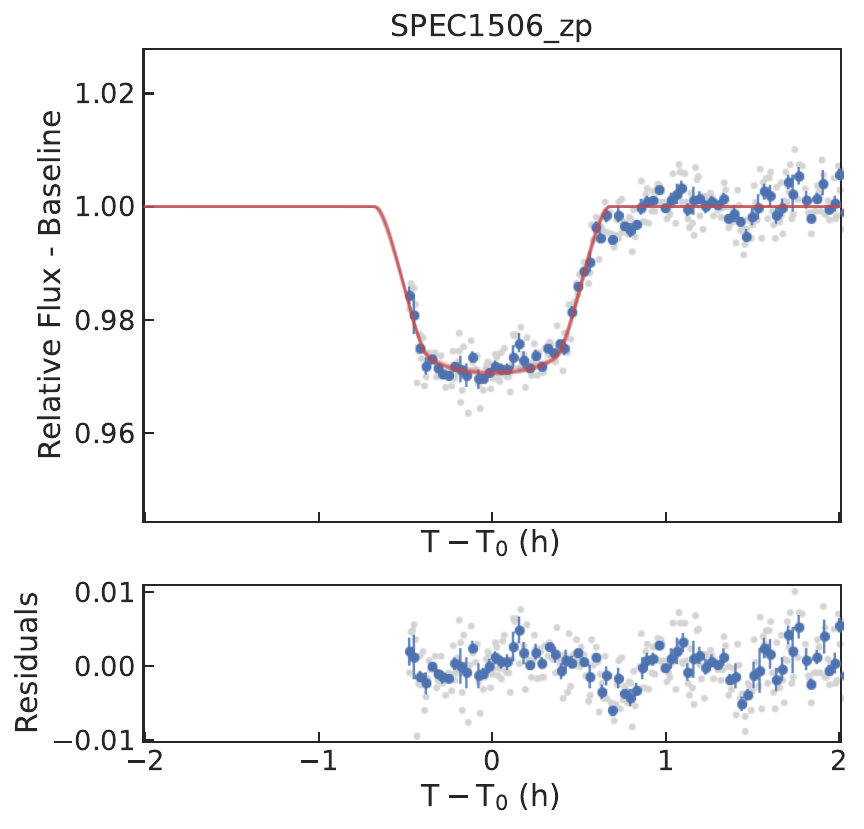}}
    \qquad
    \subfloat[]{\includegraphics[scale=0.4]{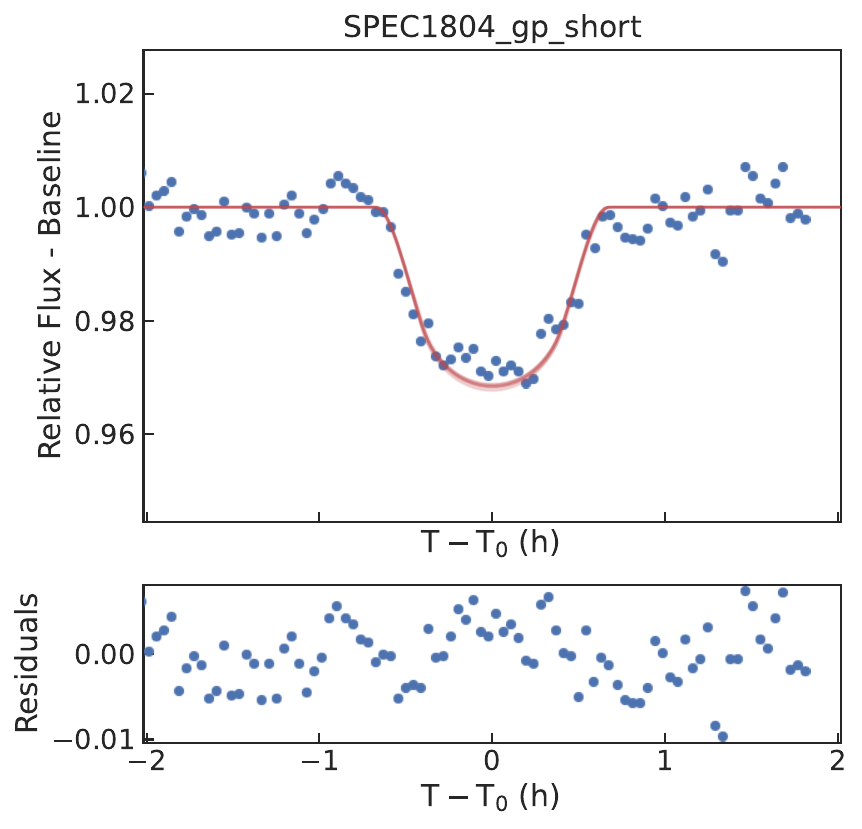}}
    \qquad
    \subfloat[]{\includegraphics[scale=0.4]{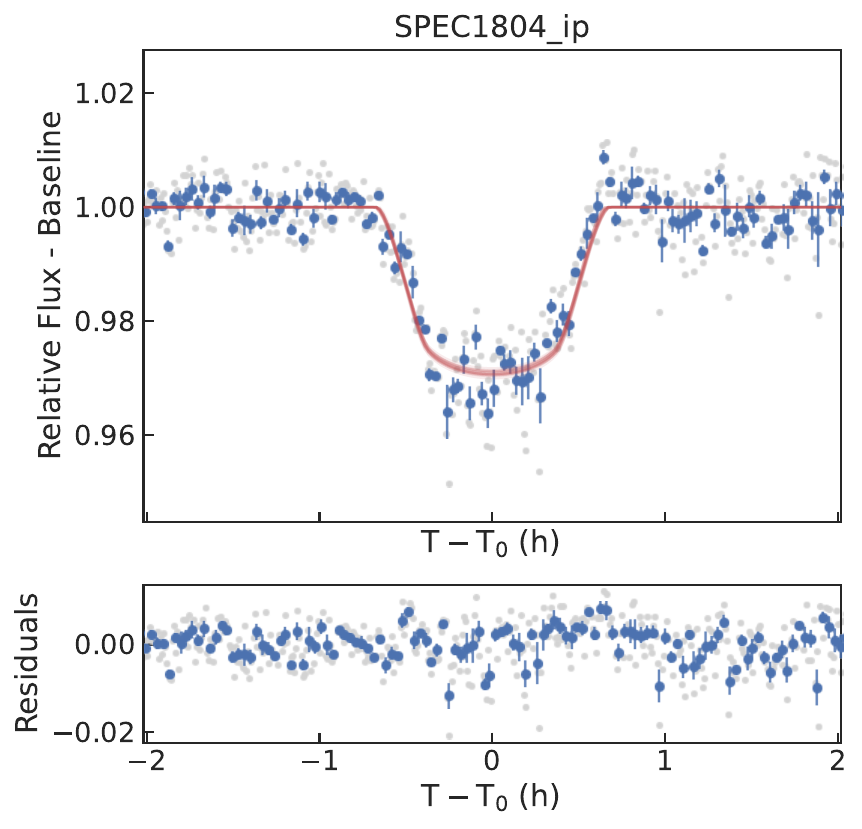}}
   \caption{Model fits to \textbf{(a)} \ngts\ data (binned to 2 minutes), \textbf{(b)} \tess\ 2 minute data, \textbf{(c)} \speculooss\ $I+z'$ band data, \textbf{(d)} \speculooss\ $z'$ band data, \textbf{(e)} \speculooss\ $g'$ band data, \textbf{(f)} \speculooss\ $g'$ band data. All model plots are produced with \textsc{allesfitter}. All plots show the inputted data in grey and the 'phased average' flux in blue, with 20 model lines sampled from the posterior.}
    \label{fig:28_model_fit_new}
\end{figure*}

\begin{figure*}
    \centering
    \qquad
    \subfloat[]{\includegraphics[scale=0.4]{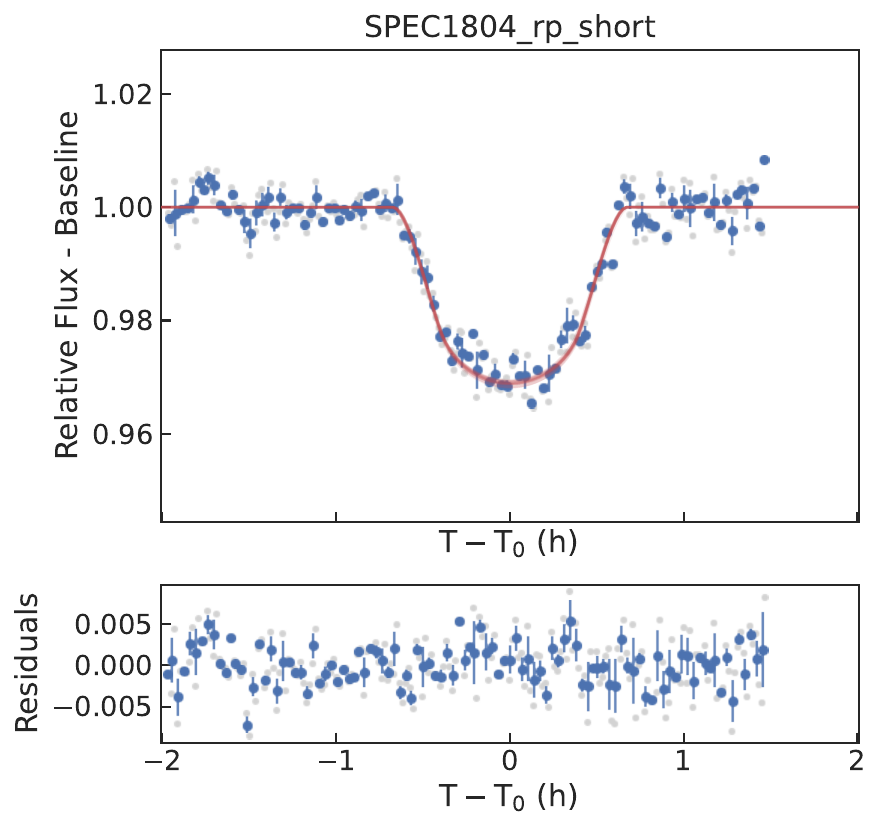}}
    \qquad
    \subfloat[]{\includegraphics[scale=0.4]{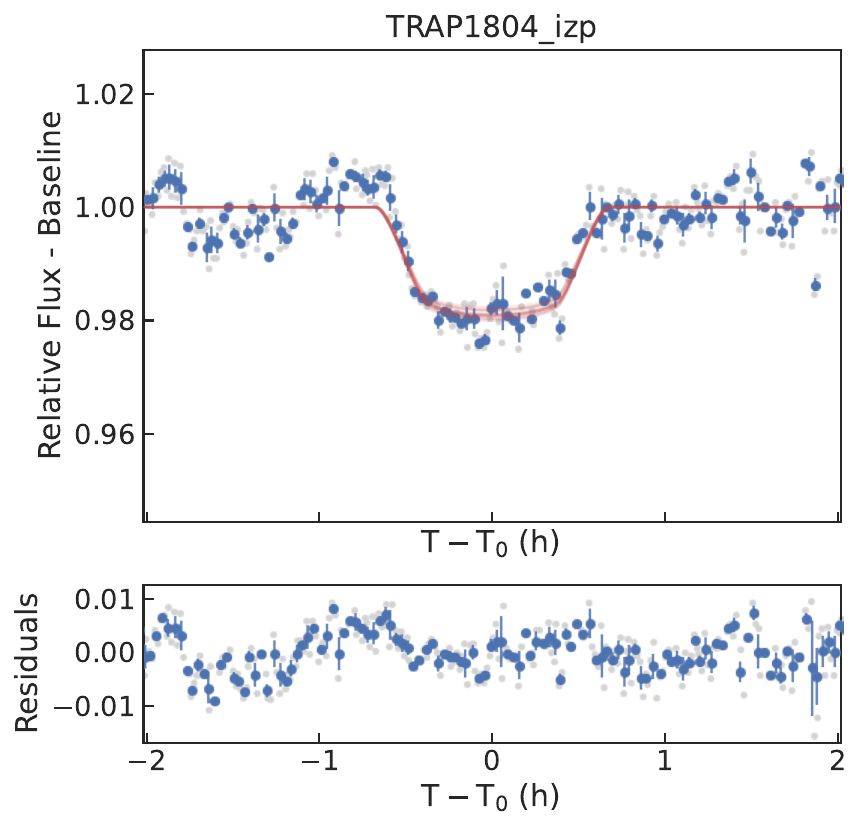}}
    \qquad
    \subfloat[]{\includegraphics[scale=0.4]{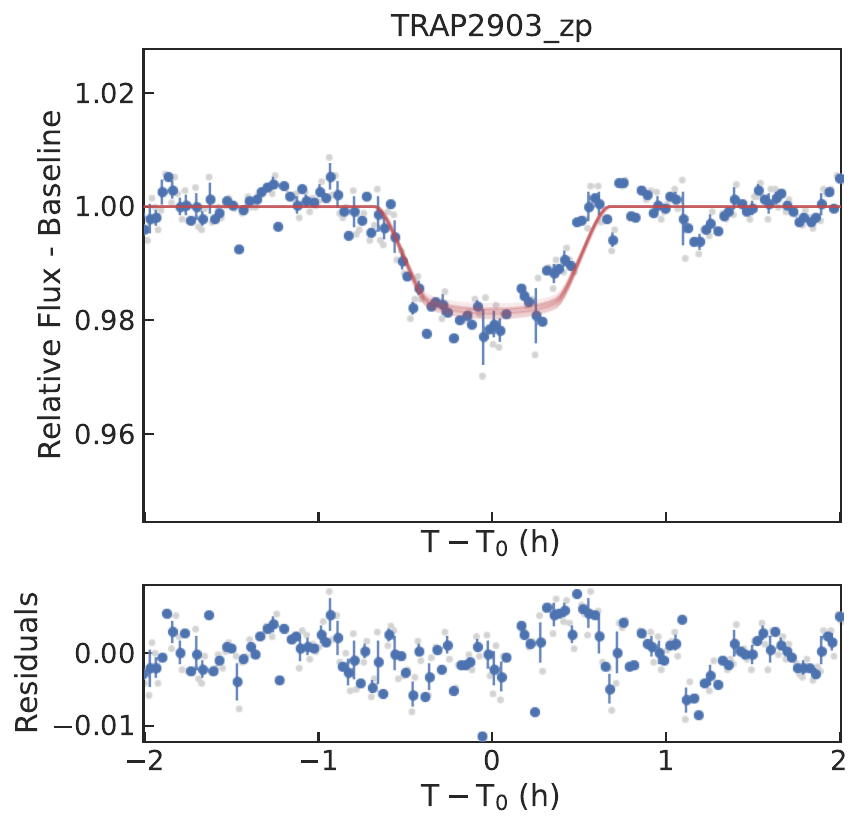}}
    \qquad
    \subfloat[]{\includegraphics[scale=0.4]{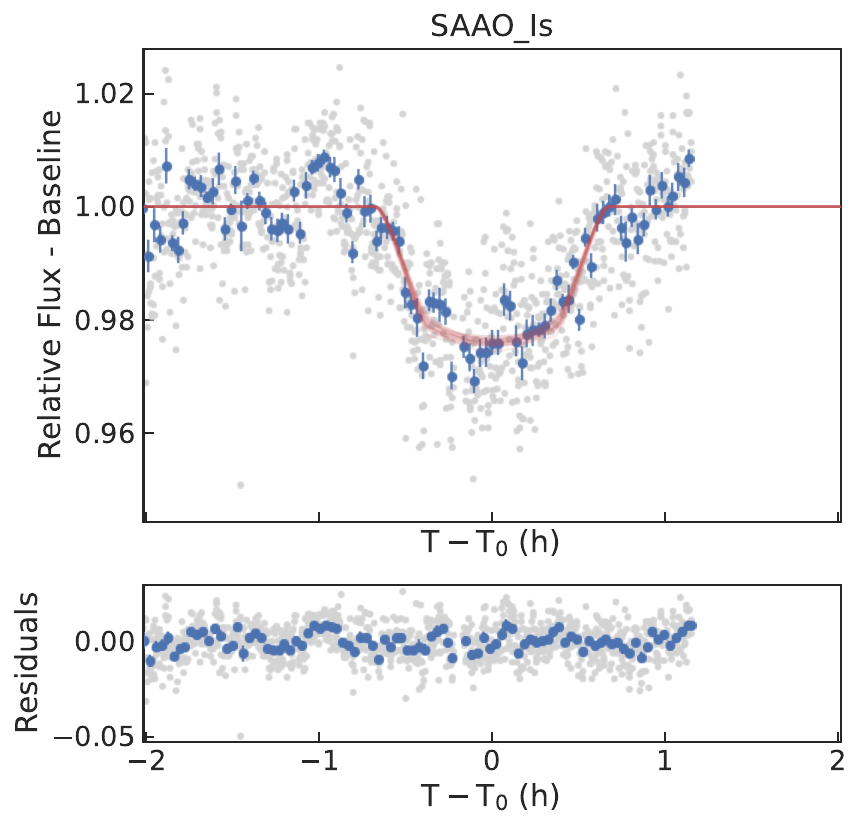}}
    \qquad
    \subfloat[]{\includegraphics[scale=0.4]{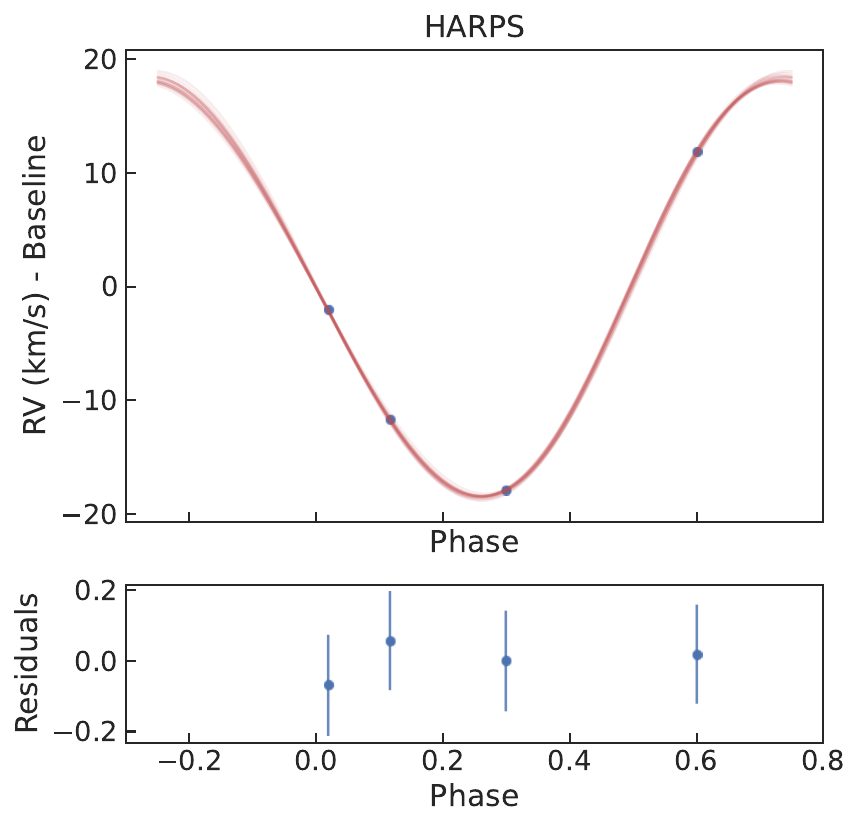}}
    \qquad
    \subfloat[]{\includegraphics[scale=0.4]{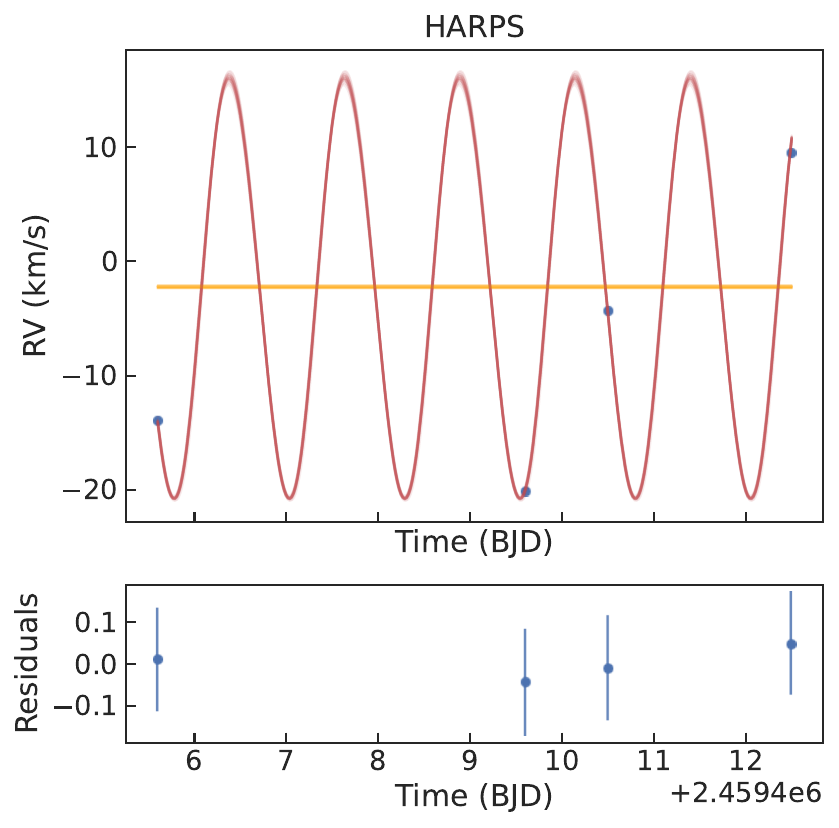}}
    \caption{Model fits to \textbf{(a)} \speculooss\ $i'$ band data, \textbf{(b)} \speculooss\ $r'$ band data, \textbf{(c)} \trappists\ $I+z'$ band data, \textbf{(d)} \trappists\ $z'$ band data, \textbf{(e)}  \saao\ $I$ band data and \textbf{(f)} phasefolded \harps\ data \textbf{(g)} timeseries \harps\ data. All model plots are produced with \textsc{allesfitter}. All plots show the inputted data in grey and the 'phased average' flux in blue, with 20 model lines sampled from the posterior.}
    \label{fig:28_model_fit_new_cont}
\end{figure*}

The derived parameters in Table \ref{tab:derived_parameters}. The corner plot for the derived values from this fit are also in the appendix in Figure \ref{fig:derived_corner}. Plots showing the model fits to the various sources of data can be seen in Figures \ref{fig:28_model_fit_new} and \ref{fig:28_model_fit_new_cont}.

The radius of \systemt\ was therefore found to be $0.95\pm0.05$ \rjup, with a mass of $69.0^{+5.3}_{-4.8}$ \mjup. These clearly put \systemt\ within the brown dwarf regime. The derived eccentricity is $0.0404_{-0.0101}^{+0.0067}$, meaning the orbit is close to circular, but still has some eccentricity.

In order to test if the eccentricity is significant, we repeated the model fit with the eccentricity parameters ($\sqrt{e_b} \cos{\omega_b}$ and $\sqrt{e_b} \sin{\omega_b}$) fixed to 0, which fixes the eccentricity of the system to 0. We kept all other priors and parameter spaces the same.

Using the Bayesian evidence calculated during the \textsc{dns} fit for both the eccentric and circular model, we calculate a logarithmic Bayes factor of 4.5. Using the criteria from \citet{jeffreys98theory} and \citet{kass}, this means there is strong evidence for the eccentricity to be a significant measurement, but not definitive. However, due to the Bayes factor being close to the limit of decisive (greater than 4.6), and the fact there is still strong evidence for some level of eccentricity, we assume the eccentric model to be the most likely. We perform other tests within Section \ref{sec:ecc} of this paper.

\begin{table*}
\caption{The fitted parameters for \systemt, produced by \textsc{Allesfitter}.}              
\label{tab:parameters}      
\centering
\begin{tabular}{l c c c} 
\hline\hline
parameter & symbol & prior, parameter space & value\\ 
\hline
\multicolumn{4}{c}{\textbf{Fitted parameters}} \\ 
\hline 

\textbf{Limb Darkening Parameters}\\
\ngts\ LDC 1 & $q_{1;\mathrm{NGTS}}$ & 0.4544, $\mathcal{N}(0.4544, 0.0088)$ &$ 0.4517_{-0.0078}^{+0.0074   }$\\ 
\ngts\ LDC 2 &$q_{2;\mathrm{NGTS}}$& 0.2901, $\mathcal{N}(0.2901, 0.0022)$ &$ 0.2900_{-0.0019}^{+0.0018 }$\\ 

\tess\ Sector 38 LDC 1 & $q_{1;\mathrm{TESS}}$ & 0.3379, $\mathcal{N}(0.3379, 0.0054)$ &$ 0.3383_{-0.0043}^{+0.0046 }$\\ 
\tess\ Sector 38 LDC 2 & $q_{2;\mathrm{TESS}}$ & 0.2640, $\mathcal{N}(0.2640, 0.0019)$ &$ 0.2638_{-0.0017}^{+0.0017}$\\

\speculooss\ I+z' band LDC 1 &$q_{1;\mathrm{SPEC_izp}}$&0.2284, $\mathcal{N}(0.2284,0.0037)$&$0.2290_{-0.0033}^{+0.0032}$\\
\speculooss\ I+z' band LDC 2 &$q_{2;\mathrm{SPEC_izp}}$&0.2199, $\mathcal{N}(0.2199, 0.0020)$&$0.2195_{-0.0017}^{+0.0016}$\\

\speculooss\ z' band LDC 1 &$q_{1;\mathrm{SPEC_zp}}$&0.2015, $\mathcal{N}(0.2015, 0.0031)$&$0.2015_{-0.0026}^{+0.0026}$\\
\speculooss\ z' band LDC 2 &$q_{2;\mathrm{SPEC_zp}}$&0.2027, $\mathcal{N}(0.2027, 0.0019)$&$0.2030_{-0.0017}^{+0.0016}$\\

\speculooss\ g' band LDC 1 &$q_{1;\mathrm{SPEC_gp}}$&0.8945, $\mathcal{N}(0.8945, 0.0143)$&$0.8911_{-0.0118}^{+0.0118}$\\
\speculooss\ g' band LDC 2 &$q_{2;\mathrm{SPEC_gp}}$&0.3271, $\mathcal{N}(0.3271, 0.0016)$&$0.3270_{-0.0014}^{+0.0014}$\\

\speculooss\ i' band LDC 1 &$q_{1;\mathrm{SPEC_ip}}$&0.3531, $\mathcal{N}(0.3531, 0.0070)$&$0.3550_{-0.0059}^{+0.0059}$\\
\speculooss\ i' band LDC 2 &$q_{2;\mathrm{SPEC_ip}}$&0.2678, $\mathcal{N}(0.2678, 0.0024)$&$0.2676_{-0.0019}^{+0.0020}$\\

\speculooss\ r' band LDC 1 &$q_{1;\mathrm{SPEC_rp}}$&0.7794, $\mathcal{N}(0.7794, 0.0151)$&$0.7790_{-0.0129}^{+0.0127}$\\
\speculooss\ r' band LDC 2 &$q_{2;\mathrm{SPEC_rp}}$&0.3161, $\mathcal{N}(0.3161, 0.0020)$&$0.3162_{-0.0018}^{+0.0018}$\\

\trappists\ I+z' band LDC 1 &$q_{1;\mathrm{TRAP_izp}}$&0.2284, $\mathcal{N}(0.2284, 0.0074)$&$0.2293_{-0.0062}^{+0.0067}$\\
\trappists\ I+z' band LDC 2 &$q_{2;\mathrm{TRAP_izp}}$&0.2199, $\mathcal{N}(0.2199, 0.0040)$&$0.2199_{-0.0034}^{+0.0034}$\\

\trappists\ z' band LDC 1 &$q_{1;\mathrm{TRAP_zp}}$&0.2015, $\mathcal{N}(0.2015, 0.0062)$&$0.2020_{-0.0054}^{+0.0053}$\\
\trappists\ z' band LDC 2 &$q_{2;\mathrm{TRAP_zp}}$&0.2027, $\mathcal{N}(0.2027, 0.0038)$&$0.2030_{-0.0033}^{+0.0032}$\\

\saao\ I band LDC 1 & $q_{1;\mathrm{SAAO_{I}}}$ & 0.5, $\mathcal{U}(0.0, 1.0)$ &$ 0.2191_{-0.0892}^{+0.1091  }$\\ 
\saao\ I band LDC 2 &$q_{2;\mathrm{SAAO_{I}}}$& 0.5, $\mathcal{U}(0.0, 1.0)$ &$ 0.6158_{-0.3336}^{+0.2615}$\\ \\

\textbf{System Parameters}\\
Radius ratio & $R_b / R_\star$ & 0.116, $\mathcal{U}(0.0, 0.3)$ &$ 0.1667_{-0.0010}^{+0.0011 }$\\ 
Scaled summed radius &$(R_\star + R_b) / a_b$ & 0.16,$\mathcal{U}(0.0, 0.3)$ &$ 0.1579_{-0.0024}^{+0.0027 }$\\ 
Cosine inclination & $\cos{i_b}$ & 0, $\mathcal{U}(0.0, 0.5)$ &$ 0.0808_{-0.0030}^{+0.0032}$\\ 
Epoch ($\mathrm{BJD}$) & $T_{0;b}$ & 2459360.305123, $\mathcal{U}(2459350, 2459370)$ &$2458953.9685_{-0.0001}^{+0.0001 }$\\
Orbital Period (d) & $P_{orb}$ & 1.25412, $\mathcal{U}(1.245, 1.265)$ &$ 1.2541_{-0.0000}^{+0.0000}$\\ 
RV Semi-amplitude ($\mathrm{km/s})$ & $K_b$ & 15, $\mathcal{U}(12, 22)$ &$ 18.4107_{-0.1659}^{+0.2367 }$\\ 
$\sqrt{e_b} \cos{\omega_b}$ & $f_{c}$& 0, $\mathcal{U}(-1, 1)$ &$ -0.0289_{-0.0173}^{+0.0236 }$\\ 
$\sqrt{e_b} \sin{\omega_b}$ & $f_{s}$& 0, $\mathcal{U}(-1, 1)$ &$ -0.1981_{-0.0156}^{+0.0267}$\\ 
\\

\textbf{Dilution values}\\

\ngts\ Dilution & $D_{\mathrm{NGTS}}$ & 0.4364, $\mathcal{N}(0.4364, 0.0186)$ &$ 0.4222_{-0.0111}^{+0.0095 }$\\ 
\tess\ Dilution & $D_{\mathrm{TESS}}$ & 0.4466,$\mathcal{N}(0.4466, 0.0156)$ &$ 0.4145_{-0.0097}^{+0.0089}$\\ 
\speculooss\ I+z' Dilution & $D_{\mathrm{SPEC_izp}}$ & - & 0.0 (fixed) \\ 
\speculooss\ z' Dilution & $D_{\mathrm{SPEC_zp}}$ & - & 0.0 (fixed) \\ 
\speculooss\ g' Dilution & $D_{\mathrm{SPEC_gp}}$ & - & 0.0 (fixed) \\ 
\speculooss\ i' Dilution & $D_{\mathrm{SPEC_ip}}$ & 0, $\mathcal{U}(0, 1)$ &$ 0.0226_{-0.0145}^{+0.0191}$\\ 
\speculooss\ r' Dilution & $D_{\mathrm{SPEC_rp}}$ & - & 0.0 (fixed) \\ 
\trappists\ I+z' Dilution & $D_{\mathrm{TRAP_izp}}$ & 0, $\mathcal{U}(0, 1)$ &$ 0.3599_{-0.0265}^{+0.0268 }$\\ 
\trappists\ z' Dilution & $D_{\mathrm{TRAP_zp}}$ & 0, $\mathcal{U}(0, 1)$ &$ 0.3648_{-0.0262}^{+0.0268 }$\\ 
\saao\ I band Dilution & $D_{\mathrm{SAAO_{I}}}$ & 0, $\mathcal{U}(0, 1)$ &$ 0.1791_{-0.0269}^{+0.0276}$\\ \\

\textbf{White noise}\\
- &$\ln{\sigma_\mathrm{NGTS}}$ & -5, $\mathcal{U}(-10,-1)$ &$ -5.0915_{-0.0073}^{+0.0073}$\\ 
- &$\ln{\sigma_\mathrm{TESS}}$ & -5, $\mathcal{U}(-10, -1)$ &$ -5.1962_{-0.0078}^{+0.0081 }$\\ 
- &$\ln{\sigma_\mathrm{SPEC_izp}}$ & -5, $\mathcal{U}(-10, -1)$ &$ -5.5845_{-0.0231}^{+0.0230}$\\ 
- &$\ln{\sigma_\mathrm{SPEC_zp}}$ & -5, $\mathcal{U}(-10, -1)$ &$ -5.6716_{-0.0328}^{+0.0334 }$\\ 
- &$\ln{\sigma_\mathrm{SPEC_gp}}$ & -5, $\mathcal{U}(-10, -1)$ &$ -5.6037_{-0.0617}^{+0.0672 }$\\ 
- &$\ln{\sigma_\mathrm{SPEC_ip}}$ & -5, $\mathcal{U}(-10, -1)$ &$ -5.2349_{-0.0304}^{+0.0304 }$\\ 
- &$\ln{\sigma_\mathrm{SPEC_rp}}$ & -5, $\mathcal{U}(-10, -1)$ &$ -5.6961_{-0.0435}^{+0.0429 }$\\ 
- &$\ln{\sigma_\mathrm{TRAP_izp}}$ & -5, $\mathcal{U}(-10, -1)$ &$ -5.4060_{-0.0335}^{+0.0345}$\\ 
- &$\ln{\sigma_\mathrm{TRAP_zp}}$ & -5, $\mathcal{U}(-10, -1)$ &$ -5.6070_{-0.0411}^{+0.0433 }$\\ 
- &$\ln{\sigma_\mathrm{SAAO_{I}}}$ & -5, $\mathcal{U}(-10, -1)$ &$-4.6564_{-0.0172}^{+0.0174}$\\ 
-  ($\mathrm{km/s}$) &$\ln{\sigma_\mathrm{jitter}} (RV_\mathrm{HARPS})$ & -2, $\mathcal{U}(-4, -1)$ &$ -2.1615_{-0.6698}^{+0.5819}$\\ \\
\textbf{Baseline parameters for GP}\\
$\ln{\sigma}_{NGTS}$&$\mathrm{offset (NGTS)}$ & -4.6161, $\mathcal{N}(-4.6161, 0.0587)$ &$ 4.5927_{-0.0359}^{+0.0386}$\\ 
$\ln{\rho}_{NGTS}$&$\mathrm{offset (NGTS)}$ &  -1.4840, $\mathcal{N}(-1.4840, 0.0764)$ &$ 1.6515_{-0.0577}^{+0.0497}$\\ 
$\ln{\sigma}_{TESS}$&$\mathrm{offset (TESS)}$ &  -4.4589, $\mathcal{N}(-4.4589, 0.1135)$ &$ 4.4356_{-0.0582}^{+0.0633 }$\\ 
$\ln{\rho}_{TESS}$&$\mathrm{offset (TESS)}$ &  -0.9483, $\mathcal{N}(-0.9483, 0.1049)$ &$ 0.9319_{-0.0678}^{+0.0697 }$\\ 
\hline
\end{tabular}
\end{table*}

\begin{table*}
\caption{The final \textsc{dns} derived parameters for \systemt, produced by \textsc{Allesfitter}. These values are used in future analysis.}              
\label{tab:derived_parameters}      
\centering   
\begin{tabular}{l c}          
\hline
\hline
Parameter & Value \\ 
\hline 
\multicolumn{2}{c}{\textit{Derived parameters}} \\ 
\hline 
Host radius over semi-major axis b; $R_\star/a_\mathrm{b}$&$0.1353_{-0.0020}^{+0.0023}$\\
Semi-major axis b over host radius; $a_\mathrm{b}/R_\star$&$7.3872_{-0.1210}^{+0.1127}$\\
Companion radius b over semi-major axis b; $R_\mathrm{b}/a_\mathrm{b}$&$0.0226_{-0.0004}^{+0.0005}$\\
Companion radius b; $R_\mathrm{b}$ ($\mathrm{R_{jup}}$)&$0.9534_{-0.0484}^{+0.0488}$\\
Semi-major axis b; $a_\mathrm{b}$ ($\mathrm{R_{\odot}}$)&$4.3387_{-0.2282}^{+0.2313}$\\
Semi-major axis b; $a_\mathrm{b}$ (AU)&$0.0202_{-0.0011}^{+0.0011}$\\
Inclination b; $i_\mathrm{b}$ (deg)&$85.3628_{-0.1863}^{+0.1735}$\\
Eccentricity b; $e_\mathrm{b}$&$0.0404_{-0.0101}^{+0.0067}$\\
Argument of periastron b; $w_\mathrm{b}$ (deg)&$261.7232_{-4.9514}^{+6.7190}$\\
Mass ratio b; $q_\mathrm{b}$&$0.1181_{-0.0069}^{+0.0077}$\\
Companion mass b; $M_\mathrm{b}$ ($\mathrm{M_{jup}}$)&$69.0125_{-4.8181}^{+5.3172}$\\
Companion mass b; $M_\mathrm{b}$ ($\mathrm{M_{\odot}}$)&$0.0659_{-0.0046}^{+0.0051}$\\
Impact parameter b; $b_\mathrm{tra;b}$&$0.6200_{-0.01367}^{+0.0147}$\\
Total transit duration b; $T_\mathrm{tot;b}$ (h)&$1.3404_{-0.0093}^{+0.0105}$\\
Full-transit duration b; $T_\mathrm{full;b}$ (h)&$0.7537_{-0.0135}^{+0.0117}$\\
Host density from orbit b; $\rho_\mathrm{\star;b}$ (cgs)&$4.8485_{-0.2344}^{+0.2253}$\\
Companion density b; $\rho_\mathrm{b}$ (cgs)&$98.7270_{-18.6470}^{+23.9807}$\\
Companion surface gravity b; $g_\mathrm{b}$ (cgs)&$210796.5463_{-8754.0340}^{+7782.8359}$\\
Equilibrium temperature b; $T_\mathrm{eq;b}$ (K)&$863.2442_{-12.5882}^{+13.1585}$\\
Transit depth (undil.) b; $\delta_\mathrm{tr; undil; b; NGTS_2}$ (ppt)&$30.2106_{-0.7502}^{+0.7684}$\\
Transit depth (dil.) b; $\delta_\mathrm{tr; dil; b; NGTS_2}$ (ppt)&$17.4391_{-0.2668}^{+0.3714}$\\
Transit depth (undil.) b; $\delta_\mathrm{tr; undil; b; TESS_2MIN}$ (ppt)&$29.8102_{-0.6463}^{+0.6466}$\\
Transit depth (dil.) b; $\delta_\mathrm{tr; dil; b; TESS_2MIN}$ (ppt)&$17.4538_{-0.2484}^{+0.2612}$\\
Transit depth (undil.) b; $\delta_\mathrm{tr; undil; b; SPEC1506_izp}$ (ppt)&$29.4026_{-0.3047}^{+0.3149}$\\
Transit depth (dil.) b; $\delta_\mathrm{tr; dil; b; SPEC1506_izp}$ (ppt)&$29.4026_{-0.3047}^{+0.3149}$\\
Transit depth (undil.) b; $\delta_\mathrm{tr; undil; b; SPEC1506_zp}$ (ppt)&$29.2768_{-0.3085}^{+0.3060}$\\
Transit depth (dil.) b; $\delta_\mathrm{tr; dil; b; SPEC1506_zp}$ (ppt)&$29.2768_{-0.3085}^{+0.3060}$\\
Transit depth (undil.) b; $\delta_\mathrm{tr; undil; b; SPEC1804_gp_short}$ (ppt)&$31.5803_{-0.3317}^{+0.3587}$\\
Transit depth (dil.) b; $\delta_\mathrm{tr; dil; b; SPEC1804_gp_short}$ (ppt)&$31.5803_{-0.3317}^{+0.3587}$\\
Transit depth (undil.) b; $\delta_\mathrm{tr; undil; b; SPEC1804_ip}$ (ppt)&$29.8533_{-0.6928}^{+0.7533}$\\
Transit depth (dil.) b; $\delta_\mathrm{tr; dil; b; SPEC1804_ip}$ (ppt)&$29.1418_{-0.5315}^{+0.4886}$\\
Transit depth (undil.) b; $\delta_\mathrm{tr; undil; b; SPEC1804_rp_short}$ (ppt)&$31.2219_{-0.3301}^{+0.3354}$\\
Transit depth (dil.) b; $\delta_\mathrm{tr; dil; b; SPEC1804_rp_short}$ (ppt)&$31.2219_{-0.3301}^{+0.3354}$\\
Transit depth (undil.) b; $\delta_\mathrm{tr; undil; b; TRAP1804_izp}$ (ppt)&$29.3713_{-1.6165}^{+1.7601}$\\
Transit depth (dil.) b; $\delta_\mathrm{tr; dil; b; TRAP1804_izp}$ (ppt)&$18.7979_{-0.7499}^{+0.7569}$\\
Transit depth (undil.) b; $\delta_\mathrm{tr; undil; b; TRAP2903_zp}$ (ppt)&$29.3008_{-1.6831}^{+1.7784}$\\
Transit depth (dil.) b; $\delta_\mathrm{tr; dil; b; TRAP2903_zp}$ (ppt)&$18.5946_{-0.7835}^{+0.8035}$\\
Transit depth (undil.) b; $\delta_\mathrm{tr; undil; b; SAAO_Is}$ (ppt)&$29.4383_{-1.3570}^{+1.4229}$\\
Transit depth (dil.) b; $\delta_\mathrm{tr; dil; b; SAAO_Is}$ (ppt)&$24.2032_{-0.8746}^{+0.7457}$\\
Limb darkening; $u_\mathrm{1; NGTS_2}$&$0.3897_{-0.0045}^{+0.0045}$\\
Limb darkening; $u_\mathrm{2; NGTS_2}$&$0.2823_{-0.0035}^{+0.0033}$\\
Limb darkening; $u_\mathrm{1; TESS_2MIN}$&$0.3069_{-0.0029}^{+0.0029}$\\
Limb darkening; $u_\mathrm{2; TESS_2MIN}$&$0.2748_{-0.0028}^{+0.0027}$\\
Limb darkening; $u_\mathrm{1; SPEC1506_izp}$&$0.2100_{-0.0022}^{+0.0022}$\\
Limb darkening; $u_\mathrm{2; SPEC1506_izp}$&$0.2684_{-0.0026}^{+0.0026}$\\
Limb darkening; $u_\mathrm{1; SPEC1506_zp}$&$0.1822_{-0.0020}^{+0.0019}$\\
Limb darkening; $u_\mathrm{2; SPEC1506_zp}$&$0.2666_{-0.0023}^{+0.0024}$\\
Limb darkening; $u_\mathrm{1; SPEC1804_gp_short}$&$0.6173_{-0.0050}^{+0.0053}$\\
Limb darkening; $u_\mathrm{2; SPEC1804_gp_short}$&$0.3265_{-0.0033}^{+0.0034}$\\
Limb darkening; $u_\mathrm{1; SPEC1804_ip}$&$0.3188_{-0.0035}^{+0.0038}$\\
Limb darkening; $u_\mathrm{2; SPEC1804_ip}$&$0.2768_{-0.0035}^{+0.0035}$\\
Limb darkening; $u_\mathrm{1; SPEC1804_rp_short}$&$0.5583_{-0.0057}^{+0.0054}$\\
Limb darkening; $u_\mathrm{2; SPEC1804_rp_short}$&$0.3244_{-0.0042}^{+0.0044}$\\
Limb darkening; $u_\mathrm{1; TRAP1804_izp}$&$0.2107_{-0.0049}^{+0.0048}$\\
Limb darkening; $u_\mathrm{2; TRAP1804_izp}$&$0.2682_{-0.0046}^{+0.0047}$\\
Limb darkening; $u_\mathrm{1; TRAP2903_zp}$&$0.1823_{-0.0039}^{+0.0042}$\\
Limb darkening; $u_\mathrm{2; TRAP2903_zp}$&$0.2668_{-0.0045}^{+0.0047}$\\
Limb darkening; $u_\mathrm{1; SAAO_Is}$&$0.5458_{-0.2874}^{+0.2411}$\\
Limb darkening; $u_\mathrm{2; SAAO_Is}$&$-0.1005_{-0.2210}^{+0.3080}$\\
\hline
\end{tabular}
\end{table*}

\section{Discussion}\label{sec:discussion}

\subsection{Population Comparisons: Bridging The Gap}\label{sec:massrad}

\begin{figure*}
    \centering
    \includegraphics[width=0.8\linewidth]{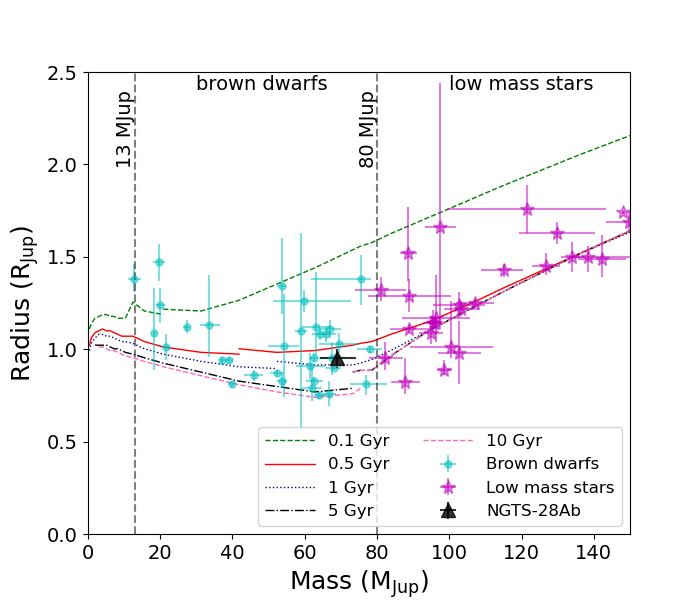}
    \caption{Mass-radius plot for the known transiting brown dwarfs (blue circles) and low mass stars (magenta stars), with \systemt\ (black triangle), adapted from Figure 10 in \citet{grieves21}. The vertical dashed lines show the upper and lower boundaries of the brown dwarf regime. The lines plotted along the graph show the expected radii of objects at all masses for 0.1 Gyr (green dashed), 0.5 Gyr (red solid), 1 Gyr (blue dotted), 5 Gyr (black dashed-dotted) and 10 Gyr (pink dashed) objects. These isochrones are from \citealt{baraffe03, baraffe15}. The \citet{baraffe03} isochrones are based on the \textsc{BT-Cond} models and \citet{baraffe15} improve on that with the inclusion of \textsc{BT-Settl} models \citep{btsettl}. The lowest masses of the isochrones are taken from \citet{baraffe03}. The disconnect in the isochrone lines is due to using the both models. Objects on this plot are listed within Table \ref{tab:full_obj_lists}. RIK-72b \citep{rik72b} and the binary system discovered by \citet{stassun06} are not included due to their youth and large radii.}
\label{fig:massradius}
\end{figure*}

\systemt\ has a mass of $69.0^{+5.3}_{-4.8}$ \mjup\ and a radius of $0.95\pm0.05$ \rjup, putting it at the upper boundary of the brown dwarf regime. The small radius also suggests that it is a brown dwarf, as most M-dwarfs are predicted by model isochrones to have radii above this value, although there is often a large scatter in the radii of late M-dwarfs (e.g. \citealt{parsons18}). Figure \ref{fig:massradius}, shows the mass-radius plot for known transiting objects within 12-150 \mjup, along with the \citealt{baraffe03, baraffe15} isochrones for 0.1, 0.5, 1, 5 and 10 Gyr ages. Objects plotted within Figure \ref{fig:massradius} are shown in Table \ref{tab:full_obj_lists}. Figure \ref{fig:massradius} and Table \ref{tab:full_obj_lists} have been adapted and updated from \citet{grieves21}. RIK 72B \citep{rik72b} and 2MASS J05352184–0546085 \citep{stassun06} are not included within this distribution due to their youth and hence large radii.

Figure \ref{fig:massradius} shows \systemt\ (plotted as the black triangle) clearly lies within the brown dwarf regime, with a system mass ratio of $0.1181_{-0.0077}^{+0.0069}$. \citet{bowler20} discuss the population statistics of brown dwarfs and hot Jupiters, splitting them based on mass and mass ratio. They found that high mass ratios (greater than 0.01) are more likely to have high eccentricities and indicate stellar formation mechanisms \citep{bowler20}. The high mass ratio of this system would make \systemt\ more likely to have formed via stellar formation mechanisms. 

\systemt\ sits between the 0.5 and 1 Gyr isochrones on Figure \ref{fig:massradius}. Although this is in agreement with the ages from SED modelling of \systemA\ and \systemB\ within errors, the large range of ages estimated for the system indicate it has a likely age range of >0.5~Gyrs.

Low mass M dwarfs have a wide distribution of radii at the low mass end with \citet{parsons18} determining that only 25 per cent of the M dwarfs they observed had radii consistent with models, while 12 per cent were inflated. It is not yet known whether this scatter continues into the brown dwarf regime, although the recent discovery of high mass inflated brown dwarfs indicates it may very well do (e.g. \citealt{19b, casewell20, schaffenroth}).

We determined the metallicity of \systemA\ using \textsc{specmatch-emp} and \textsc{ariadne}, determining a value of [Fe/H]=$-0.14_{-0.17}^{+0.16}$. Figure \ref{fig:massfeh} shows the mass and metallicity of the known late M dwarf and brown dwarf systems in Table \ref{tab:full_obj_lists}. \systemt\ is not an outlier in metallicity within the population for brown dwarfs and low mass stars.

Using \textsc{ariadne}, we determined an effective temperatures of $3626_{-44}^{+47}$~K for \systemA\ and $3441_{-52}^{+70}$~K for \systemB. \systemB\ has a lower temperature than \systemA\ and it is a redder object. Both objects have early M spectral types. Figure \ref{fig:massteff} shows the distribution of objects from Table \ref{tab:full_obj_lists} based on effective temperature. It also shows there is a lack of closely transiting objects around K dwarfs, with \systemA\ as one of the hottest M Dwarf hosts to transiting BDs, lying close to this gap. A similar gap is seen with eclipsing binaries, where tidal forces cause the companions of K-type stars to spiral inward, as discussed in \cite{ktype}.

\subsection{Rotation and Age}\label{sec:rotation}

As stated, Figure \ref{fig:massradius} shows \systemt\ has mass and radius values which are consistent with the 0.5-1 Gyr isochrone from \citealt{baraffe03,baraffe15}. This is also consistent with the age estimates of \systemA\ from \textsc{ariadne}. A range of ages is not unexpected due to the difficulty in determining ages of low mass stars. Other checks we performed to test the age of the system are discussed in this section.

 We performed a check to confirm whether \systemA\ and \systemB\ are part of a known young cluster or association. The \textsc{BANYAN} $\sum$ online tool \citep{banyan} was used. \textsc{BANYAN} $\sum$ uses RA, dec, proper motions and parallax to calculate cluster membership probabilities for various young clusters (up to 800 Myrs) within 150 pc, as well as the probability of the object being field age ($\sim$5 Gyr). For both stars, \textsc{BANYAN} $\sum$ estimates a 99.9 per cent probability of their being field objects, and are therefore unlikely to be members of young clusters. This high probability of both \systemA\ and \systemB\ being field age is consistent with the estimates from \textsc{ariadne}.

Another check for youth is whether \systemt\ is in spin-orbit sychronisation with \systemA. Spin-orbit synchronisation is expected for an object of this orbital radius/period. Figure \ref{fig:tess_full} shows the full \tess\ data for \system\ in the top panel. From the bottom of Figure \ref{fig:tess_full}, an apparent signature due to rotation can be seen, along with the transits and a flare. If it is believed to be the rotation of \systemA\ alone, it appears the rotation period of the object is very close to, but not quite equal to, the orbital period of \systemt. However, it is likely that the lightcurve modulation shows a composite effect of both \systemA\ and \systemB.

From the large uncertainties in the ages and the high probability of the objects being field age, it is likely that \systemt\ is old. If we make the assumption that the modulation seen in the \tess\ and \ngts\ data is from \systemA\ alone, we can calculate the rotation period using a Lomb-Scargle (LS) periodogram (\citealt{lomb,scargle}). We used \textsc{astropy}'s \textsc{LombScargle} package (\textsc{astropy}: \citealt{astropy1,astropy2,astropy3}), using a period range between 0.9 and 2 d. The errors associated with the period were calculated by randomly sampling the time-flux array for \tess\ and \ngts, allowing for duplication of values, and repeating the LS process for each random sample, known as `bootstrapping'. We repeated the process 1000 times and the standard deviation of the resultant periods was used as the error on the period. We determined the rotation period to be $1.42 \pm 0.02$~d from the \tess\ data and $1.42 \pm 0.11$~d from the \ngts\ data. Unfortunately, due to the high level of blending in the photometry from \tess\ and \ngts, we are unable to conclusively say if this rotation period of \systemA\ is accurate or is a modulation effect caused by the blended \systemB. This uncertainty means we cannot use this rotation period for gyrochronology or to determine any level of spin-orbit synchronisation accurately. Therefore, the mass-radius relation and \textsc{ariadne} is used to provide a best estimate for the age of \systemt\ of $>$0.5~Gyrs.

\subsection{System Eccentricity and Tidal Effects}\label{sec:ecc}

The eccentricity of \systemt\ is $0.0404_{-0.0101}^{+0.0067}$. To test if the eccentricity is a significant detection, we use the method from \citet{Lucy71}, where the detected eccentricity is only significant if the eccentricity value is 2.45$\sigma_{e}$ above the errors:

\begin{equation}
    e=\hat{e}, \text{ if } \hat{e} > 2.45\sigma_{e} ,
\end{equation}
\begin{equation}
    e=0, \text{ if } \hat{e} \le 2.45\sigma_{e} ,
\end{equation}

Where $\hat{e}$ is estimated eccentricity and $\sigma_{e}$ is the standard error on the value. Using this relationship from \citet{Lucy71}, our eccentricity is a significant detection. 

We expect \systemt\ to be circularised when considering the tidal circularisation timescale. Tidal circularisation is where tidal interactions between a star and it's companion cause the orbit to become less eccentric, first discussed in \citet{zahn89}. From there, objects then undergo rotational and orbital synchronisation. To calculate an estimate for the circularisation timescale we use Equations \eqref{eq:circstar}, \eqref{eq:circb} and \eqref{eq:circe} from \citet{jackson08}.

\begin{equation}\label{eq:circstar}
    \frac{1}{\tau_{circ,\star}} =  \frac{171}{16} \sqrt{\frac{G}{M_{\star}}} \frac{R_{\star}^{5} M_{b}}{Q_{\star}} a^{\frac{-13}{2}} ,
\end{equation}
\begin{equation}\label{eq:circb}
    \frac{1}{\tau_{circ,b}} =  \frac{63}{4} \frac{\sqrt{G M_{\star}^{3}} R_{b}^{5}}{Q_{b} M_{b}} a^{\frac{-13}{2}} ,
\end{equation}
\begin{equation}\label{eq:circe}
    \frac{1}{\tau_{e}} =  \frac{1}{\tau_{circ,\star}} + \frac{1}{\tau_{circ,b}} ,
\end{equation}

Where $\tau_{circ,\star}$, $\tau_{circ,b}$ and $\tau_{e}$ are the circularisation timescales for the star, companion and system respectively. $G$ is the gravitational constant, $M_{\star}$ and $M_{b}$ are the masses of the star and companion respectively, $a$ is the orbital radius and $Q_{\star}$ and $Q_{b}$ are the tidal quality factors. This will estimate a circularisation timescale with assumptions such as the orbit is not tidally locked, having a period less than 10 d and the eccentricity being low \citep{jackson08}. 

Using our results and a value of $10^{5}$ for $Q_{\star}$ (as suggested for M dwarfs in \citealt{heller10,heller11,mardling04}), we estimated the circularisation timescale for \systemt\ for a range of $Q_{b}$ values ($10^{4.5}$ to $10^{6}$) following the methods in \citet{carmichael20} and \citet{19b}. We use a range of values as the tidal quality factor is not well constrained for brown dwarfs. The results can be seen in Table \ref{tab:circ_time}. From this test, \systemt\ should have a circularised orbit due to the low timescales, regardless of $Q_{b}$ value. This is in disagreement to our significant eccentricity detection. With more radial velocity points we may be able to confirm whether the eccentricity is real.

\begin{table}
\caption{Estimated values for tidal circularisation timescale for \systemt. These values show a range of possible timescales for $Q_{b}$ values from $10^{4.5}$ to $10^{6}$. }              
\label{tab:circ_time}      
\centering   
\begin{tabular}{l c} 
\hline\hline  
$Q_{b}$ & $\tau_{e}$ (MYr)\\
\hline
$10^{4.5}$&0.96\\
$10^{4.75}$&0.97\\
$10^{5}$&0.98\\
$10^{5.25}$&0.99\\
$10^{5.5}$&0.99\\
$10^{5.75}$&0.99\\
$10^{6}$&1.00\\
\hline
\end{tabular}
\end{table}

\begin{figure*}
    \centering
    \subfloat[]{\includegraphics[width=0.48\linewidth]{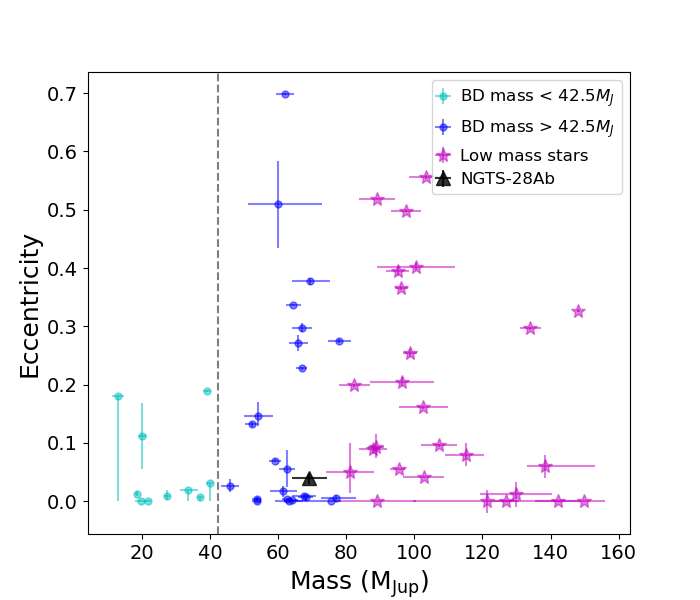}}
    \qquad
    \subfloat[]{\includegraphics[width=0.48\linewidth]{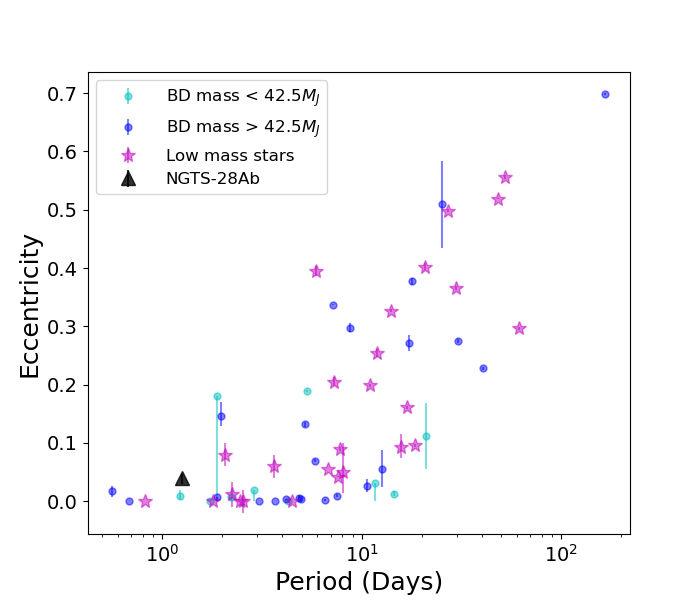}}
    \caption{\textbf{(a):} Mass-Eccentricity plot, with a vertical dashed line marking 42.5 \mjup. \textbf{(b):} Period-Eccentricity plot. Both figures show the objects from Table \ref{tab:full_obj_lists} with \systemt\ plotted as an black triangle. Low mass stars are plotted as magenta stars. BDs with masses above 42.5 \mjup\ are dark blue circles and BDs with below 42.5 \mjup\ are light blue circles. RIK-72b \citep{rik72b} and the binary system discovered by \citet{stassun06} are not included due to their youth.}
    \label{fig:ecc_plots}
\end{figure*}

How the object sits within the brown dwarf/low mass star population is important. Figure \ref{fig:ecc_plots}, shows the mass-eccentricity relation for all objects from Table \ref{tab:full_obj_lists}. In Figure \ref{fig:ecc_plots}, the objects which are more massive than 42.5 \mjup\ (shown by the dotted line), show a flat eccentricity distribution with increasing mass. This is in agreement with \citet{mage14} and \citet{grieves17}. \systemt\ lies comfortably within this flat distribution.

From the period-eccentricity plot in Figure \ref{fig:ecc_plots}, it is clear that objects with the highest eccentricities also tend to have long periods. \systemt\ is one of the shortest period transiting brown dwarfs within the desert.

\section{Conclusion}

\systemt\ is a $69.0^{+5.3}_{-4.8}$ \mjup, $0.95\pm0.05$ \rjup brown dwarf, transiting within the sparsely populated brown dwarf desert \citep{grether06}. Therefore, \systemt\ provides a new opportunity to gain a deeper understanding of this region. The brown dwarf desert is believed to be caused by the minimal overlap of the tail of distributions between planetary and stellar formation mechanisms (\citealt{grieves21,mage14}). Therefore, it is important to model parameters that can indicate the formation of each object. 

We used spectral fitting and SED fitting methods to analyse the host star \systemA\ as well as the probable common proper motion companion \systemB\ \citep{multiplicity}. We were then able to use the global modelling method \textsc{allesfitter} (\citealt{allesfitter-code, allesfitter-paper}) to model the parameters of \systemt. Through this we found \systemt\ is likely to have an age $>$0.5~Gyrs and has a high mass ratio of $0.1181_{-0.0077}^{+0.0069}$, meaning it likely formed via stellar formation mechanisms, with a low eccentricity value of $0.0404_{-0.0101}^{+0.0067}$.

We found that \systemt\ is orbiting one of the hottest M~dwarfs within the transiting BD desert. It is also found to be one of the shortest period transiting brown dwarfs within the desert.

\section*{Acknowledgements}

This work is based on data collected under the \ngts\ project at the ESO La Silla Paranal Observatory. The \ngts\ facility is operated by the consortium institutes with support from the UK Science and Technology Facilities Council (STFC) under projects ST/M001962/1, ST/S002642/1 and ST/W003163/1. This study is based on observations collected at the European Southern Observatory under ESO programme 105.20G9. This paper includes public data collected by the TESS mission. Funding for the TESS mission is provided by the NASA Explorer Program. This paper uses observations made at the South African Astronomical Observatory (SAAO). This work has made use of data from the European Space Agency (ESA) mission {\it Gaia} (\url{https://www.cosmos.esa.int/gaia}), processed by the {\it Gaia} Data Processing and Analysis Consortium (DPAC, \url{https://www.cosmos.esa.int/web/gaia/dpac/consortium}). Funding for the DPAC has been provided by national institutions, in particular, the institutions participating in the {\it Gaia} Multilateral Agreement.

This publication makes use of data products from the Two Micron All Sky Survey, which is a joint project of the University of Massachusetts and the Infrared Processing and Analysis Center/California Institute of Technology, funded by the National Aeronautics and Space Administration and the National Science Foundation. The Pan-STARRS1 Surveys (PS1) and the PS1 public science archive have been made possible through contributions by the Institute for Astronomy, the University of Hawaii, the Pan-STARRS Project Office, the Max-Planck Society and its participating institutes, the Max Planck Institute for Astronomy, Heidelberg and the Max Planck Institute for Extraterrestrial Physics, Garching, The Johns Hopkins University, Durham University, the University of Edinburgh, the Queen's University Belfast, the Harvard-Smithsonian Center for Astrophysics, the Las Cumbres Observatory Global Telescope Network Incorporated, the National Central University of Taiwan, the Space Telescope Science Institute, the National Aeronautics and Space Administration under Grant No. NNX08AR22G issued through the Planetary Science Division of the NASA Science Mission Directorate, the National Science Foundation Grant No. AST-1238877, the University of Maryland, Eotvos Lorand University (ELTE), the Los Alamos National Laboratory, and the Gordon and Betty Moore Foundation.

Some of the Observations in the paper made use of the High-Resolution Imaging instrument Zorro. Zorro was funded by the NASA Exoplanet Exploration Program and built at the NASA Ames Research Center by Steve B. Howell, Nic Scott, Elliott P. Horch, and Emmett Quigley. Zorro was mounted on the Gemini South telescope of the international Gemini Observatory, a program of NSF’s NOIRLab, which is managed by the Association of Universities for Research in Astronomy (AURA) under a cooperative agreement with the National Science Foundation. on behalf of the Gemini partnership: the National Science Foundation (United States), National Research Council (Canada), Agencia Nacional de Investigación y Desarrollo (Chile), Ministerio de Ciencia, Tecnología e Innovación (Argentina), Ministério da Ciência, Tecnologia, Inovações e Comunicações (Brazil), and Korea Astronomy and Space Science Institute (Republic of Korea).

BH is supported by an STFC studentship (ST/S505511/1 and ST/T506242/1).
SLC acknowledges support from an STFC Ernest Rutherford Fellowship (ST/R003726/1).
The contribution of FB, ML, OT and SU has been carried out within the framework of the NCCR PlanetS supported by the Swiss National Science Foundation under grants 51NF40\_182901 and 51NF40\_205606. ML acknowledges support of the Swiss National Science Foundation under grant number PCEFP2\_194576. The contribution of EMB has been supported by STFC through the consolidated grant ST/W001136/1.
CAC acknowledges that this research was carried out at the Jet Propulsion Laboratory, California Institute of Technology, under a contract with NASA (80NM0018D0004).
JSJ greatfully acknowledges support by FONDECYT grant 1201371 and from the ANID BASAL projects ACE210002 and FB210003.
The postdoctoral fellowship of KB is funded by F.R.S.-FNRS grant T.0109.20 and by the Francqui Foundation.
This publication benefits from the support of the French Community of Belgium in the context of the FRIA Doctoral Grant awarded to MT.
MG is F.R.S.-FNRS Research Director and EJ is F.R.S.-FNRS Senior Research Associate. F.J.P acknowledges financial support from the grant CEX2021-001131-S
funded by MCIN/AEI/ 10.13039/501100011033 and through projects
PID2019-109522GB-C52 and PID2022-137241NB-C43. 
This research is in part funded by the European Union's Horizon 2020 research and innovation programme (grant's agreement n$^{\circ}$ 803193/BEBOP), from the MERAC foundation, and from the Science and Technology Facilities Council (STFC; grant n$^\circ$ ST/S00193X/1, and ST/W000385/1).
Based on data collected by the SPECULOOS-South Observatory at the ESO Paranal Observatory in Chile.The ULiege's contribution to SPECULOOS has received funding from the European Research Council under the European Union's Seventh Framework Programme (FP/2007-2013) (grant Agreement n$^\circ$ 336480/SPECULOOS), from the Balzan Prize and Francqui Foundations, from the Belgian Scientific Research Foundation (F.R.S.-FNRS; grant n$^\circ$ T.0109.20), from the University of Liege, and from the ARC grant for Concerted Research Actions financed by the Wallonia-Brussels Federation. This work is supported by a grant from the Simons Foundation (PI Queloz, grant number 327127).
Based on data collected by the TRAPPIST-South telescope at the ESO La Silla Observatory. TRAPPIST is funded by the Belgian Fund for Scientific Research (Fond National de la Recherche Scientifique, FNRS) under the grant PDR T.0120.21, with the participation of the Swiss National Science Fundation (SNF).

\section*{Data Availability}

The \tess\ data is available via the MAST (MikulskiArchive for Space Telescopes) portal at \url{https://mast.stsci.edu/portal/Mashup/Clients/Mast/Portal.html}. Data can also be downloaded from TFOP for NGTS-28A (TOI-6110) at \url{https://exofop.ipac.caltech.edu/tess/target.php?id=7439480}. Public NGTS and HARPS data is available in the ESO arhive at \url{http://archive.eso.org/eso/eso_archive_main.html}. SAAO data is available on its public archive at \url{https://ssda.saao.ac.za/}. The other data within this article will be shared on reasonable request to the corresponding author.

\textsc{specmatch-emp}, \textsc{ariadne} and \textsc{Allesfitter} are open-source and public software.



\bibliographystyle{mnras}
\bibliography{bibliography.bib} 

\begin{thebibliography}{}
\makeatletter
\relax
\def\mn@urlcharsother{\let\do\@makeother \do\$\do\&\do\#\do\^\do\_\do\%\do\~}
\def\mn@doi{\begingroup\mn@urlcharsother \@ifnextchar [ {\mn@doi@}
  {\mn@doi@[]}}
\def\mn@doi@[#1]#2{\def\@tempa{#1}\ifx\@tempa\@empty \href
  {http://dx.doi.org/#2} {doi:#2}\else \href {http://dx.doi.org/#2} {#1}\fi
  \endgroup}
\def\mn@eprint#1#2{\mn@eprint@#1:#2::\@nil}
\def\mn@eprint@arXiv#1{\href {http://arxiv.org/abs/#1} {{\tt arXiv:#1}}}
\def\mn@eprint@dblp#1{\href {http://dblp.uni-trier.de/rec/bibtex/#1.xml}
  {dblp:#1}}
\def\mn@eprint@#1:#2:#3:#4\@nil{\def\@tempa {#1}\def\@tempb {#2}\def\@tempc
  {#3}\ifx \@tempc \@empty \let \@tempc \@tempb \let \@tempb \@tempa \fi \ifx
  \@tempb \@empty \def\@tempb {arXiv}\fi \@ifundefined
  {mn@eprint@\@tempb}{\@tempb:\@tempc}{\expandafter \expandafter \csname
  mn@eprint@\@tempb\endcsname \expandafter{\@tempc}}}

\bibitem[\protect\citeauthoryear{Acton et~al.,}{Acton et~al.}{2021}]{19b}
Acton J.~S.,  et~al., 2021, \mn@doi [\mnras] {10.1093/mnras/stab1459}, 505,
  2741–2752

\bibitem[\protect\citeauthoryear{{Allard}, {Homeier}  \& {Freytag}}{{Allard}
  et~al.}{2012}]{btsettl}
{Allard} F.,  {Homeier} D.,   {Freytag} B.,  2012, \mn@doi [Philosophical
  Transactions of the Royal Society of London Series A]
  {10.1098/rsta.2011.0269}, \href
  {https://ui.adsabs.harvard.edu/abs/2012RSPTA.370.2765A} {370, 2765}

\bibitem[\protect\citeauthoryear{{Alves} et~al.,}{{Alves}
  et~al.}{2022}]{alves22}
{Alves} D.~R.,  et~al., 2022, \mn@doi [\mnras] {10.1093/mnras/stac2884}, \href
  {https://ui.adsabs.harvard.edu/abs/2022MNRAS.tmp.2658A} {}

\bibitem[\protect\citeauthoryear{{Artigau} et~al.,}{{Artigau}
  et~al.}{2021}]{toi1278b}
{Artigau} {\'E}.,  et~al., 2021, \mn@doi [\aj] {10.3847/1538-3881/ac096d},
  \href {https://ui.adsabs.harvard.edu/abs/2021AJ....162..144A} {162, 144}

\bibitem[\protect\citeauthoryear{{Astropy Collaboration} et~al.,}{{Astropy
  Collaboration} et~al.}{2013}]{astropy1}
{Astropy Collaboration} et~al., 2013, \mn@doi [\aap]
  {10.1051/0004-6361/201322068}, \href
  {http://adsabs.harvard.edu/abs/2013A%26A...558A..33A} {558, A33}

\bibitem[\protect\citeauthoryear{{Astropy Collaboration} et~al.,}{{Astropy
  Collaboration} et~al.}{2018}]{astropy2}
{Astropy Collaboration} et~al., 2018, \mn@doi [\aj] {10.3847/1538-3881/aabc4f},
  \href {https://ui.adsabs.harvard.edu/abs/2018AJ....156..123A} {156, 123}

\bibitem[\protect\citeauthoryear{{Astropy Collaboration} et~al.,}{{Astropy
  Collaboration} et~al.}{2022}]{astropy3}
{Astropy Collaboration} et~al., 2022, \mn@doi [apj] {10.3847/1538-4357/ac7c74},
  \href {https://ui.adsabs.harvard.edu/abs/2022ApJ...935..167A} {935, 167}

\bibitem[\protect\citeauthoryear{{Baraffe}, {Chabrier}, {Allard}  \&
  {Hauschildt}}{{Baraffe} et~al.}{2002}]{baraffe02}
{Baraffe} I.,  {Chabrier} G.,  {Allard} F.,   {Hauschildt} P.~H.,  2002,
  \mn@doi [\aap] {10.1051/0004-6361:20011638}, \href
  {https://ui.adsabs.harvard.edu/abs/2002A&A...382..563B} {382, 563}

\bibitem[\protect\citeauthoryear{{Baraffe}, {Chabrier}, {Barman}, {Allard}  \&
  {Hauschildt}}{{Baraffe} et~al.}{2003}]{baraffe03}
{Baraffe} I.,  {Chabrier} G.,  {Barman} T.~S.,  {Allard} F.,   {Hauschildt}
  P.~H.,  2003, \mn@doi [\aap] {10.1051/0004-6361:20030252}, \href
  {https://ui.adsabs.harvard.edu/abs/2003A&A...402..701B} {402, 701}

\bibitem[\protect\citeauthoryear{{Baraffe}, {Homeier}, {Allard}  \&
  {Chabrier}}{{Baraffe} et~al.}{2015}]{baraffe15}
{Baraffe} I.,  {Homeier} D.,  {Allard} F.,   {Chabrier} G.,  2015, \mn@doi
  [\aap] {10.1051/0004-6361/201425481}, \href
  {https://ui.adsabs.harvard.edu/abs/2015A&A...577A..42B} {577, A42}

\bibitem[\protect\citeauthoryear{{Baranne} et~al.,}{{Baranne}
  et~al.}{1996}]{Baranne1996}
{Baranne} A.,  et~al., 1996, \aaps, \href
  {https://ui.adsabs.harvard.edu/abs/1996A&AS..119..373B} {119, 373}

\bibitem[\protect\citeauthoryear{{Barkaoui} et~al.,}{{Barkaoui}
  et~al.}{2023}]{barkaoui23}
{Barkaoui} K.,  et~al., 2023, \mn@doi [\aap] {10.1051/0004-6361/202346838},
  \href {https://ui.adsabs.harvard.edu/abs/2023A&A...677A..38B} {677, A38}

\bibitem[\protect\citeauthoryear{{Bate}, {Bonnell}  \& {Bromm}}{{Bate}
  et~al.}{2002}]{bate02}
{Bate} M.~R.,  {Bonnell} I.~A.,   {Bromm} V.,  2002, \mn@doi [\mnras]
  {10.1046/j.1365-8711.2002.05539.x}, \href
  {https://ui.adsabs.harvard.edu/abs/2002MNRAS.332L..65B} {332, L65}

\bibitem[\protect\citeauthoryear{{Bayliss} et~al.,}{{Bayliss}
  et~al.}{2017}]{epic201702477b}
{Bayliss} D.,  et~al., 2017, \mn@doi [\aj] {10.3847/1538-3881/153/1/15}, \href
  {https://ui.adsabs.harvard.edu/abs/2017AJ....153...15B} {153, 15}

\bibitem[\protect\citeauthoryear{{Beatty} et~al.,}{{Beatty}
  et~al.}{2007}]{hat205}
{Beatty} T.~G.,  et~al., 2007, \mn@doi [\apj] {10.1086/518413}, \href
  {https://ui.adsabs.harvard.edu/abs/2007ApJ...663..573B} {663, 573}

\bibitem[\protect\citeauthoryear{{Beatty} et~al.,}{{Beatty}
  et~al.}{2014}]{beatty14_1b}
{Beatty} T.~G.,  et~al., 2014, \mn@doi [\apj] {10.1088/0004-637X/783/2/112},
  \href {https://ui.adsabs.harvard.edu/abs/2014ApJ...783..112B} {783, 112}

\bibitem[\protect\citeauthoryear{{Benni} et~al.,}{{Benni} et~al.}{2021}]{gpx1b}
{Benni} P.,  et~al., 2021, \mn@doi [\mnras] {10.1093/mnras/stab1567}, \href
  {https://ui.adsabs.harvard.edu/abs/2021MNRAS.505.4956B} {505, 4956}

\bibitem[\protect\citeauthoryear{{Bonomo} et~al.,}{{Bonomo}
  et~al.}{2015}]{kepler39b}
{Bonomo} A.~S.,  et~al., 2015, \mn@doi [\aap] {10.1051/0004-6361/201323042},
  \href {https://ui.adsabs.harvard.edu/abs/2015A&A...575A..85B} {575, A85}

\bibitem[\protect\citeauthoryear{{Bouchy} et~al.,}{{Bouchy}
  et~al.}{2011}]{corot15b}
{Bouchy} F.,  et~al., 2011, \mn@doi [\aap] {10.1051/0004-6361/201015276}, \href
  {https://ui.adsabs.harvard.edu/abs/2011A&A...525A..68B} {525, A68}

\bibitem[\protect\citeauthoryear{{Bowler}, {Blunt}  \& {Nielsen}}{{Bowler}
  et~al.}{2020}]{bowler20}
{Bowler} B.~P.,  {Blunt} S.~C.,   {Nielsen} E.~L.,  2020, \mn@doi [\aj]
  {10.3847/1538-3881/ab5b11}, \href
  {https://ui.adsabs.harvard.edu/abs/2020AJ....159...63B} {159, 63}

\bibitem[\protect\citeauthoryear{{Broeg}, {Fern{\'a}ndez}  \&
  {Neuh{\"a}user}}{{Broeg} et~al.}{2005}]{broeg05}
{Broeg} C.,  {Fern{\'a}ndez} M.,   {Neuh{\"a}user} R.,  2005, \mn@doi
  [Astronomische Nachrichten] {10.1002/asna.200410350}, \href
  {https://ui.adsabs.harvard.edu/abs/2005AN....326..134B} {326, 134}

\bibitem[\protect\citeauthoryear{Burgasser}{Burgasser}{2008}]{burgasser}
Burgasser A.~J.,  2008, Physics Today, 61, 70

\bibitem[\protect\citeauthoryear{{Burrows} \& {Liebert}}{{Burrows} \&
  {Liebert}}{1993}]{burrows93}
{Burrows} A.,  {Liebert} J.,  1993, \mn@doi [Reviews of Modern Physics]
  {10.1103/RevModPhys.65.301}, \href
  {https://ui.adsabs.harvard.edu/abs/1993RvMP...65..301B} {65, 301}

\bibitem[\protect\citeauthoryear{{Carmichael}, {Latham}  \&
  {Vanderburg}}{{Carmichael} et~al.}{2019}]{carmichael19}
{Carmichael} T.~W.,  {Latham} D.~W.,   {Vanderburg} A.~M.,  2019, \mn@doi [\aj]
  {10.3847/1538-3881/ab245e}, \href
  {https://ui.adsabs.harvard.edu/abs/2019AJ....158...38C} {158, 38}

\bibitem[\protect\citeauthoryear{{Carmichael} et~al.,}{{Carmichael}
  et~al.}{2020}]{carmichael20}
{Carmichael} T.~W.,  et~al., 2020, \mn@doi [\aj] {10.3847/1538-3881/ab9b84},
  \href {https://ui.adsabs.harvard.edu/abs/2020AJ....160...53C} {160, 53}

\bibitem[\protect\citeauthoryear{Carmichael et~al.,}{Carmichael
  et~al.}{2021}]{carmichael21}
Carmichael T.~W.,  et~al., 2021, \mn@doi [\aj] {10.3847/1538-3881/abd4e1}, 161,
  97

\bibitem[\protect\citeauthoryear{{Carmichael} et~al.,}{{Carmichael}
  et~al.}{2022}]{carmichael22}
{Carmichael} T.~W.,  et~al., 2022, \mn@doi [\mnras] {10.1093/mnras/stac1666},
  \href {https://ui.adsabs.harvard.edu/abs/2022MNRAS.514.4944C} {514, 4944}

\bibitem[\protect\citeauthoryear{{Casewell} et~al.,}{{Casewell}
  et~al.}{2020a}]{wd1032}
{Casewell} S.~L.,  et~al., 2020a, \mn@doi [\mnras] {10.1093/mnras/staa1608},
  \href {https://ui.adsabs.harvard.edu/abs/2020MNRAS.497.3571C} {497, 3571}

\bibitem[\protect\citeauthoryear{{Casewell}, {Debes}, {Braker}, {Cushing},
  {Mace}, {Marley}  \& {Kirkpatrick}}{{Casewell} et~al.}{2020b}]{casewell20}
{Casewell} S.~L.,  {Debes} J.,  {Braker} I.~P.,  {Cushing} M.~C.,  {Mace} G.,
  {Marley} M.~S.,   {Kirkpatrick} J.~D.,  2020b, \mn@doi [\mnras]
  {10.1093/mnras/staa3184}, \href
  {https://ui.adsabs.harvard.edu/abs/2020MNRAS.499.5318C} {499, 5318}

\bibitem[\protect\citeauthoryear{Castelli \& Kurucz}{Castelli \&
  Kurucz}{2004}]{ck04}
Castelli F.,  Kurucz R.,  2004, Proceedings of the International Astronomical
  Union

\bibitem[\protect\citeauthoryear{{Chambers} et~al.,}{{Chambers}
  et~al.}{2016}]{panstarrs1}
{Chambers} K.~C.,  et~al., 2016, arXiv e-prints, \href
  {https://ui.adsabs.harvard.edu/abs/2016arXiv161205560C} {p. arXiv:1612.05560}

\bibitem[\protect\citeauthoryear{{Chaturvedi}, {Chakraborty}, {Anandarao},
  {Roy}  \& {Mahadevan}}{{Chaturvedi} et~al.}{2016}]{j2343}
{Chaturvedi} P.,  {Chakraborty} A.,  {Anandarao} B.~G.,  {Roy} A.,
  {Mahadevan} S.,  2016, \mn@doi [\mnras] {10.1093/mnras/stw1560}, \href
  {https://ui.adsabs.harvard.edu/abs/2016MNRAS.462..554C} {462, 554}

\bibitem[\protect\citeauthoryear{{Chaturvedi}, {Sharma}, {Chakraborty},
  {Anandarao}  \& {Prasad}}{{Chaturvedi} et~al.}{2018}]{ktype}
{Chaturvedi} P.,  {Sharma} R.,  {Chakraborty} A.,  {Anandarao} B.~G.,
  {Prasad} N. J.~S.~S.~V.,  2018, \mn@doi [\aj] {10.3847/1538-3881/aac5de},
  \href {https://ui.adsabs.harvard.edu/abs/2018AJ....156...27C} {156, 27}

\bibitem[\protect\citeauthoryear{{Ciardi}, {Beichman}, {Horch}  \&
  {Howell}}{{Ciardi} et~al.}{2015}]{dilution}
{Ciardi} D.~R.,  {Beichman} C.~A.,  {Horch} E.~P.,   {Howell} S.~B.,  2015,
  \mn@doi [\apj] {10.1088/0004-637X/805/1/16}, \href
  {https://ui.adsabs.harvard.edu/abs/2015ApJ...805...16C} {805, 16}

\bibitem[\protect\citeauthoryear{Coppejans et~al.,}{Coppejans
  et~al.}{2013}]{saao}
Coppejans R.,  et~al., 2013, \mn@doi [\pasp] {10.1086/672156}, 125, 976

\bibitem[\protect\citeauthoryear{Csizmadia}{Csizmadia}{2016}]{NLTTrad}
Csizmadia S.,  2016, in , The {CoRoT} Legacy Book.
{EDP} Sciences, p.~143, \mn@doi{10.1051/978-2-7598-1876-1.c036}, \url
  {https://doi.org/10.1051\%2F978-2-7598-1876-1.c036}

\bibitem[\protect\citeauthoryear{{Csizmadia} et~al.,}{{Csizmadia}
  et~al.}{2015}]{corot33b}
{Csizmadia} S.,  et~al., 2015, \mn@doi [\aap] {10.1051/0004-6361/201526763},
  \href {https://ui.adsabs.harvard.edu/abs/2015A&A...584A..13C} {584, A13}

\bibitem[\protect\citeauthoryear{{Dall}, {Santos}, {Arentoft}, {Bedding}  \&
  {Kjeldsen}}{{Dall} et~al.}{2006}]{Dall06}
{Dall} T.~H.,  {Santos} N.~C.,  {Arentoft} T.,  {Bedding} T.~R.,   {Kjeldsen}
  H.,  2006, \mn@doi [\aap] {10.1051/0004-6361:20065021}, \href
  {https://ui.adsabs.harvard.edu/abs/2006A&A...454..341D} {454, 341}

\bibitem[\protect\citeauthoryear{{David}, {Hillenbrand}, {Gillen}, {Cody},
  {Howell}, {Isaacson}  \& {Livingston}}{{David} et~al.}{2019}]{rik72b}
{David} T.~J.,  {Hillenbrand} L.~A.,  {Gillen} E.,  {Cody} A.~M.,  {Howell}
  S.~B.,  {Isaacson} H.~T.,   {Livingston} J.~H.,  2019, \mn@doi [\apj]
  {10.3847/1538-4357/aafe09}, \href
  {https://ui.adsabs.harvard.edu/abs/2019ApJ...872..161D} {872, 161}

\bibitem[\protect\citeauthoryear{{Deleuil} et~al.,}{{Deleuil}
  et~al.}{2008}]{corot3b}
{Deleuil} M.,  et~al., 2008, \mn@doi [\aap] {10.1051/0004-6361:200810625},
  \href {https://ui.adsabs.harvard.edu/abs/2008A&A...491..889D} {491, 889}

\bibitem[\protect\citeauthoryear{Delrez et~al.,}{Delrez
  et~al.}{2018}]{delrez2018}
Delrez L.,  et~al., 2018, in Ground-based and Airborne Telescopes VII. pp
  446--466

\bibitem[\protect\citeauthoryear{{D{\'\i}az} et~al.,}{{D{\'\i}az}
  et~al.}{2013}]{koi205b}
{D{\'\i}az} R.~F.,  et~al., 2013, \mn@doi [\aap] {10.1051/0004-6361/201321124},
  \href {https://ui.adsabs.harvard.edu/abs/2013A&A...551L...9D} {551, L9}

\bibitem[\protect\citeauthoryear{{D{\'\i}az} et~al.,}{{D{\'\i}az}
  et~al.}{2014}]{koi189b}
{D{\'\i}az} R.~F.,  et~al., 2014, \mn@doi [\aap] {10.1051/0004-6361/201424406},
  \href {https://ui.adsabs.harvard.edu/abs/2014A&A...572A.109D} {572, A109}

\bibitem[\protect\citeauthoryear{{Dransfield} et~al.,}{{Dransfield}
  et~al.}{2024}]{dransfield24}
{Dransfield} G.,  et~al., 2024, \mn@doi [\mnras] {10.1093/mnras/stad1439},
  \href {https://ui.adsabs.harvard.edu/abs/2024MNRAS.527...35D} {527, 35}

\bibitem[\protect\citeauthoryear{{Eaton}, {Draper}, {Allan}, {Naylor}, {Mukai},
  {Currie}  \& {McCaughrean}}{{Eaton} et~al.}{2014}]{Currie2014}
{Eaton} N.,  {Draper} P.~W.,  {Allan} A.,  {Naylor} T.,  {Mukai} K.,  {Currie}
  M.~J.,   {McCaughrean} M.,  2014, {PHOTOM: Photometry of digitized images},
  Astrophysics Source Code Library, record ascl:1405.013 (\mn@eprint {ascl}
  {1405.013})

\bibitem[\protect\citeauthoryear{{Flewelling} et~al.,}{{Flewelling}
  et~al.}{2020}]{panstarrs2}
{Flewelling} H.~A.,  et~al., 2020, \mn@doi [\apjs] {10.3847/1538-4365/abb82d},
  \href {https://ui.adsabs.harvard.edu/abs/2020ApJS..251....7F} {251, 7}

\bibitem[\protect\citeauthoryear{{Foreman-Mackey}, {Hogg}, {Lang}  \&
  {Goodman}}{{Foreman-Mackey} et~al.}{2013}]{emcee}
{Foreman-Mackey} D.,  {Hogg} D.~W.,  {Lang} D.,   {Goodman} J.,  2013, \mn@doi
  [\pasp] {10.1086/670067}, \href
  {https://ui.adsabs.harvard.edu/abs/2013PASP..125..306F} {125, 306}

\bibitem[\protect\citeauthoryear{{Furlan} \& {Howell}}{{Furlan} \&
  {Howell}}{2017}]{furlan17}
{Furlan} E.,  {Howell} S.~B.,  2017, \mn@doi [\aj] {10.3847/1538-3881/aa7b70},
  \href {https://ui.adsabs.harvard.edu/abs/2017AJ....154...66F} {154, 66}

\bibitem[\protect\citeauthoryear{{Gagn{\'e}} et~al.,}{{Gagn{\'e}}
  et~al.}{2018}]{banyan}
{Gagn{\'e}} J.,  et~al., 2018, \mn@doi [\apj] {10.3847/1538-4357/aaae09}, \href
  {https://ui.adsabs.harvard.edu/abs/2018ApJ...856...23G} {856, 23}

\bibitem[\protect\citeauthoryear{{Gaia Collaboration} et~al.,}{{Gaia
  Collaboration} et~al.}{2016}]{gaia}
{Gaia Collaboration} et~al., 2016, \mn@doi [\aap]
  {10.1051/0004-6361/201629272}, \href
  {https://ui.adsabs.harvard.edu/abs/2016A&A...595A...1G} {595, A1}

\bibitem[\protect\citeauthoryear{{Gaia Collaboration} et~al.,}{{Gaia
  Collaboration} et~al.}{2022}]{gaiadr3}
{Gaia Collaboration} et~al., 2022, arXiv e-prints, \href
  {https://ui.adsabs.harvard.edu/abs/2022arXiv220800211G} {p. arXiv:2208.00211}

\bibitem[\protect\citeauthoryear{{Garcia}, {Timmermans}, {Pozuelos}, {Ducrot},
  {Gillon}, {Delrez}, {Wells}  \& {Jehin}}{{Garcia} et~al.}{2022}]{garcia22}
{Garcia} L.~J.,  {Timmermans} M.,  {Pozuelos} F.~J.,  {Ducrot} E.,  {Gillon}
  M.,  {Delrez} L.,  {Wells} R.~D.,   {Jehin} E.,  2022, \mn@doi [\mnras]
  {10.1093/mnras/stab3113}, \href
  {https://ui.adsabs.harvard.edu/abs/2022MNRAS.509.4817G} {509, 4817}

\bibitem[\protect\citeauthoryear{{Gill} et~al.,}{{Gill} et~al.}{2020}]{tic2310}
{Gill} S.,  et~al., 2020, \mn@doi [\mnras] {10.1093/mnras/staa1248}, \href
  {https://ui.adsabs.harvard.edu/abs/2020MNRAS.495.2713G} {495, 2713}

\bibitem[\protect\citeauthoryear{{Gill} et~al.,}{{Gill}
  et~al.}{2022}]{tic320687387b}
{Gill} S.,  et~al., 2022, \mn@doi [\mnras] {10.1093/mnras/stac798}, \href
  {https://ui.adsabs.harvard.edu/abs/2022MNRAS.513.1785G} {513, 1785}

\bibitem[\protect\citeauthoryear{{Gillen}, {Hillenbrand}, {David}, {Aigrain},
  {Rebull}, {Stauffer}, {Cody}  \& {Queloz}}{{Gillen} et~al.}{2017}]{ad3116b}
{Gillen} E.,  {Hillenbrand} L.~A.,  {David} T.~J.,  {Aigrain} S.,  {Rebull} L.,
   {Stauffer} J.,  {Cody} A.~M.,   {Queloz} D.,  2017, \mn@doi [\apj]
  {10.3847/1538-4357/aa84b3}, \href
  {https://ui.adsabs.harvard.edu/abs/2017ApJ...849...11G} {849, 11}

\bibitem[\protect\citeauthoryear{{Gillon}, {Jehin}, {Magain}, {Chantry},
  {Hutsem{\'e}kers}, {Manfroid}, {Queloz}  \& {Udry}}{{Gillon}
  et~al.}{2011}]{gillon11}
{Gillon} M.,  {Jehin} E.,  {Magain} P.,  {Chantry} V.,  {Hutsem{\'e}kers} D.,
  {Manfroid} J.,  {Queloz} D.,   {Udry} S.,  2011, in European Physical Journal
  Web of Conferences. p. 06002 (\mn@eprint {arXiv} {1101.5807}),
  \mn@doi{10.1051/epjconf/20101106002}

\bibitem[\protect\citeauthoryear{{Grether} \& {Lineweaver}}{{Grether} \&
  {Lineweaver}}{2006}]{grether06}
{Grether} D.,  {Lineweaver} C.~H.,  2006, \mn@doi [\apj] {10.1086/500161},
  \href {https://ui.adsabs.harvard.edu/abs/2006ApJ...640.1051G} {640, 1051}

\bibitem[\protect\citeauthoryear{Grieves et~al.,}{Grieves
  et~al.}{2017}]{grieves17}
Grieves N.,  et~al., 2017, \mn@doi [\mnras] {10.1093/mnras/stx334}, 467, 4264

\bibitem[\protect\citeauthoryear{Grieves et~al.,}{Grieves
  et~al.}{2021}]{grieves21}
Grieves N.,  et~al., 2021, \mn@doi [\aap] {10.1051/0004-6361/202141145}, 652,
  A127

\bibitem[\protect\citeauthoryear{{G{\"u}nther} \& {Daylan}}{{G{\"u}nther} \&
  {Daylan}}{2019}]{allesfitter-code}
{G{\"u}nther} M.~N.,  {Daylan} T.,  2019, {Allesfitter: Flexible Star and
  Exoplanet Inference From Photometry and Radial Velocity}, Astrophysics Source
  Code Library (\mn@eprint {ascl} {1903.003})

\bibitem[\protect\citeauthoryear{{G{\"u}nther} \& {Daylan}}{{G{\"u}nther} \&
  {Daylan}}{2021}]{allesfitter-paper}
{G{\"u}nther} M.~N.,  {Daylan} T.,  2021, \mn@doi [\apjs]
  {10.3847/1538-4365/abe70e}, \href
  {https://ui.adsabs.harvard.edu/abs/2021ApJS..254...13G} {254, 13}

\bibitem[\protect\citeauthoryear{{G{\"u}nther} et~al.,}{{G{\"u}nther}
  et~al.}{2018}]{ngts3Ab}
{G{\"u}nther} M.~N.,  et~al., 2018, \mn@doi [\mnras] {10.1093/mnras/sty1193},
  \href {https://ui.adsabs.harvard.edu/abs/2018MNRAS.478.4720G} {478, 4720}

\bibitem[\protect\citeauthoryear{{Hauschildt}, {Allard}  \&
  {Baron}}{{Hauschildt} et~al.}{1999}]{btnextgen}
{Hauschildt} P.~H.,  {Allard} F.,   {Baron} E.,  1999, \mn@doi [\apj]
  {10.1086/306745}, \href
  {https://ui.adsabs.harvard.edu/abs/1999ApJ...512..377H} {512, 377}

\bibitem[\protect\citeauthoryear{{Heller}, {Jackson}, {Barnes}, {Greenberg}  \&
  {Homeier}}{{Heller} et~al.}{2010}]{heller10}
{Heller} R.,  {Jackson} B.,  {Barnes} R.,  {Greenberg} R.,   {Homeier} D.,
  2010, \mn@doi [\aap] {10.1051/0004-6361/200912826}, \href
  {https://ui.adsabs.harvard.edu/abs/2010A&A...514A..22H} {514, A22}

\bibitem[\protect\citeauthoryear{{Heller}, {Leconte}  \& {Barnes}}{{Heller}
  et~al.}{2011}]{heller11}
{Heller} R.,  {Leconte} J.,   {Barnes} R.,  2011, \mn@doi [\aap]
  {10.1051/0004-6361/201015809}, \href
  {https://ui.adsabs.harvard.edu/abs/2011A&A...528A..27H} {528, A27}

\bibitem[\protect\citeauthoryear{Hodžić et~al.,}{Hodžić
  et~al.}{2018}]{wasp128b}
Hodžić V.,  et~al., 2018, \mn@doi [\mnras] {10.1093/mnras/sty2512}, 481, 5091

\bibitem[\protect\citeauthoryear{Howell \& Furlan}{Howell \&
  Furlan}{2022}]{howell22}
Howell S.~B.,  Furlan E.,  2022, \mn@doi [Frontiers in Astronomy and Space
  Sciences] {10.3389/fspas.2022.871163}, 9

\bibitem[\protect\citeauthoryear{{Howell}, {Everett}, {Sherry}, {Horch}  \&
  {Ciardi}}{{Howell} et~al.}{2011}]{howell11}
{Howell} S.~B.,  {Everett} M.~E.,  {Sherry} W.,  {Horch} E.,   {Ciardi} D.~R.,
  2011, \mn@doi [\aj] {10.1088/0004-6256/142/1/19}, \href
  {https://ui.adsabs.harvard.edu/abs/2011AJ....142...19H} {142, 19}

\bibitem[\protect\citeauthoryear{{Husser}, {Wende-von Berg}, {Dreizler},
  {Homeier}, {Reiners}, {Barman}  \& {Hauschildt}}{{Husser}
  et~al.}{2013}]{phoenix}
{Husser} T.~O.,  {Wende-von Berg} S.,  {Dreizler} S.,  {Homeier} D.,  {Reiners}
  A.,  {Barman} T.,   {Hauschildt} P.~H.,  2013, \mn@doi [\aap]
  {10.1051/0004-6361/201219058}, \href
  {https://ui.adsabs.harvard.edu/abs/2013A&A...553A...6H} {553, A6}

\bibitem[\protect\citeauthoryear{{Irwin} et~al.,}{{Irwin} et~al.}{2010}]{nltt}
{Irwin} J.,  et~al., 2010, \mn@doi [\apj] {10.1088/0004-637X/718/2/1353}, \href
  {https://ui.adsabs.harvard.edu/abs/2010ApJ...718.1353I} {718, 1353}

\bibitem[\protect\citeauthoryear{{Irwin} et~al.,}{{Irwin} et~al.}{2018}]{lp261}
{Irwin} J.~M.,  et~al., 2018, \mn@doi [\aj] {10.3847/1538-3881/aad9a3}, \href
  {https://ui.adsabs.harvard.edu/abs/2018AJ....156..140I} {156, 140}

\bibitem[\protect\citeauthoryear{{Jackman} et~al.,}{{Jackman}
  et~al.}{2019}]{ngts7ab}
{Jackman} J. A.~G.,  et~al., 2019, \mn@doi [\mnras] {10.1093/mnras/stz2496},
  \href {https://ui.adsabs.harvard.edu/abs/2019MNRAS.489.5146J} {489, 5146}

\bibitem[\protect\citeauthoryear{{Jackson}, {Greenberg}  \& {Barnes}}{{Jackson}
  et~al.}{2008}]{jackson08}
{Jackson} B.,  {Greenberg} R.,   {Barnes} R.,  2008, \mn@doi [\apj]
  {10.1086/529187}, \href
  {https://ui.adsabs.harvard.edu/abs/2008ApJ...678.1396J} {678, 1396}

\bibitem[\protect\citeauthoryear{Jeffreys}{Jeffreys}{1998}]{jeffreys98theory}
Jeffreys H.,  1998, The Theory of Probability.
Oxford Classic Texts in the Physical Sciences, OUP Oxford, \url
  {https://books.google.co.uk/books?id=vh9Act9rtzQC}

\bibitem[\protect\citeauthoryear{{Jehin} et~al.,}{{Jehin}
  et~al.}{2011}]{jehin11}
{Jehin} E.,  et~al., 2011, The Messenger, \href
  {https://ui.adsabs.harvard.edu/abs/2011Msngr.145....2J} {145, 2}

\bibitem[\protect\citeauthoryear{{Jehin} et~al.,}{{Jehin}
  et~al.}{2018}]{jehin18}
{Jehin} E.,  et~al., 2018, \mn@doi [The Messenger] {10.18727/0722-6691/5105},
  \href {https://ui.adsabs.harvard.edu/abs/2018Msngr.174....2J} {174, 2}

\bibitem[\protect\citeauthoryear{Jenkins et~al.,}{Jenkins
  et~al.}{2015}]{jenkins15}
Jenkins J.~S.,  et~al., 2015, \mn@doi [\mnras] {10.1093/mnras/stv1596}, 453,
  1439

\bibitem[\protect\citeauthoryear{{Jenkins} et~al.,}{{Jenkins}
  et~al.}{2016}]{spoc}
{Jenkins} J.~M.,  et~al., 2016, in {Chiozzi} G.,  {Guzman} J.~C.,  eds,
  Society of Photo-Optical Instrumentation Engineers (SPIE) Conference Series
  Vol. 9913, Software and Cyberinfrastructure for Astronomy IV. p. 99133E,
  \mn@doi{10.1117/12.2233418}

\bibitem[\protect\citeauthoryear{{Johnson} et~al.,}{{Johnson}
  et~al.}{2011}]{lhs6343}
{Johnson} J.~A.,  et~al., 2011, \mn@doi [\apj] {10.1088/0004-637X/730/2/79},
  \href {https://ui.adsabs.harvard.edu/abs/2011ApJ...730...79J} {730, 79}

\bibitem[\protect\citeauthoryear{Kass \& Raftery}{Kass \& Raftery}{1995}]{kass}
Kass R.~E.,  Raftery A.~E.,  1995, \mn@doi [Journal of the American Statistical
  Association] {10.1080/01621459.1995.10476572}, 90, 773

\bibitem[\protect\citeauthoryear{Kipping}{Kipping}{2013}]{kipping13}
Kipping D.~M.,  2013, \mn@doi [\mnras] {10.1093/mnras/stt1435}, 435, 2152

\bibitem[\protect\citeauthoryear{{Kurucz}}{{Kurucz}}{1993}]{kurucz}
{Kurucz} R.,  1993, ATLAS9 Stellar Atmosphere Programs and 2 km/s grid. Kurucz
  CD-ROM No. 13. Cambridge, \href
  {https://ui.adsabs.harvard.edu/abs/1993KurCD..13.....K} {13}

\bibitem[\protect\citeauthoryear{{Latham}, {Mazeh}, {Stefanik}, {Mayor}  \&
  {Burki}}{{Latham} et~al.}{1989}]{rvbd}
{Latham} D.~W.,  {Mazeh} T.,  {Stefanik} R.~P.,  {Mayor} M.,   {Burki} G.,
  1989, \mn@doi [\nat] {10.1038/339038a0}, \href
  {https://ui.adsabs.harvard.edu/abs/1989Natur.339...38L} {339, 38}

\bibitem[\protect\citeauthoryear{{Lomb}}{{Lomb}}{1976}]{lomb}
{Lomb} N.~R.,  1976, \mn@doi [\apss] {10.1007/BF00648343}, \href
  {https://ui.adsabs.harvard.edu/abs/1976Ap&SS..39..447L} {39, 447}

\bibitem[\protect\citeauthoryear{{Lucy} \& {Sweeney}}{{Lucy} \&
  {Sweeney}}{1971}]{Lucy71}
{Lucy} L.~B.,  {Sweeney} M.~A.,  1971, \mn@doi [\aj] {10.1086/111159}, \href
  {https://ui.adsabs.harvard.edu/abs/1971AJ.....76..544L} {76, 544}

\bibitem[\protect\citeauthoryear{Ma \& Ge}{Ma \& Ge}{2014}]{mage14}
Ma B.,  Ge J.,  2014, \mn@doi [\mnras] {10.1093/mnras/stu134}, 439, 2781

\bibitem[\protect\citeauthoryear{{Mardling} \& {Lin}}{{Mardling} \&
  {Lin}}{2004}]{mardling04}
{Mardling} R.~A.,  {Lin} D.~N.~C.,  2004, \mn@doi [\apj] {10.1086/423794},
  \href {https://ui.adsabs.harvard.edu/abs/2004ApJ...614..955M} {614, 955}

\bibitem[\protect\citeauthoryear{Marley et~al.,}{Marley
  et~al.}{2021}]{marley21}
Marley M.~S.,  et~al., 2021, \mn@doi [\apj] {10.3847/1538-4357/ac141d}, 920, 85

\bibitem[\protect\citeauthoryear{{Maxted}}{{Maxted}}{2016}]{ellc}
{Maxted} P.~F.~L.,  2016, \mn@doi [\aap] {10.1051/0004-6361/201628579}, \href
  {https://ui.adsabs.harvard.edu/abs/2016A&A...591A.111M} {591, A111}

\bibitem[\protect\citeauthoryear{{Mayor} et~al.,}{{Mayor} et~al.}{2003}]{harps}
{Mayor} M.,  et~al., 2003, The Messenger, \href
  {https://ui.adsabs.harvard.edu/abs/2003Msngr.114...20M} {114, 20}

\bibitem[\protect\citeauthoryear{{Mireles} et~al.,}{{Mireles}
  et~al.}{2020}]{toi694}
{Mireles} I.,  et~al., 2020, \mn@doi [\aj] {10.3847/1538-3881/aba526}, \href
  {https://ui.adsabs.harvard.edu/abs/2020AJ....160..133M} {160, 133}

\bibitem[\protect\citeauthoryear{{Moutou} et~al.,}{{Moutou}
  et~al.}{2013}]{koi415b}
{Moutou} C.,  et~al., 2013, \mn@doi [\aap] {10.1051/0004-6361/201322201}, \href
  {https://ui.adsabs.harvard.edu/abs/2013A&A...558L...6M} {558, L6}

\bibitem[\protect\citeauthoryear{{Mugrauer}, {R{\"u}ck}  \&
  {Michel}}{{Mugrauer} et~al.}{2023}]{multiplicity}
{Mugrauer} M.,  {R{\"u}ck} J.,   {Michel} K.~U.,  2023, \mn@doi [arXiv
  e-prints] {10.48550/arXiv.2306.02452}, \href
  {https://ui.adsabs.harvard.edu/abs/2023arXiv230602452M} {p. arXiv:2306.02452}

\bibitem[\protect\citeauthoryear{{Nakajima}, {Oppenheimer}, {Kulkarni},
  {Golimowski}, {Matthews}  \& {Durrance}}{{Nakajima} et~al.}{1995}]{gliese}
{Nakajima} T.,  {Oppenheimer} B.~R.,  {Kulkarni} S.~R.,  {Golimowski} D.~A.,
  {Matthews} K.,   {Durrance} S.~T.,  1995, \mn@doi [\nat] {10.1038/378463a0},
  \href {https://ui.adsabs.harvard.edu/abs/1995Natur.378..463N} {378, 463}

\bibitem[\protect\citeauthoryear{{Nefs} et~al.,}{{Nefs} et~al.}{2013}]{wts19g}
{Nefs} S.~V.,  et~al., 2013, \mn@doi [\mnras] {10.1093/mnras/stt405}, \href
  {https://ui.adsabs.harvard.edu/abs/2013MNRAS.431.3240N} {431, 3240}

\bibitem[\protect\citeauthoryear{{Nowak} et~al.,}{{Nowak}
  et~al.}{2017}]{cww89ab1}
{Nowak} G.,  et~al., 2017, \mn@doi [\aj] {10.3847/1538-3881/aa5cb6}, \href
  {https://ui.adsabs.harvard.edu/abs/2017AJ....153..131N} {153, 131}

\bibitem[\protect\citeauthoryear{{Ofir}, {Gandolfi}, {Buchhave}, {Lacy},
  {Hatzes}  \& {Fridlund}}{{Ofir} et~al.}{2012}]{kic1571}
{Ofir} A.,  {Gandolfi} D.,  {Buchhave} L.,  {Lacy} C.~H.~S.,  {Hatzes} A.~P.,
  {Fridlund} M.,  2012, \mn@doi [\mnras] {10.1111/j.1745-3933.2011.01191.x},
  \href {https://ui.adsabs.harvard.edu/abs/2012MNRAS.423L...1O} {423, L1}

\bibitem[\protect\citeauthoryear{{Palle} et~al.,}{{Palle}
  et~al.}{2021}]{toi263b2}
{Palle} E.,  et~al., 2021, \mn@doi [\aap] {10.1051/0004-6361/202039937}, \href
  {https://ui.adsabs.harvard.edu/abs/2021A&A...650A..55P} {650, A55}

\bibitem[\protect\citeauthoryear{Parsons et~al.,}{Parsons
  et~al.}{2018}]{parsons18}
Parsons S.~G.,  et~al., 2018, \mn@doi [\mnras] {10.1093/mnras/sty2345}, 481,
  1083

\bibitem[\protect\citeauthoryear{Parviainen \& Aigrain}{Parviainen \&
  Aigrain}{2015}]{ldtk}
Parviainen H.,  Aigrain S.,  2015, \mn@doi [MNRAS] {10.1093/mnras/stv1857},
  453, 3821

\bibitem[\protect\citeauthoryear{{Parviainen} et~al.,}{{Parviainen}
  et~al.}{2020}]{toi263b1}
{Parviainen} H.,  et~al., 2020, \mn@doi [\aap] {10.1051/0004-6361/201935958},
  \href {https://ui.adsabs.harvard.edu/abs/2020A&A...633A..28P} {633, A28}

\bibitem[\protect\citeauthoryear{{Pecaut} \& {Mamajek}}{{Pecaut} \&
  {Mamajek}}{2013}]{mamajek}
{Pecaut} M.~J.,  {Mamajek} E.~E.,  2013, \mn@doi [\apjs]
  {10.1088/0067-0049/208/1/9}, \href
  {https://ui.adsabs.harvard.edu/abs/2013ApJS..208....9P} {208, 9}

\bibitem[\protect\citeauthoryear{{Pepe}, {Mayor}, {Galland}, {Naef}, {Queloz},
  {Santos}, {Udry}  \& {Burnet}}{{Pepe} et~al.}{2002}]{Pepe2002}
{Pepe} F.,  {Mayor} M.,  {Galland} F.,  {Naef} D.,  {Queloz} D.,  {Santos}
  N.~C.,  {Udry} S.,   {Burnet} M.,  2002, \mn@doi [\aap]
  {10.1051/0004-6361:20020433}, \href
  {https://ui.adsabs.harvard.edu/abs/2002A&A...388..632P} {388, 632}

\bibitem[\protect\citeauthoryear{{Persson} et~al.,}{{Persson}
  et~al.}{2019}]{epic212036875b1}
{Persson} C.~M.,  et~al., 2019, \mn@doi [\aap] {10.1051/0004-6361/201935505},
  \href {https://ui.adsabs.harvard.edu/abs/2019A&A...628A..64P} {628, A64}

\bibitem[\protect\citeauthoryear{{Pont}, {Melo}, {Bouchy}, {Udry}, {Queloz},
  {Mayor}  \& {Santos}}{{Pont} et~al.}{2005a}]{ogle122b}
{Pont} F.,  {Melo} C.~H.~F.,  {Bouchy} F.,  {Udry} S.,  {Queloz} D.,  {Mayor}
  M.,   {Santos} N.~C.,  2005a, \mn@doi [\aap] {10.1051/0004-6361:200500025},
  \href {https://ui.adsabs.harvard.edu/abs/2005A&A...433L..21P} {433, L21}

\bibitem[\protect\citeauthoryear{{Pont}, {Bouchy}, {Melo}, {Santos}, {Mayor},
  {Queloz}  \& {Udry}}{{Pont} et~al.}{2005b}]{ogle106b}
{Pont} F.,  {Bouchy} F.,  {Melo} C.,  {Santos} N.~C.,  {Mayor} M.,  {Queloz}
  D.,   {Udry} S.,  2005b, \mn@doi [\aap] {10.1051/0004-6361:20052771}, \href
  {https://ui.adsabs.harvard.edu/abs/2005A&A...438.1123P} {438, 1123}

\bibitem[\protect\citeauthoryear{{Pont} et~al.,}{{Pont}
  et~al.}{2006}]{ogle123b}
{Pont} F.,  et~al., 2006, \mn@doi [\aap] {10.1051/0004-6361:20053692}, \href
  {https://ui.adsabs.harvard.edu/abs/2006A&A...447.1035P} {447, 1035}

\bibitem[\protect\citeauthoryear{{Psaridi} et~al.,}{{Psaridi}
  et~al.}{2022}]{psaridi22}
{Psaridi} A.,  et~al., 2022, \mn@doi [\aap] {10.1051/0004-6361/202243454},
  \href {https://ui.adsabs.harvard.edu/abs/2022A&A...664A..94P} {664, A94}

\bibitem[\protect\citeauthoryear{{Queloz} et~al.,}{{Queloz}
  et~al.}{2001}]{queloz01}
{Queloz} D.,  et~al., 2001, \mn@doi [\aap] {10.1051/0004-6361:20011308}, \href
  {https://ui.adsabs.harvard.edu/abs/2001A&A...379..279Q} {379, 279}

\bibitem[\protect\citeauthoryear{Raghavan et~al.,}{Raghavan
  et~al.}{2010}]{Raghavan10}
Raghavan D.,  et~al., 2010, \mn@doi [Astrophysical Journal, Supplement Series]
  {10.1088/0067-0049/190/1/1}, 190, 1

\bibitem[\protect\citeauthoryear{{Rajpurohit}, {Reyl{\'e}}, {Allard},
  {Homeier}, {Schultheis}, {Bessell}  \& {Robin}}{{Rajpurohit}
  et~al.}{2013}]{spec}
{Rajpurohit} A.~S.,  {Reyl{\'e}} C.,  {Allard} F.,  {Homeier} D.,  {Schultheis}
  M.,  {Bessell} M.~S.,   {Robin} A.~C.,  2013, \mn@doi [\aap]
  {10.1051/0004-6361/201321346}, \href
  {https://ui.adsabs.harvard.edu/abs/2013A&A...556A..15R} {556, A15}

\bibitem[\protect\citeauthoryear{{Rebolo}, {Zapatero Osorio}  \&
  {Mart{\'\i}n}}{{Rebolo} et~al.}{1995}]{teide}
{Rebolo} R.,  {Zapatero Osorio} M.~R.,   {Mart{\'\i}n} E.~L.,  1995, \mn@doi
  [\nat] {10.1038/377129a0}, \href
  {https://ui.adsabs.harvard.edu/abs/1995Natur.377..129R} {377, 129}

\bibitem[\protect\citeauthoryear{Ricker et~al.,}{Ricker et~al.}{2014}]{tess}
Ricker G.~R.,  et~al., 2014, \mn@doi [Journal of Astronomical Telescopes,
  Instruments, and Systems] {10.1117/1.JATIS.1.1.014003}, 1, 1

\bibitem[\protect\citeauthoryear{{STScI Development Team}}{{STScI Development
  Team}}{2013}]{pysyn}
{STScI Development Team} 2013, {pysynphot: Synthetic photometry software
  package}, Astrophysics Source Code Library, record ascl:1303.023 (\mn@eprint
  {ascl} {1303.023})

\bibitem[\protect\citeauthoryear{{Santos} et~al.,}{{Santos}
  et~al.}{2002}]{santos02}
{Santos} N.~C.,  et~al., 2002, \mn@doi [\aap] {10.1051/0004-6361:20020876},
  \href {https://ui.adsabs.harvard.edu/abs/2002A&A...392..215S} {392, 215}

\bibitem[\protect\citeauthoryear{{Scargle}}{{Scargle}}{1982}]{scargle}
{Scargle} J.~D.,  1982, \mn@doi [\apj] {10.1086/160554}, \href
  {https://ui.adsabs.harvard.edu/abs/1982ApJ...263..835S} {263, 835}

\bibitem[\protect\citeauthoryear{{Schaffenroth} et~al.,}{{Schaffenroth}
  et~al.}{2021}]{schaffenroth}
{Schaffenroth} V.,  et~al., 2021, \mn@doi [\mnras] {10.1093/mnras/staa3661},
  \href {https://ui.adsabs.harvard.edu/abs/2021MNRAS.501.3847S} {501, 3847}

\bibitem[\protect\citeauthoryear{{Science Software Branch at STScI}}{{Science
  Software Branch at STScI}}{2012}]{pyraf}
{Science Software Branch at STScI} 2012, {PyRAF: Python alternative for IRAF},
  Astrophysics Source Code Library, record ascl:1207.011 (\mn@eprint {ascl}
  {1207.011})

\bibitem[\protect\citeauthoryear{Scott et~al.,}{Scott et~al.}{2021}]{scott21}
Scott N.~J.,  et~al., 2021, \mn@doi [Frontiers in Astronomy and Space Sciences]
  {10.3389/fspas.2021.716560}, 8

\bibitem[\protect\citeauthoryear{{Sebastian} et~al.,}{{Sebastian}
  et~al.}{2021}]{sebastian21}
{Sebastian} D.,  et~al., 2021, \mn@doi [\aap] {10.1051/0004-6361/202038827},
  \href {https://ui.adsabs.harvard.edu/abs/2021A&A...645A.100S} {645, A100}

\bibitem[\protect\citeauthoryear{{Shporer} et~al.,}{{Shporer}
  et~al.}{2017}]{k276}
{Shporer} A.,  et~al., 2017, \mn@doi [\apjl] {10.3847/2041-8213/aa8bff}, \href
  {https://ui.adsabs.harvard.edu/abs/2017ApJ...847L..18S} {847, L18}

\bibitem[\protect\citeauthoryear{{Siverd} et~al.,}{{Siverd}
  et~al.}{2012}]{kelt1b}
{Siverd} R.~J.,  et~al., 2012, \mn@doi [\apj] {10.1088/0004-637X/761/2/123},
  \href {https://ui.adsabs.harvard.edu/abs/2012ApJ...761..123S} {761, 123}

\bibitem[\protect\citeauthoryear{{Skrutskie} et~al.,}{{Skrutskie}
  et~al.}{2006}]{2mass}
{Skrutskie} M.~F.,  et~al., 2006, \mn@doi [\aj] {10.1086/498708}, \href
  {https://ui.adsabs.harvard.edu/abs/2006AJ....131.1163S} {131, 1163}

\bibitem[\protect\citeauthoryear{{Speagle}}{{Speagle}}{2020}]{dynesty}
{Speagle} J.~S.,  2020, \mn@doi [\mnras] {10.1093/mnras/staa278}, \href
  {https://ui.adsabs.harvard.edu/abs/2020MNRAS.493.3132S} {493, 3132}

\bibitem[\protect\citeauthoryear{{Stassun}, {Mathieu}  \& {Valenti}}{{Stassun}
  et~al.}{2006}]{stassun06}
{Stassun} K.~G.,  {Mathieu} R.~D.,   {Valenti} J.~A.,  2006, \mn@doi [\nat]
  {10.1038/nature04570}, \href
  {https://ui.adsabs.harvard.edu/abs/2006Natur.440..311S} {440, 311}

\bibitem[\protect\citeauthoryear{{Tal-Or} et~al.,}{{Tal-Or}
  et~al.}{2013}]{corot1011}
{Tal-Or} L.,  et~al., 2013, \mn@doi [\aap] {10.1051/0004-6361/201220862}, \href
  {https://ui.adsabs.harvard.edu/abs/2013A&A...553A..30T} {553, A30}

\bibitem[\protect\citeauthoryear{Tamuz, Mazeh  \& Zucker}{Tamuz
  et~al.}{2005}]{sysrem}
Tamuz O.,  Mazeh T.,   Zucker S.,  2005, \mn@doi [\mnras]
  {10.1111/j.1365-2966.2004.08585.x}, 356, 1466

\bibitem[\protect\citeauthoryear{{Thorngren}, {Fortney}, {Lopez}, {Berger}  \&
  {Huber}}{{Thorngren} et~al.}{2021}]{thorngren}
{Thorngren} D.~P.,  {Fortney} J.~J.,  {Lopez} E.~D.,  {Berger} T.~A.,   {Huber}
  D.,  2021, \mn@doi [\apjl] {10.3847/2041-8213/abe86d}, \href
  {https://ui.adsabs.harvard.edu/abs/2021ApJ...909L..16T} {909, L16}

\bibitem[\protect\citeauthoryear{{Tilbrook} et~al.,}{{Tilbrook}
  et~al.}{2021}]{tilbrook}
{Tilbrook} R.~H.,  et~al., 2021, \mn@doi [\mnras] {10.1093/mnras/stab815},
  \href {https://ui.adsabs.harvard.edu/abs/2021MNRAS.504.6018T} {504, 6018}

\bibitem[\protect\citeauthoryear{{Tody}}{{Tody}}{1986}]{tody86}
{Tody} D.,  1986, in {Crawford} D.~L.,  ed.,  Society of Photo-Optical
  Instrumentation Engineers (SPIE) Conference Series Vol. 627, Instrumentation
  in astronomy VI. p.~733, \mn@doi{10.1117/12.968154}

\bibitem[\protect\citeauthoryear{{Tody}}{{Tody}}{1993}]{tody93}
{Tody} D.,  1993, in {Hanisch} R.~J.,  {Brissenden} R.~J.~V.,   {Barnes} J.,
  eds,  Astronomical Society of the Pacific Conference Series Vol. 52,
  Astronomical Data Analysis Software and Systems II. p.~173

\bibitem[\protect\citeauthoryear{{Triaud} et~al.,}{{Triaud}
  et~al.}{2013}]{wasp30b2}
{Triaud} A.~H.~M.~J.,  et~al., 2013, \mn@doi [\aap]
  {10.1051/0004-6361/201219643}, \href
  {https://ui.adsabs.harvard.edu/abs/2013A&A...549A..18T} {549, A18}

\bibitem[\protect\citeauthoryear{{Triaud} et~al.,}{{Triaud}
  et~al.}{2017}]{triaud17}
{Triaud} A. H.~M.~J.,  et~al., 2017, \mn@doi [\aap]
  {10.1051/0004-6361/201730993}, \href
  {https://ui.adsabs.harvard.edu/abs/2017A&A...608A.129T} {608, A129}

\bibitem[\protect\citeauthoryear{{Triaud} et~al.,}{{Triaud}
  et~al.}{2020}]{triaud20}
{Triaud} A. H.~M.~J.,  et~al., 2020, \mn@doi [\nat]
  {10.1038/s41550-020-1018-2}, \href
  {https://ui.adsabs.harvard.edu/abs/2020NatAs...4..650T} {4, 650}

\bibitem[\protect\citeauthoryear{{Vines} \& {Jenkins}}{{Vines} \&
  {Jenkins}}{2022}]{ariadne}
{Vines} J.~I.,  {Jenkins} J.~S.,  2022, \mn@doi [\mnras]
  {10.1093/mnras/stac956}, \href
  {https://ui.adsabs.harvard.edu/abs/2022MNRAS.tmp..920V} {}

\bibitem[\protect\citeauthoryear{{Wheatley} et~al.,}{{Wheatley}
  et~al.}{2018}]{NGTS}
{Wheatley} P.~J.,  et~al., 2018, \mn@doi [\mnras] {10.1093/mnras/stx2836},
  \href {https://ui.adsabs.harvard.edu/abs/2018MNRAS.475.4476W} {475, 4476}

\bibitem[\protect\citeauthoryear{{Whitworth}}{{Whitworth}}{2018}]{whitworth18}
{Whitworth} A.,  2018, arXiv e-prints, \href
  {https://ui.adsabs.harvard.edu/abs/2018arXiv181106833W} {p. arXiv:1811.06833}

\bibitem[\protect\citeauthoryear{{Yee}, {Petigura}  \& {von Braun}}{{Yee}
  et~al.}{2017}]{specmatch}
{Yee} S.~W.,  {Petigura} E.~A.,   {von Braun} K.,  2017, \mn@doi [\apj]
  {10.3847/1538-4357/836/1/77}, \href
  {https://ui.adsabs.harvard.edu/abs/2017ApJ...836...77Y} {836, 77}

\bibitem[\protect\citeauthoryear{{Zahn} \& {Bouchet}}{{Zahn} \&
  {Bouchet}}{1989}]{zahn89}
{Zahn} J.~P.,  {Bouchet} L.,  1989, \aap, \href
  {https://ui.adsabs.harvard.edu/abs/1989A&A...223..112Z} {223, 112}

\bibitem[\protect\citeauthoryear{{Zhou} et~al.,}{{Zhou}
  et~al.}{2014}]{hats016b}
{Zhou} G.,  et~al., 2014, \mn@doi [\mnras] {10.1093/mnras/stt2100}, \href
  {https://ui.adsabs.harvard.edu/abs/2014MNRAS.437.2831Z} {437, 2831}

\bibitem[\protect\citeauthoryear{{Zhou} et~al.,}{{Zhou} et~al.}{2019}]{hats70b}
{Zhou} G.,  et~al., 2019, \mn@doi [\aj] {10.3847/1538-3881/aaf1bb}, \href
  {https://ui.adsabs.harvard.edu/abs/2019AJ....157...31Z} {157, 31}

\bibitem[\protect\citeauthoryear{{{\v{S}}ubjak} et~al.,}{{{\v{S}}ubjak}
  et~al.}{2020}]{toi503b}
{{\v{S}}ubjak} J.,  et~al., 2020, \mn@doi [\aj] {10.3847/1538-3881/ab7245},
  \href {https://ui.adsabs.harvard.edu/abs/2020AJ....159..151S} {159, 151}

\bibitem[\protect\citeauthoryear{{von Boetticher} et~al.,}{{von Boetticher}
  et~al.}{2017}]{j055557ab}
{von Boetticher} A.,  et~al., 2017, \mn@doi [\aap]
  {10.1051/0004-6361/201731107}, \href
  {https://ui.adsabs.harvard.edu/abs/2017A&A...604L...6V} {604, L6}

\bibitem[\protect\citeauthoryear{{von Boetticher} et~al.,}{{von Boetticher}
  et~al.}{2019}]{j0954}
{von Boetticher} A.,  et~al., 2019, \mn@doi [\aap]
  {10.1051/0004-6361/201834539}, \href
  {https://ui.adsabs.harvard.edu/abs/2019A&A...625A.150V} {625, A150}

\makeatother
\end{thebibliography}


\appendix
\section{Extra Data, figures and tables}\label{appendix:section}
\subsection{Full Data Tables}\label{appendix:tables}
\begin{table*}
\caption{Parameters of all objects used within the population analysis. RIK-72b \citep{rik72b} and the binary system discovered by \citet{stassun06} are not included due to their youth and large radii. Table has been adapted and updated from \citet{grieves21}. Sources: (1): \citet{hats70b}, (2): \citet{toi1278b}, (3): \citet{gpx1b}, (4): \citet{kepler39b}, (5): \citet{corot3b}, (6): \citet{kelt1b}, (7): \citet{beatty14_1b}, (8): \citet{nltt}, (9): \citet{NLTTrad}, (10): \citet{wasp128b}, (11): \citet{cww89ab1}, (12): \citet{carmichael19}, (13): \citet{koi205b}, (14):  \citet{carmichael20}, (15): \citet{epic212036875b1}, (16): \citet{toi503b}, (17): \citet{carmichael21}, (18): \citet{ad3116b}, (19): \citet{corot33b}, (20): \citet{toi263b1}, (21): \citet{toi263b2}, (22): \citet{koi415b}, (23): \citet{wasp30b2}, (24): \citet{lhs6343}, (25): \citet{corot15b}, (26): \citet{carmichael22}, (27): \citet{psaridi22}, (28): \citet{epic201702477b}, (29): \citet{lp261}, (30): \citet{19b}, (31): \citet{ngts7ab}, (32): \citet{grieves21}, (33): \citet{koi189b}, (34): \citet{j055557ab}, (35): \citet{j0954}, (36): \citet{ogle123b}, (37): \citet{toi694}, (38): \citet{tic320687387b},(39):\citet{ogle122b}, (40): \citet{k276}, (41): \citet{corot1011}, (42): \citet{j2343}, (43): \citet{hats016b}, (44): \citet{ogle106b}, (45): \citet{hat205}, (46): \citet{tic2310}, (47): \citet{kic1571}, (48): \citet{wts19g}.}             
\label{tab:full_obj_lists}      
\centering   
\begin{tabular}{>{\bfseries}l l l l l l l l l l}
\hline\hline                        
Object & \textbf{P [d]} & \textbf{M$_{2}$ [\mjup]} & \textbf{R$_{2}$ [\rjup]} & \textbf{\teff\ [K] } & \textbf{M$_{1}$ [\msun]} & \textbf{R$_{1}$ [\rsun]} & \textbf{ecc} & \textbf{[Fe/H]} & \textbf{Source}\\
\hline
HATS-70b&1.89&$12.9^{+1.8}_{-1.6}$&$1.38^{+0.08}_{-0.07}$&$7930^{+630}_{-820}$&$1.78^{+0.12}_{-0.12}$&$1.88^{+0.06}_{-0.07}$&<$0.18$&$0.04^{+0.10}_{-0.11}$& (1)\\
TOI-1278b&14.48&$18.5^{+0.5}_{-0.5}$&$1.09^{+0.24}_{-0.20}$&$3799^{+42}_{-42}$&$0.55^{+0.02}_{-0.02}$&$0.57^{+0.01}_{-0.01}$&$0.013^{+0.004}_{-0.004}$&$-0.01^{+0.28}_{-0.28}$&(2)\\
GPX-1b&1.74&$19.7^{+1.6}_{-1.6}$&$1.47^{+0.10}_{-0.10}$&$7000^{+200}_{-200}$&$1.68^{+0.10}_{-0.10}$&$1.56^{+0.10}_{-0.10}$&$0$ (fixed)&$0.35^{+0.10}_{-0.10}$&(3)\\
Kepler-39b&21.09&$20.1^{+1.3}_{-1.2}$&$1.24^{+0.09}_{-0.10}$&$6350^{+100}_{-100}$&$1.29^{+0.06}_{-0.07}$&$1.40^{+0.10}_{-0.10}$&$0.112^{+0.057}_{-0.057}$&$0.10^{+0.14}_{-0.14}$&(4)\\
CoRoT-3b&4.26&$21.7^{+1.0}_{-1.0}$&$1.01^{+0.07}_{-0.07}$&$6740^{+140}_{-140}$&$1.37^{+0.09}_{-0.09}$&$1.56^{+0.09}_{-0.09}$&$0$ (fixed) &$-0.02^{+0.06}_{-0.06}$ &(5)\\
KELT-1b&1.22&$27.4^{+0.9}_{-0.9}$&$1.12^{+0.04}_{-0.03}$&$6516^{+49}_{-49}$&$1.34^{+0.06}_{-0.06}$&$1.47^{+0.05}_{-0.04}$&$0.010^{+0.010}_{-0.007}$&$0.05^{+0.08}_{-0.08}$& (6)(7)\\
NLTT41135b&2.89&$33.7^{+2.8}_{-2.6}$&$1.13^{+0.27}_{-0.17}$&$3230^{+130}_{-130}$&$0.19^{+0.03}_{-0.02}$&$0.21^{+0.02}_{-0.01}$&$< 0.02$&$0$ (fixed)&(8)(9)\\
WASP-128b&2.21&$37.2^{+0.8}_{-0.9}$&$0.94^{+0.02}_{-0.02}$&$5950^{+50}_{-50}$&$1.16^{+0.04}_{-0.04}$&$1.15^{+0.02}_{-0.02}$&$<0.007$&$0.01^{+0.12}_{-0.12}$&(10)\\
CWW89Ab&5.29&$39.2^{+1.1}_{-1.1}$&$0.94^{+0.02}_{-0.02}$&$5755^{+49}_{-49}$&$1.10^{+0.05}_{-0.05}$&$1.03^{+0.02}_{-0.02}$&$0.189^{+0.002}_{-0.002}$&$0.20^{+0.09}_{-0.09}$ & (11)(12)\\
KOI-205b&11.72&$39.9^{+1.0}_{-1.0}$&$0.81^{+0.02}_{-0.02}$&$5237^{+60}_{-60}$&$0.93^{+0.03}_{-0.03}$&$0.84^{+0.02}_{-0.02}$&$<0.031$&$0.14^{+0.12}_{-0.12}$&(13)\\
TOI-1406b&10.57&$46.0^{+2.6}_{-2.7}$&$0.86^{+0.03}_{-0.03}$&$6290^{+100}_{-100}$&$1.18^{+0.08}_{-0.09}$&$1.35^{+0.03}_{-0.03}$&$0.026^{+0.013}_{-0.010}$&$-0.08^{+0.09}_{-0.09}$& (14)\\
EPIC212036875b&5.17&$52.3^{+1.9}_{-1.9}$&$0.87^{+0.02}_{-0.02}$&$6238^{+59}_{-60}$&$1.29^{+0.07}_{-0.06}$&$1.50^{+0.03}_{-0.03}$&$0.132^{+0.004}_{-0.004}$&$0.01^{+0.10}_{-0.10}$&(12)(15)\\
TOI-503b&3.68&$53.7^{+1.2}_{-1.2}$&$1.34^{+0.26}_{-0.15}$&$7650^{+140}_{-160}$&$1.80^{+0.06}_{-0.06}$&$1.70^{+0.05}_{-0.04}$&$0$ (fixed) &$0.30^{+0.08}_{-0.09}$&(16)\\
TOI-852b&4.95&$53.7^{+1.4}_{-1.3}$&$0.83^{+0.04}_{-0.04}$&$5768^{+84}_{-81}$&$1.32^{+0.05}_{-0.04}$&$1.71^{+0.04}_{-0.04}$&$0.004^{+0.004}_{-0.003}$&$0.33^{+0.09}_{-0.09}$&(17)\\
AD3116b&1.98&$54.2^{+4.3}_{-4.3}$&$1.02^{+0.28}_{-0.28}$&$3184^{+29}_{-29}$&$0.28^{+0.02}_{-0.02}$&$0.29^{+0.08}_{-0.08}$&$0.146^{+0.024}_{-0.016}$&$0$ (fixed)&(18)\\
CoRoT-33b&5.82&$59.0^{+1.8}_{-1.7}$&$1.10^{+0.53}_{-0.53}$&$5225^{+80}_{-80}$&$0.86^{+0.04}_{-0.04}$&$0.94^{+0.14}_{-0.08}$&$0.070^{+0.002}_{-0.002}$&$0.44^{+0.10}_{-0.10}$&(19)\\
TOI-811b&25.17&$59.9^{+13.0}_{-8.6}$&$1.26^{+0.06}_{-0.06}$&$6107^{+77}_{-77}$&$1.32^{+0.05}_{-0.07}$&$1.27^{+0.06}_{-0.09}$&$0.509^{+0.075}_{-0.075}$&$0.40^{+0.07}_{-0.09}$&(17)\\
TOI-263b&0.56&$61.6^{+4.0}_{-4.0}$&$0.91^{+0.07}_{-0.07}$&$3471^{+33}_{-33}$&$0.44^{+0.04}_{-0.04}$&$0.44^{+0.03}_{-0.03}$&$0.017^{+0.009}_{-0.010}$&$0.00^{+0.10}_{-0.10}$& (20)(21)\\
KOI-415b&166.79&$62.1^{+2.7}_{-2.7}$&$0.79^{+0.12}_{-0.07}$&$5810^{+80}_{-80}$&$0.94^{+0.06}_{-0.06}$&$1.25^{+0.15}_{-0.10}$&$0.698^{+0.002}_{-0.002}$&$-0.24^{+0.11}_{-0.11}$&(22)\\
WASP-30b&4.16&$62.5^{+1.2}_{-1.2}$&$0.95^{+0.03}_{-0.02}$&$6202^{+42}_{-51}$&$1.25^{+0.03}_{-0.04}$&$1.39^{+0.03}_{-0.03}$&$<0.004$&$0.08^{+0.07}_{-0.05}$&(23)\\
LHS6343c&12.71&$62.7^{+2.4}_{-2.4}$&$0.83^{+0.02}_{-0.02}$&$3130^{+20}_{-20}$&$0.37^{+0.01}_{-0.01}$&$0.38^{+0.01}_{-0.01}$&$0.056^{+0.032}_{-0.032}$&$0.04^{+0.08}_{-0.08}$&(24)\\
CoRoT-15b&3.06&$63.3^{+4.1}_{-4.1}$&$1.12^{+0.30}_{-0.15}$&$6350^{+200}_{-200}$&$1.32^{+0.12}_{-0.12}$&$1.46^{+0.31}_{-0.14}$&$0$ (fixed)&$0.10^{+0.20}_{-0.20}$&(25)\\
TOI-569b&6.56&$64.1^{+1.9}_{-1.4}$&$0.75^{+0.02}_{-0.02}$&$5768^{+110}_{-92}$&$1.21^{+0.05}_{-0.05}$&$1.48^{+0.03}_{-0.03}$&$0.002^{+0.002}_{-0.001}$&$0.29^{+0.09}_{-0.08}$&(14)\\
TOI-2119b&7.20&$64.4^{+2.3}_{-2.2}$&$1.08^{+0.03}_{-0.03}$&$3621^{+48}_{-46}$&$0.53^{+0.02}_{-0.02}$&$0.50^{+0.02}_{-0.02}$&$0.337^{+0.002}_{-0.001}$&$0.06^{+0.08}_{-0.08}$&(26)\\
TOI-1982b&17.17&$65.9^{+2.8}_{-2.7}$&$1.08^{+0.04}_{-0.04}$&$6325^{+110}_{-110}$&$1.41^{+0.08}_{-0.08}$&$1.51^{+0.05}_{-0.05}$&$0.272^{+0.014}_{-0.014}$&$-0.10^{+0.09}_{-0.09}$&(27)\\
NGTS-28Ab&1.25&$69.0^{+5.3}_{-4.8}$&$0.95\pm0.05$&$3626^{+47}_{-44}$&$0.56^{+0.02}_{-0.02}$&$0.59^{+0.03}_{-0.03}$&$0.040^{+0.007}_{-0.010}$&$-0.14^{+0.16}_{-0.17}$&This work\\
EPIC201702477b&40.74&$66.9^{+1.7}_{-1.7}$&$0.76^{+0.07}_{-0.07}$&$5517^{+70}_{-70}$&$0.87^{+0.03}_{-0.03}$&$0.90^{+0.06}_{-0.06}$&$0.228^{+0.003}_{-0.003}$&$-0.16^{+0.05}_{-0.05}$&(28)\\
TOI-629b&8.72&$67.0^{+3.0}_{-3.0}$&$1.11^{+0.05}_{-0.05}$&$9100^{+200}_{-200}$&$2.16^{+0.13}_{-0.13}$&$2.37^{+0.11}_{-0.11}$&$0.298^{+0.008}_{-0.008}$&$0.10^{+0.15}_{-0.15}$&(27)\\
TOI-2543b&7.54&$67.6^{+3.5}_{-3.5}$&$0.95^{+0.09}_{-0.09}$&$6060^{+82}_{-82}$&$1.29^{+0.08}_{-0.08}$&$1.86^{+0.15}_{-0.15}$&$0.009^{+0.003}_{-0.002}$&$-0.28^{+0.10}_{-0.10}$& (27)\\
LP261-75b&1.88&$68.1^{+2.1}_{-2.1}$&$0.90^{+0.01}_{-0.01}$&$3100^{+50}_{-50}$&$0.30^{+0.02}_{-0.02}$&$0.31^{+0.00}_{-0.00}$&$< 0.007$&-&(29)\\
NGTS-19b&17.84&$69.5^{+5.7}_{-5.4}$&$1.03^{+0.06}_{-0.05}$&$4716^{+39}_{-28}$&$0.81^{+0.04}_{-0.04}$&$0.90^{+0.04}_{-0.04}$&$0.377^{+0.006}_{-0.006}$&$0.11^{+0.07}_{-0.07}$&(30)\\
NGTS-7Ab&0.68&$75.5^{+3.0}_{-13.7}$&$1.38^{+0.13}_{-0.14}$&$3359^{+106}_{-89}$&$0.48^{+0.03}_{-0.12}$&$0.61^{+0.06}_{-0.06}$&$0$ (fixed)&$0$ (fixed)&(31)\\
TOI-148b&4.87&$77.1^{+5.8}_{-4.6}$&$0.81^{+0.05}_{-0.06}$&$5990^{+140}_{-140}$&$0.97^{+0.12}_{-0.09}$&$1.20^{+0.07}_{-0.07}$&$0.005^{+0.006}_{-0.004}$&$-0.24^{+0.25}_{-0.25}$&(32)\\
KOI-189b&30.36&$78.0^{+3.4}_{-3.4}$&$1.00^{+0.02}_{-0.02}$&$4952^{+40}_{-40}$&$0.76^{+0.05}_{-0.05}$&$0.73^{+0.02}_{-0.02}$&$0.275^{+0.004}_{-0.004}$&$-0.12^{+0.10}_{-0.10}$&(33)\\
TOI-587b&8.04&$81.1^{+7.1}_{-7.0}$&$1.32^{+0.07}_{-0.06}$&$9800^{+200}_{-200}$&$2.33^{+0.12}_{-0.12}$&$2.01^{+0.09}_{-0.09}$&$0.051^{+0.049}_{-0.036}$&$0.08^{+0.11}_{-0.12}$&(32)\\
TOI-746b&10.98&$82.2^{+4.9}_{-4.4}$&$0.95^{+0.09}_{-0.06}$&$5690^{+140}_{-140}$&$0.94^{+0.09}_{-0.08}$&$0.97^{+0.04}_{-0.03}$&$0.199^{+0.003}_{-0.003}$&$-0.02^{+0.23}_{-0.23}$&(32)\\
EBLM-J0555-57Ab&7.76&$87.9^{+4.0}_{-4.0}$&$0.82^{+0.13}_{-0.06}$&$6368^{+124}_{-124}$&$1.18^{+0.08}_{-0.08}$&$1.00^{+0.14}_{-0.07}$&$0.090^{+0.004}_{-0.004}$&$-0.04^{+0.14}_{-0.14}$&(34)(35)\\
TOI-681b&15.78&$88.7^{+2.5}_{-2.3}$&$1.52^{+0.25}_{-0.15}$&$7440^{+150}_{-140}$&$1.54^{+0.06}_{-0.05}$&$1.47^{+0.04}_{-0.04}$&$0.093^{+0.022}_{-0.019}$&$-0.08^{+0.05}_{-0.05}$&(32)\\
OGLE-TR-123b&1.80&$89.0^{+11.5}_{-11.5}$&$1.29^{+0.09}_{-0.09}$&$6700^{+300}_{-300}$&$1.29^{+0.26}_{-0.26}$&$1.55^{+0.10}_{-0.10}$&$0$ (fixed)&-&(36)\\
TOI-694b&48.05&$89.0^{+5.3}_{-5.3}$&$1.11^{+0.02}_{-0.02}$&$5496^{+87}_{-81}$&$0.97^{+0.05}_{-0.04}$&$1.00^{+0.01}_{-0.01}$&$0.519^{+0.001}_{-0.001}$&$0.21^{+0.08}_{-0.08}$&(37)\\
KOI-607b&5.89&$95.1^{+3.3}_{-3.4}$&$1.09^{+0.09}_{-0.06}$&$5418^{+87}_{-85}$&$0.99^{+0.05}_{-0.05}$&$0.92^{+0.03}_{-0.03}$&$0.395^{+0.009}_{-0.009}$&$0.38^{+0.08}_{-0.09}$& (12)\\
J1219-39b&6.76&$95.4^{+1.9}_{-2.5}$&$1.14^{+0.07}_{-0.05}$&$5412^{+81}_{-65}$&$0.83^{+0.03}_{-0.03}$&$0.81^{+0.04}_{-0.02}$&$0.055^{+0.000}_{-0.000}$&$-0.21^{+0.07}_{-0.08}$&(23)\\
TIC-320687387 B&29.77&$96.2^{+1.9}_{-2.0}$&$1.14^{+0.02}_{-0.02}$&$5780^{+80}_{-80}$&$1.08^{+0.03}_{-0.03}$&$1.16^{+0.02}_{-0.02}$&$0.366^{+0.003}_{-0.003}$&$0.30^{+0.08}_{-0.08}$&(38)\\
OGLE-TR-122b&7.27&$96.4^{+9.4}_{-9.4}$&$1.17^{+0.23}_{-0.13}$&$5700^{+300}_{-300}$&$0.98^{+0.14}_{-0.14}$&$1.05^{+0.20}_{-0.09}$&$0.205^{+0.008}_{-0.008}$&$0.15^{+0.36}_{-0.36}$&(39)\\
TOI-1213b&27.22&$97.5^{+4.4}_{-4.2}$&$1.66^{+0.78}_{-0.55}$&$5590^{+150}_{-150}$&$0.99^{+0.07}_{-0.06}$&$0.99^{+0.04}_{-0.04}$&$0.498^{+0.003}_{-0.002}$&$0.25^{+0.13}_{-0.14}$&(32)\\
K2-76b&11.99&$98.7^{+2.0}_{-2.0}$&$0.89^{+0.05}_{-0.03}$&$5747^{+70}_{-64}$&$0.96^{+0.03}_{-0.03}$&$1.17^{+0.06}_{-0.03}$&$0.255^{+0.007}_{-0.007}$&$0.01^{+0.04}_{-0.04}$&(40)\\
CoRoT-101186644&20.68&$100.6^{+11.5}_{-11.5}$&$1.01^{+0.25}_{-0.06}$&$6090^{+200}_{-200}$&$1.20^{+0.20}_{-0.20}$&$1.07^{+0.07}_{-0.07}$&$0.402^{+0.006}_{-0.006}$&$0.20^{+0.20}_{-0.20}$&(41)\\
J2343+29Ab&16.95&$102.7^{+7.3}_{-7.3}$&$1.24^{+0.07}_{-0.07}$&$5150^{+90}_{-60}$&$0.86^{+0.10}_{-0.10}$&$0.85^{+0.05}_{-0.06}$&$0.161^{+0.002}_{-0.003}$&$0.07^{+0.01}_{-0.17}$&(42)\\
EBLM-J0954-23Ab&7.57&$102.8^{+5.9}_{-6.0}$&$0.98^{+0.17}_{-0.17}$&$6406^{+124}_{-124}$&$1.17^{+0.08}_{-0.08}$&$1.23^{+0.17}_{-0.17}$&$0.042^{+0.001}_{-0.001}$&$-0.01^{+0.14}_{-0.14}$&(35)\\
KOI-686b&52.51&$103.4^{+4.8}_{-4.8}$&$1.22^{+0.04}_{-0.04}$&$5834^{+100}_{-100}$&$0.98^{+0.07}_{-0.07}$&$1.04^{+0.03}_{-0.03}$&$0.556^{+0.004}_{-0.004}$&$-0.06^{+0.13}_{-0.13}$&(33)\\
TIC-220568520b&18.56&$107.2^{+5.2}_{-5.2}$&$1.25^{+0.02}_{-0.02}$&$5589^{+81}_{-81}$&$1.03^{+0.04}_{-0.04}$&$1.01^{+0.01}_{-0.01}$&$0.096^{+0.003}_{-0.003}$&$0.26^{+0.07}_{-0.07}$&(37)\\
\multicolumn{10}{l}{}
\end{tabular}
\end{table*}

\begin{table*}
\centering
\contcaption{}
\begin{tabular}{>{\bfseries}l l l l l l l l l l}
HATS550-016B&2.05&$115.2^{+5.2}_{-6.3}$&$1.43^{+0.03}_{-0.04}$&$6420^{+90}_{-90}$&$0.97^{+0.05}_{-0.06}$&$1.22^{+0.02}_{-0.03}$&$0.080^{+0.020}_{-0.020}$&$-0.60^{+0.06}_{-0.06}$&(43)\\
OGLE-TR-106b&2.54&$121.5^{+22.0}_{-22.0}$&$1.76^{+0.13}_{-0.13}$&-&-&$1.31^{+0.09}_{-0.09}$&$0.000^{+0.020}_{-0.020}$&-&(44)\\
EBLM-J1431-11Ab&4.45&$126.9^{+3.8}_{-3.9}$&$1.45^{+0.07}_{-0.05}$&$6161^{+124}_{-124}$&$1.20^{+0.06}_{-0.06}$&$1.11^{+0.04}_{-0.03}$&$0$ (fixed)&$0.15^{+0.14}_{-0.14}$&(35)\\
HAT-TR-205-013B&2.23&$129.9^{+10.5}_{-10.5}$&$1.63^{+0.06}_{-0.06}$&$6295^{+200}_{-200}$&$1.04^{+0.13}_{-0.13}$&$1.28^{+0.04}_{-0.04}$&$0.012^{+0.021}_{-0.021}$&-&(45)\\
TIC-231005575b&61.78&$134.1^{+3.1}_{-3.1}$&$1.50^{+0.08}_{-0.08}$&$5500^{+85}_{-85}$&$1.05^{+0.04}_{-0.04}$&$0.99^{+0.05}_{-0.05}$&$0.298^{+0.001}_{-0.004}$&$-0.44^{+0.06}_{-0.06}$&(46)\\
HATS551-021B&3.64&$138.3^{+14.7}_{-5.2}$&$1.50^{+0.06}_{-0.08}$&$6670^{+220}_{-220}$&$1.10^{+0.10}_{-0.10}$&$1.20^{+0.08}_{-0.01}$&$0.060^{+0.020}_{-0.020}$&$-0.40^{+0.10}_{-0.10}$&(43)\\
EBLM-J2017+02Ab&0.82&$142.2^{+6.6}_{-6.7}$&$1.49^{+0.13}_{-0.10}$&$6161^{+124}_{-124}$&$1.11^{+0.07}_{-0.07}$&$1.20^{+0.08}_{-0.05}$&$0$ (fixed)&$-0.07^{+0.14}_{-0.14}$&(35)\\
KIC-1571511B&14.02&$148.1^{+0.5}_{-0.4}$&$1.74^{+0.00}_{-0.01}$&$6195^{+50}_{-50}$&$1.27^{+0.04}_{-0.03}$&$1.34^{+0.01}_{-0.01}$&$0.327^{+0.003}_{-0.003}$&$0.37^{+0.08}_{-0.08}$&(47)\\
WTS-19G-4-02069B&2.44&$149.8^{+6.3}_{-6.3}$&$1.69^{+0.06}_{-0.06}$&$3300^{+140}_{-140}$&$0.53^{+0.02}_{-0.02}$&$0.51^{+0.01}_{-0.01}$&$0$ (fixed)&-&(48)\\
\hline
\end{tabular}
\label{tab:full_objs_lists_cont}
\end{table*}

\subsection{Corner Plots}\label{appendix:cplots}

\begin{figure*}
    \centering
    \includegraphics[width=\linewidth, trim = 0cm 0cm 0cm 0.1cm, clip]{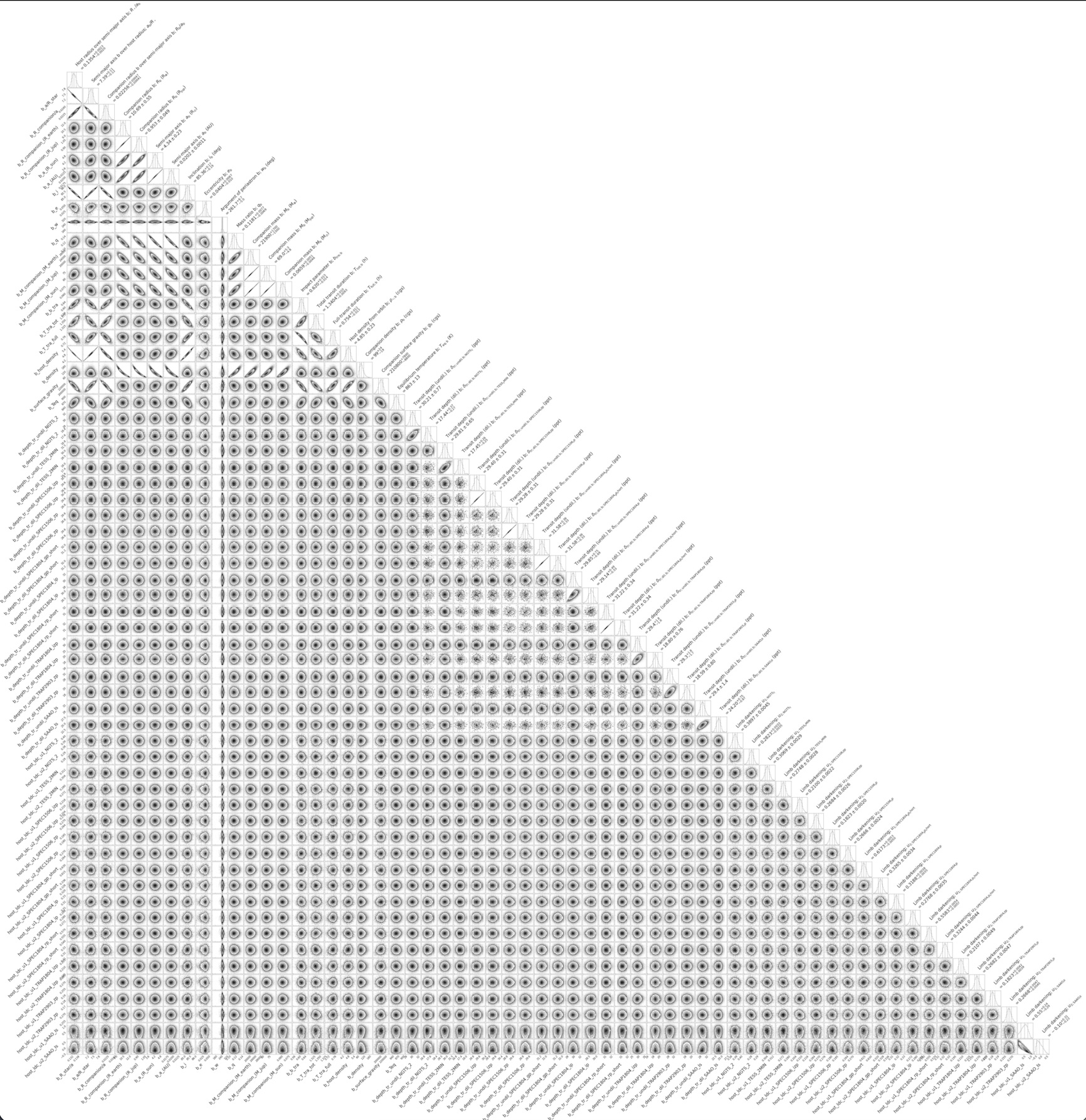}
    \caption{The corner plot for the derived values in the final fit.}
\label{fig:derived_corner}
\end{figure*}

\subsection{Brown Dwarf Desert Population Plots}\label{appendix:bdplots}

\begin{figure}
    \centering
    \includegraphics[width=\linewidth]{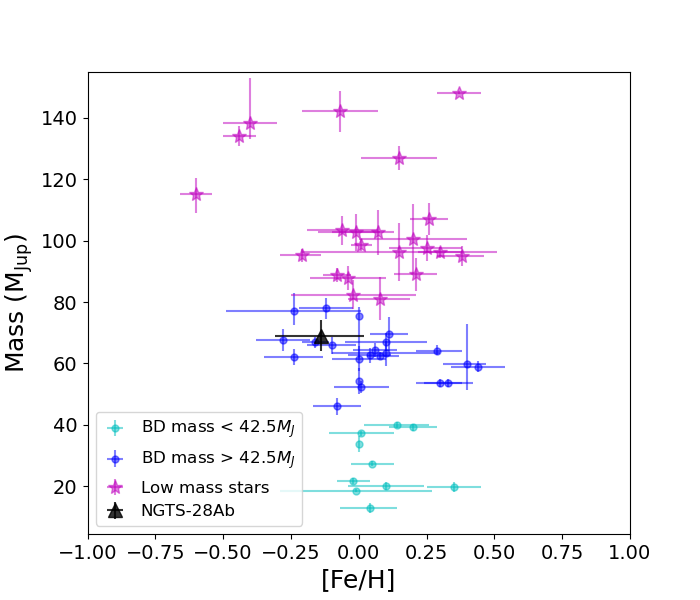}
    \caption{Mass-Metallicity plot of objects in \ref{tab:full_obj_lists}. The metallicity is for the host star in each system. \systemt\ is plotted as a black triangle. Low mass stars are plotted as magenta stars. BDs with masses above 42.5 \mjup\ are dark blue circles and BDs with below 42.5 \mjup\ are light blue circles. RIK-72b \citep{rik72b} and the binary system discovered by \citet{stassun06} are not included due to their youth.}
\label{fig:massfeh}
\end{figure}

\begin{figure}
    \centering
    \includegraphics[width=\linewidth]{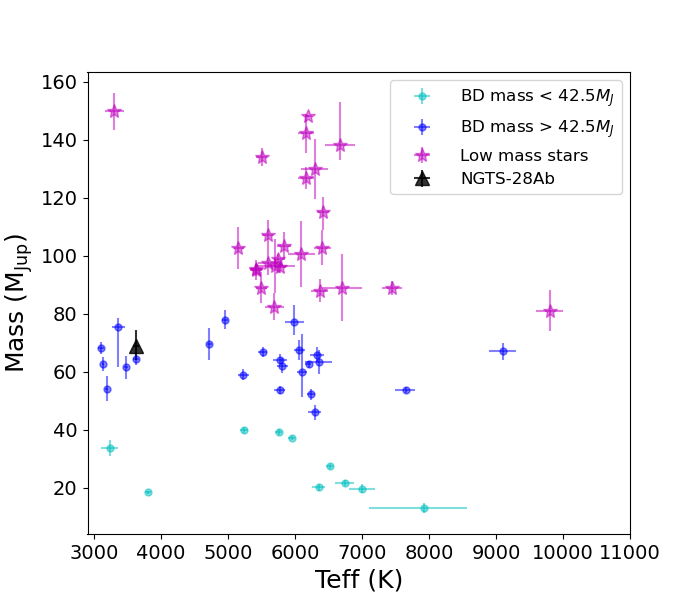}
    \caption{Mass-\teff\ plot of objects in \ref{tab:full_obj_lists}. The \teff\ values are for the host star in each system. \systemt\ is plotted as a black triangle. Low mass stars are plotted as magenta stars. BDs with masses above 42.5 \mjup\ are dark blue circles and BDs with below 42.5 \mjup\ are light blue circles. RIK-72b \citep{rik72b} and the binary system discovered by \citet{stassun06} are not included due to their youth.}
\label{fig:massteff}
\end{figure}

\bsp	
\label{lastpage}
\end{document}